\documentclass[journal=jacsat,manuscript=article]{achemso}
\usepackage{graphicx}
\usepackage{subfigure}
\usepackage{booktabs}

\usepackage{dcolumn}
\usepackage{enumitem} 

\usepackage{bm}
\usepackage{mathrsfs}
\usepackage{amsmath}
\usepackage{mathtools}
\usepackage{physics}
\usepackage{braket}

\usepackage{caption}
\usepackage{subcaption}
\usepackage{tablefootnote}
\usepackage[font={footnotesize}]{caption}

\usepackage[hidelinks]{hyperref}
\usepackage{cleveref}

\usepackage[]{xcolor}
\usepackage{soul}

\usepackage{pdfpages}

\author{Yassir El Moutaoukal}
\affiliation{Department of Chemistry, Norwegian University of Science and Technology, 7491 Trondheim, Norway}
\author{Rosario R. Riso}
\affiliation{Department of Chemistry, Norwegian University of Science and Technology, 7491 Trondheim, Norway}
\author{Matteo Castagnola}
\affiliation{Department of Chemistry, Norwegian University of Science and Technology, 7491 Trondheim, Norway}
\author{Henrik Koch}
\affiliation{Department of Chemistry, Norwegian University of Science and Technology, 7491 Trondheim, Norway}
\email{henrik.koch@ntnu.no}
\title[]
  {Towards polaritonic molecular orbitals for large molecular systems}
\abbreviations{ab-initio QED, SC-QED-HF}
\begin{document}
\begin{abstract}
A comprehensive theoretical understanding of electron-photon correlation is essential for describing the reshaping of molecular orbitals in quantum electrodynamics (QED) environments.
The strong coupling QED Hartree-Fock (SC-QED-HF) theory tackles these aspects by providing consistent molecular orbitals in the strong coupling regime. The previous implementation, however, has significant convergence issues that limit the applicability. In this work we introduce two second-order algorithms that significantly reduce the computational requirements, thereby enhancing the modeling of large molecular systems in QED environments. Furthermore, the implementation will enable the development of correlated methods based on a reliable molecular orbital framework.
\end{abstract}

\section{1. Introduction}
Strong coupling between electromagnetic vacuum fluctuations and molecular systems has been suggested to be an innovative way to non-invasively engineer molecular properties\cite{Ebbesen_1,wang2019cavity,singlet_fission,Ramezani:17,coles2014polariton}.
To effectively achieve light-matter strong coupling, different optical devices able to spatially confine electromagnetic fields have been designed\cite{Bragg,whispering_2,graphene_cavity,plasmon_cavity,chiral_cavity_0}. 
The strong coupling regime is unlocked once the coherent energy exchange rate between the electromagnetic field and the molecular system exceeds the dissipation processes\cite{strong_coupling,strong_coupling_2,strong_coupling_3}.
This interaction leads to the formation of polaritons, which mix photonic and molecular degrees of freedom\cite{Herrera_Spano,ebbesen2016hybrid,molecular_polaritons,molecular_polaritons_2}.

While new experimental studies keep increasing the range of possible applications\cite{ahn2023modification,biswas2024electronic,kumar2023extraordinary,li2023strong,balasubrahmaniyam2023enhanced}, a complete rationalization of the mechanisms behind these modifications is missing, underlying the pressing need for theoretical insight into the complex interplay between light and matter\cite{fregoni2022theoretical}.
In this regard, \textit{ab initio} methods that model the underlying physical processes starting from wave functions are of the utmost importance to faithfully reproduce the molecular features of the polaritons. 
The electromagnetic fields and matter degrees of freedom must be treated on the same footing by means of quantum electrodynamics (QED) theory\cite{ashida2021cavity}. 
Several \textit{ab initio} models have been proposed in the last few years to capture electron-photon correlation while keeping a polynomial scaling describing the overall complexity.
Most of the well-established quantum chemical approaches have been extended. More specifically Hartree-Fock\cite{QEDCC} (QED-HF), density functional theory\cite{QEDDFT,QEDDFT1} (QEDFT), as well as coupled cluster\cite{QEDCC,liebenthal2022equation,mordovina2020polaritonic,pavosevic2021polaritonic} (QED-CC), full configuration interaction\cite{QEDCC} (QED-CI), complete active space configuration interaction\cite{vu2024cavity} (QED-CASCI) and Møller–Plesset second order perturbation theory\cite{bauer2023perturbation} (QED-MP2). 
However, one also encounter instances where the extension of the quantum chemical concepts is nontrivial, like in the case of polaritonic molecular orbitals.

Molecular orbitals are powerful theoretical tools able to provide a description of molecular properties, for instance, the rationalization of stereoselectivity in chemical reactions.  
In quantum chemistry, the molecular orbitals are obtained by solving the Hartree-Fock method. 
However, the orbitals obtained from the straightforward generalization of Hartree-Fock to cavity environments (QED-HF) has unphysical features, notably they do not display correct intermolecular consistency and they are not origin invariant for charged systems.
A polaritonic molecular orbital theory is necessary to address these issues and provide a more accurate description of the molecular behavior under strong light-matter coupling. Several groundbreaking works have indeed demonstrated that strong light-matter coupling can change both the ground and excited state reactivity, altering the reaction kinetics \cite{baranov}, changing reactive yields, and even affecting the selectivity toward a particular product\cite{thomas2016ground}.

Recently, Riso \textit{et al.}\cite{SCQEDHF} presented the strong coupling QED Hartree-Fock (SC-QED-HF) model, the first fully consistent molecular orbital theory for QED environments. 
The approach is very promising not only because it can be used to rationalize how molecular orbitals are reshaped, but also because it can represent a valuable reference for the development of more accurate correlated approaches. 
Despite its potential, the first implementation in $e^\mathcal{T}$ program\cite{eT} has convergence difficulties that restrict its applicability.
These numerical limitations find their roots in the multicomposite nature of the wave function parametrization, which includes two classes of parameters, one accounting for the orbitals optimization and one for the electron-photon interaction.
Simultaneous optimization of these two physically different variables negatively affects the convergence.
In this work, we tackle this issue by developing two second-order algorithms. 
The new algorithms significantly speed up convergence.
These results pave the way for developing correlated methodologies and significantly increase the application range for large molecular systems.

This paper is organized as follows: in Section 2, we present a brief overview of the SC-QED-HF theory, highlighting differences between the previous implementation and the new algorithms.
In Section 3, we demonstrate the improved convergence with a set of benchmark molecules that includes organic as well as inorganic ones. 
Thereafter, we discuss the computational scaling of the improved methodology.
In the last Section we present our conclusions and future perspectives.

\section{2. Theory}
In this work, the light-matter interaction inside a cavity with quantization volume $V$ is modeled using the Pauli-Fierz Hamiltonian in the length gauge and dipole approximation, where only one effective cavity mode is considered\cite{craig1998molecular,castagnola2024polaritonic, ruggenthaler2023understanding}
\begin{equation} \label{Pauli-Fierz Hamiltonian}
    \begin{split}
        H & = \sum_{pq} h_{pq} E_{pq} + \frac{1}{2} \sum_{pqrs}g_{pqrs}e_{pqrs} + \omega  b^\dagger b  \\
        & + \frac{\lambda^2}{2}  ({\mathbf{d}}  \cdot \pmb{\epsilon} )^2 -\lambda \sqrt{\frac{\omega}{2}} ( {\mathbf{d}} \cdot \pmb{\epsilon}  ) ( b^{\dagger}+b ) .
    \end{split}
\end{equation}
In \cref{Pauli-Fierz Hamiltonian}, the bosonic operators $b^\dagger$ and $b$ respectively create and annihilate a photonic mode of the cavity with frequency $\omega$. The light-matter interaction is mediated through the photonic-bilinear term where $\pmb{\epsilon}$ is the polarization vector of the field, $\lambda$ is the coupling strength 
\begin{equation} \label{lambda coupling strength}
   \lambda \propto \sqrt{\frac{1}{V}} ,
\end{equation}
while ${\mathbf{d}}$ is the molecular dipole operator defined as 
\begin{equation}\label{dipole operator}
{\mathbf{d}} = \sum_{pq} \mathbf{d}_{pq} E_{pq} = \sum_{pq} \left(\mathbf{d}^{e}_{pq}+\frac{\mathbf{d}^{nuc}}{N_{e}}\delta_{pq}\right) E_{pq} .
\end{equation}
with $\mathbf{d}^{e}$ being the electronic dipole and $\mathbf{d}^{nuc}$ the nuclear dipole of a system of $N_{e}$ electrons.
The electronic operators $E_{pq}$ and $e_{pqrs}$ are given by
\begin{equation}
E_{pq}=\sum_{\sigma}a^{\dagger}_{p\sigma}a_{q\sigma}
\end{equation}
\begin{equation} 
e_{pqrs}=E_{pq}E_{rs}-\delta_{rq}E_{ps} ,
\end{equation}
where $a^{\dagger}_{p\sigma}$ and $a_{p\sigma}$ are the creation and annihilation operators for an electron in orbital $p$ and spin $\sigma$.
Finally, $\mathbf{d}_{pq}^e$, $h_{pq}$ and $g_{pqrs}$ are the one and two electron integrals that enter in the Pauli-Fierz Hamiltonian.
We note that the dipole self-energy (DSE) term ensures the Hamiltonian in \cref{Pauli-Fierz Hamiltonian} is bounded from below\cite{rokaj2018light}. For simplicity, in the remaining part of this work, the symbol $\sim$ denotes integrals and operators in the basis that diagonalizes $( \mathbf{d}\cdot\pmb{\epsilon})$:
\begin{equation}
    \sum_{rs}V_{rp}( \mathbf{d}\cdot\pmb{\epsilon})_{rs} V_{sq} = (\tilde{\mathbf{d}}\cdot\pmb{\epsilon})_{pp}\delta_{pq} ,
\end{equation}
\begin{equation}
    \Tilde{E}_{pq} = \sum_{rs}V_{pr}E_{rs} V_{qs} 
\end{equation}
where $\mathbf{V}$ is an orthogonal matrix.
The dipole basis is particularly suitable in the strong coupling regime as Slater determinants in that specific basis are the exact eigenstates for the Pauli-Fierz Hamiltonian in the infinite coupling limit.

\subsection{Strong coupling QED Hartree-Fock}
The SC-QED-HF method is the first QED \textit{ab-initio} framework able to provide origin independent molecular orbitals in a non-perturbative treatment that capture cavity frequency dispersion as well as being intermolecular consistent\cite{haugland2021intermolecular,SC-MP2}. 
In this approach, the wave function reads
\begin{equation} \label{SC-QED-HF Ansatz}
    \ket{\psi_{\mathrm{SC}}} = U_{\mathrm{SC}} \exp(\kappa) \prod_{i,\sigma}^{n_{occ}} a_{i\sigma}^\dagger \ket{vac} \otimes \ket{0} ,
\end{equation}
where $U_{SC}$ is the strong coupling transformation 
\begin{equation}
    U_{\mathrm{SC}} = \exp(-\frac{\lambda}{\sqrt{2\omega}} \sum_p \eta_p \Tilde{E}_{pp} ( b - b^\dagger )) ,
\end{equation}
where $\{\eta_p\}$ are orbital specific coherent state parameters. The electronic and photonic vaqua are referred respectively to as $\ket{vac}$ and $\ket{0}$.
The wave function in \cref{SC-QED-HF Ansatz} becomes increasingly accurate as $\lambda\rightarrow\infty$ because it is obtained by relaxing the infinite coupling solution to a finite strength (see Supporting information for a detailed derivation).
The molecular orbitals are optimized through a unitary transformation\cite{thouless1960stability,helgaker2013molecular} $\exp(\kappa)$, where
\begin{equation}\label{kappa}
    {\kappa} = \sum_{ai} \kappa_{ai}E_{ai}^- \ \ \ \ \ E_{ai}^-=( E_{ai} - E_{ai}) 
\end{equation}
and $a$ and $i$ denote virtual and occupied MOs.
Unlike the uncorrelated QED-HF model, the SC-QED-HF theory incorporates electron-photon correlation by dressing the electronic molecular orbitals with the photonic degrees of freedom as seen from
\begin{equation}
    \exp(-\kappa) U^\dagger_{\mathrm{SC}} a_{p\sigma}^{\dagger} U_{\mathrm{SC}} \exp(\kappa)= \sum_q a_{q\sigma}^{\dagger} \exp(\frac{\lambda \eta_q}{\sqrt{2\omega}} ( b - b^\dagger )) \exp(\pmb{\kappa})_{qp}.
\end{equation}
In the dipole basis, the Pauli-Fierz Hamiltonian can be written as
\begin{equation}\label{SC Hamiltonian}
    {H} = {H}_{e} + \omega \Big(b^\dagger-\frac{\lambda}{\sqrt{2\omega}}\sum_p(\tilde{\mathbf{d}}\cdot\pmb{\epsilon})_{pp}\tilde{E}_{pp}\Big)\Big(b-\frac{\lambda}{\sqrt{2\omega}}\sum_p(\tilde{\mathbf{d}}\cdot\pmb{\epsilon})_{pp}\tilde{E}_{pp}\Big) 
\end{equation}
and transforming it with $U_{SC}$, we obtain
\begin{equation}\label{SC Hamiltoninan}
    \begin{split}
    {H}_{\mathrm{SC}} &= U^\dagger_{\mathrm{SC}} H U_{\mathrm{SC}} \\
    & = \sum_{pq} \tilde{h}_{pq} \tilde{E}_{pq}\exp (\frac{\lambda}{\sqrt{2\omega}}(\eta_p-\eta_q)(b-b^\dagger)) \\
    &+ \frac{1}{2}\sum_{pqrs}\tilde{g}_{pqrs}\tilde{e}_{pqrs}\exp(\frac{\lambda}{\sqrt{2\omega}}(\eta_p+\eta_r-\eta_q-\eta_s)(b-b^\dagger)) \\
    & + \omega \Big(b^\dagger-\frac{\lambda}{\sqrt{2\omega}}\sum_p((\tilde{\mathbf{d}}\cdot\pmb{\epsilon})_{pp} - \eta_p)\tilde{E}_{pp}\Big)\Big(b-\frac{\lambda}{\sqrt{2\omega}}\sum_p((\tilde{\mathbf{d}}\cdot\pmb{\epsilon})_{pp} - \eta_p)\tilde{E}_{pp}\Big) .
    \end{split}
\end{equation}
This Hamiltonian differs from the Pauli-Fierz operator in \cref{SC Hamiltonian} by the $\eta$-shifting of the dipole integrals and the photonic dressing of the electronic terms.
The optimal wave function is determined by energy minimization using the gradients with respect to the parameters:
\begin{equation}
    \mathbf{E}^{\left(1\right)} = \begin{pmatrix} 
     \partial E / \partial \pmb{\kappa} \\
    {\partial E} / {\partial \pmb{\eta}}
    \end{pmatrix} \equiv \begin{pmatrix} 
     \mathbf{f}^\kappa \\
    \mathbf{f}^\eta 
    \end{pmatrix} ,
\end{equation}
where $E=\bra{\psi_{\mathrm{SC}}}H\ket{\psi_{\mathrm{SC}}}$ and $\braket{\psi_{\mathrm{SC}}|\psi_{\mathrm{SC}}}= 1$.
\\
\\
Now we obtain the gradient with respect to $\eta_r$ as
\begin{equation}\label{eta gradient}
    \begin{split}
        \Tilde{f}^{\eta}_{r} = \frac{\partial E}{\partial \eta_r} & = \frac{\lambda^2}{\omega} \sum_q {\tilde{h}}^a_{rq}\Tilde{D}_{rq} ( \eta_q - \eta_r ) - \lambda^2 \tilde{D}_{rr} ( ( \tilde{\mathbf{d}}\cdot \pmb{\epsilon} )_{rr} - \eta_r ) \\
        & + \frac{\lambda^2}{\omega} \sum_{pqt}{\tilde{g}}^a_{rpqt}\Tilde{d}_{rpqt} (  \eta_p + \eta_t - \eta_r - \eta_q) \\
        & - \lambda^2 \sum_q \Tilde{d}_{qqrr} ( ( \tilde{\mathbf{d}} \cdot \pmb{\epsilon} )_{qq} - \eta_q ) ,
    \end{split}
\end{equation}
where
\begin{equation}
    {\tilde{h}}^a_{pq} = \tilde{h}_{pq}\exp(-\frac{\lambda^2}{4\omega}(\eta_p - \eta_q)^2) ,
\end{equation}
\begin{equation}
    {\tilde{g}}^a_{pqrs} = \tilde{g}_{pqrs}\exp(-\frac{\lambda^2}{4\omega}(\eta_p+\eta_r - \eta_q-\eta_s)^2) ,
\end{equation}
are the one and two electron integrals scaled by the $\omega$-dependent Gaussian factors. The density matrix elements are given by $\tilde{D}_{pq} = \bra{\mathrm{HF}}\Tilde{E}_{pq}\ket{\mathrm{HF}}$ and $\tilde{d}_{pqrs} = \bra{\mathrm{HF}}\Tilde{e}_{pqrs}\ket{\mathrm{HF}}$.
The gradient with respect to $\tilde{D}_{pq}$ equals the SC-QED Fock matrix element in the dipole basis
\begin{equation}
    \begin{split}
        \tilde{F}_{pq}=\frac{\partial E}{\partial\tilde{D}_{pq}} & = {\tilde{h}}^a_{pq}  + \frac{1}{2} \sum_{rs} ( 2 {\tilde{g}}^a_{pqrs} - {\tilde{g}}^a_{psrq} ) \tilde{D}_{rs} + \frac{\lambda^2}{2} \delta_{pq} ( ( \tilde{\mathbf{d}} \cdot \pmb{\epsilon} )_{pp} - \eta_p )^2 \\
        & + \lambda^2 \delta_{pq}  ( ( \tilde{\mathbf{d}} \cdot \pmb{\epsilon} )_{pp} - \eta_p ) \sum_r \tilde{D}_{rr}  ( ( \tilde{\mathbf{d}} \cdot \pmb{\epsilon} )_{rr} - \eta_r ) \\
        &  - \frac{\lambda^2}{2} \tilde{D}_{qp}  ( ( \tilde{\mathbf{d}} \cdot \pmb{\epsilon} )_{pp} - \eta_p ) ( ( \tilde{\mathbf{d}} \cdot \pmb{\epsilon} )_{qq} - \eta_q ) ,
    \end{split}
\end{equation}
which is related to the non-redundant Hartree-Fock gradient in the canonical basis:
\begin{equation}\label{kappa gradient}
    f^{\kappa}_{ai} = \left(\frac{\partial E}{\partial \kappa_{ai}}\right)_{\pmb{\kappa} = \textbf{0}} = \bra{\mathrm{HF}} \big[ \braket{{H}_{\mathrm{SC}}}_0 , E^-_{ai}\big] \ket{\mathrm{HF}} = 2F_{ai} 
\end{equation}
where $\braket{{H}_{\mathrm{SC}}}_0$ is the vacuum averaged SC-transformed Hamiltonian and
\begin{equation}
    F_{pq} = \sum_{rs} V_{pr} \Tilde{F}_{rs} V_{qs} .
\end{equation}
The origin dependence of the QED-HF orbitals stems from the changes that a displacement $\mathbf{a}$ has on the dipole operator of a charged molecule
\begin{equation} \label{dipole shift}
    ( \mathbf{d} \cdot \pmb{\epsilon} )_{pq} \rightarrow  ( \mathbf{d} \cdot \pmb{\epsilon} )_{pq} + \frac{Q_{tot}}{N_{e}} ( \mathbf{a} \cdot \pmb{\epsilon} ) \delta_{pq} ,
\end{equation}
where $Q_{tot}$ is the total charge of the system. 
For SC-QED-HF, this change in the molecular dipole can be reabsorbed through an appropriate shift in the $\eta$ parameters, leading to an origin invariant Fock matrix and thus orbitals.\cite{SCQEDHF}

In the previous implementation, the orbital optimization is performed using the Roothaan-Hall self-consisted field (SCF) procedure\cite{scf1,Cancès2000} where diagonalization of the Fock matrix is performed coupled with the direct inversion in the iterative subspace (DIIS) algorithm\cite{diis1,diis2}. 
Meanwhile, the $\eta$-parameters were updated in a steepest descent fashion accelerated by the DIIS.
The numerical difficulties of the previous implementation stem from the inability to provide a proper preconditioner in the $\eta$ update. 
Specifically, optimizing the density matrix exhibits similar behavior as standard Hartree-Fock, while the gradient in the $\eta$-parameters fail to predict a reliable convergence path. A partial solution to this issue is performing very small steps in the $\{\eta_p$\}, but this significantly increases the calculation time. 

It is well-known in numerical optimization that more reliable convergence paths can be found using higher derivatives\cite{fletcher2000practical}. In the next two sections, we present two new optimization schemes exploiting parts of the Hessian matrix to improve convergence stability and speed.

\subsection{Trust region Newton-Raphson optimization}
The construction of the Hessian matrix allows us to understand how closely different parameters are interrelated by capturing the curvature of the parameter hypersurface. 
Using the second-order derivatives, a Newton-Raphson type algorithm can be developed\cite{bacskay1981quadratically}.
Since the wave function is composed of two different classes of parameters, $\kappa$ and $\eta$, the Hessian matrix contains four different blocks
\begin{equation}
    \mathbf{E}^{\left(2\right)} = \begin{pmatrix} \mathbf{E}^{\kappa\kappa} & \mathbf{E}^{{\kappa\eta}} \\
    \mathbf{E}^{{\eta\kappa}} & \mathbf{E}^{{\eta\eta}}
    \end{pmatrix} =
    \begin{pmatrix} \frac{\partial^2E}{\partial \kappa_{ai}\partial\kappa_{bj}} & \frac{\partial^2E}{\partial \kappa_{ai}\partial\eta_{r}} \\
    \frac{\partial^2E}{\partial\eta_{r}\partial \kappa_{ai}} & \frac{\partial^2E}{\partial \eta_{r}\partial\eta_{s}}
    \end{pmatrix} .
\end{equation}
The individual blocks are obtained from the following expression
\begin{equation}\label{k-k Hessian 1}
    {E}^{\kappa\kappa}_{ai,bj} = ( 1 + P_{ai,bj} ) \bra{\mathrm{HF}} \big[ \big[ \braket{{H}_{\mathrm{SC}}}_0  , E_{ai} \big] , E^-_{bj}\big] \ket{\mathrm{HF}} ,
\end{equation}
\begin{equation}\label{eta-eta Hessian 1}
    {E}^{{\eta\eta}}_{r,s} = \frac{\lambda^2}{2\omega} \bra{\mathrm{HF},0} \big[ \tilde{E}_{ss}(b-b^\dagger),\big[ \tilde{E}_{rr}(b-b^\dagger)  , {H}_{\mathrm{SC}} \big] \big] \ket{\mathrm{HF},0} ,
\end{equation}
\begin{equation}\label{kappa-eta Hessian 1}
    {E}^{{\kappa\eta}}_{ai,r} =   {E}^{{\eta\kappa}}_{r,ai} = \lambda \sqrt{\frac{2}{\omega}} \bra{\mathrm{HF},0} \big[ \big[ \tilde{E}_{rr}(b-b^\dagger)  , {H}_{\mathrm{SC}} \big] , E_{ai} \big] \ket{\mathrm{HF},0} .
\end{equation}
For each block, the derivatives are calculated at $\pmb{\kappa} = \textbf{0}$ because the Hessian is computed in the updated MO-basis. Instead, for the $\eta$-parameters, the derivatives are evaluated at the present values.
For the explicit derivation of the Hessian blocks, we refer to the Supporting information. 
Moreover, by employing a trust region (Levenberg–Marquard) approach, it is possible to enhance the robustness of the optimization by constraining each new step to stay within a trusted neighbourhood of the previous iteration\cite{hoyvik2012trust,fletcher2000practical,jensen1984direct}.
The optimization is carried out by solving in each iteration
\begin{equation} \label{level-shifted linear equations}
    (\mathbf{E}^{\left(2\right)} -\mu\mathbf{I}){\Delta}\mathbf{z} = -\mathbf{E}^{(1)}
\end{equation}
featuring the level-shifted Hessian ($\mathbf{E}^{\left(2\right)} -\mu\mathbf{I}$), the new step ${\Delta}\mathbf{z}$ and the gradient vector $\mathbf{E}^{\left(1\right)}$. 
For the derivation and the computational details of the trust region Newton-Raphson algorithm, we refer to the Supporting information.
To compute the new step, one could invert the level-shifted Hessian in \cref{level-shifted linear equations}, but this approach is computationally demanding due to the large number of the non-redundant $\kappa$-parameters. For this reason, we solve the linear system iteratively only calculating the action of the Hessian on a trial vector:
\begin{equation} \label{linear transformations}
    \pmb{\sigma} = \begin{pmatrix} \mathbf{E}^{\kappa\kappa} & \mathbf{E}^{{\kappa\eta}} \\
    \mathbf{E}^{{\eta\kappa}} & \mathbf{E}^{{\eta\eta}}
    \end{pmatrix}
    \begin{pmatrix}
        \ \Delta\pmb{\kappa} \ \ \\
        \ \Delta\pmb{\eta} \ \ 
    \end{pmatrix}
     = \begin{pmatrix} 
     \mathbf{E}^{\kappa\kappa}\Delta\pmb{\kappa} + \mathbf{E}^{{\kappa\eta}}\Delta\pmb{\eta} \\
    \mathbf{E}^{{\eta\kappa}}\Delta\pmb{\kappa} + \mathbf{E}^{{\eta\eta}}\Delta\pmb{\eta}
    \end{pmatrix} .
\end{equation}
This linear transformation approach is particularly efficient because each term in \cref{linear transformations} can be expressed in terms of gradient elements.
Moreover, our numerical investigations reveal that neglecting the mixed blocks of the Hessian gives a robust and faster converging algorithm. This implies that the coupling between the two sets of parameters is not particularly tight and relevant for the optimization. 
In this case, the linear transformation simplifies to
\begin{equation}\label{davidson}
    \pmb{\sigma} = \begin{pmatrix} 
     \mathbf{E}^{{\kappa\kappa}}\Delta\pmb{\kappa} \\
    \mathbf{E}^{{\eta\eta}}\Delta\pmb{\eta} 
    \end{pmatrix} .
\end{equation}
For the purely $\kappa$-$\kappa$ linear transformations, we have the Hartree-Fock equation in terms of the $\kappa$-gradients
\begin{equation}
    \left( \mathbf{E}^{{\kappa\kappa}}\Delta\pmb{\kappa} \right)_{ai} = f^{\kappa}_{ai}( {{h}}^{b,\Delta\kappa}_{pq} , \  {{g}}^{b,\Delta\kappa}_{pqrs})
\end{equation}
where the redefined one and two electrons integrals
\begin{equation}\label{new h}
    {\tilde{h}}^b_{pq} = {\tilde{h}}^a_{pq} + \frac{\lambda^2}{2}((\tilde{\mathbf{d}}\cdot\pmb{\epsilon})_{pp} - \eta_p)^2 \delta_{pq} ,
\end{equation}
\begin{equation} \label{new g}
    {\tilde{g}}^b_{pqrs} = {\tilde{g}}^a_{pqrs} + \lambda^2((\tilde{\mathbf{d}}\cdot\pmb{\epsilon})_{pp} - \eta_p)((\tilde{\mathbf{d}}\cdot\pmb{\epsilon})_{rr} - \eta_r) \delta_{pq}\delta_{rs} ,
\end{equation}
are rotated back to the canonical basis and then $\Delta\kappa$-transformed as follows:
\begin{equation}\label{h k transf}
    {{h}}^{b,\Delta\kappa}_{pq} = \sum_m ({\Delta\kappa}_{mp}{{h}}_{mq}^b + {\Delta\kappa}_{mq}{{h}}_{pm}^b) ,
\end{equation}
\begin{equation}\label{g k transf}
    {{g}}^{b,\Delta\kappa}_{pqrs} = \sum_m ({\Delta\kappa}_{mp}{{g}}_{mqrs}^b + {\Delta\kappa}_{mq}{{g}}_{pmrs}^b + {\Delta\kappa}_{mr}{{g}}_{pqms}^b + {\Delta\kappa}_{ms}{{g}}_{pqrm}^b) .
\end{equation}
On the other hand, for the purely $\eta$-$\eta$ linear transformations in terms of the $\eta$-gradients we have
\begin{equation}
    \left( \mathbf{E}^{{\eta\eta}} \Delta\pmb{\eta}\right)_r = \tilde{f}^{{\Delta}\eta}_r( {\tilde{h}}^{a,\eta}_{pq} , \ {\tilde{g}}^{a,\eta}_{pqrs} ) + \lambda^2 \sum_q ( \tilde{\mathbf{d}} \cdot \pmb{\epsilon} )_{qq}\Tilde{d}_{qqrr} + \lambda^2( \tilde{\mathbf{d}} \cdot \pmb{\epsilon})_{rr}\Tilde{D}_{rr} ,
\end{equation}
where the $\eta$-transformed  one and two electrons integrals are
\begin{equation}
    {\tilde{h}}^{a,\eta}_{pq} = {\tilde{h}}^a_{pq}\Big( 1-\frac{\lambda^2}{2\omega}\left( \eta_p-\eta_q \right)^2\Big) ,
\end{equation} 
\begin{equation}
    {\tilde{g}}^{a,\eta}_{pqrs}  = {\tilde{g}}^a_{pqrs}\Big( 1-\frac{\lambda^2}{2\omega}\left( \eta_p-\eta_q +\eta_r-\eta_s\right)^2\Big) .
\end{equation}
For the explicit derivation of the linear transformations comprising also the mixed parameters ones we refer the reader to the Supporting information.
\subsection{Direct inversion of the $\pmb{\eta}$-$\pmb{\eta}$ Hessian block}
The trust region Newton-Raphson approach requires an iterative algorithm in order to solve the linear equations that determine the step length. However, as shown in Section 3, the gradient based algorithm only struggles with the optimization of the $\eta$-parameters. This suggests an alternative algorithm where we only use DIIS acceleration for the density matrix and Newton-Raphson for $\{\eta_p\}$ obtained from the direct inversion of the $\eta$-$\eta$ Hessian block:
\begin{equation}
\begin{split}
\left(\frac{\partial^2E}{\partial \eta_{r}\partial\eta_{s}}\right) &= \delta_{rs}
\Bigg( \frac{\lambda^2}{\omega}\sum_q {\tilde{h}}^a_{rq}\Tilde{D}_{rq} \Big( \frac{\lambda^2}{2\omega} (\eta_r - \eta_q)^2 - 1 \Big) + \lambda^2 \Tilde{D}_{rr} \\
& \ \ \ \ \ \ \ \ \ +\frac{\lambda^2}{\omega}\sum_{pqt}{\tilde{g}}^a_{rpqt}\Tilde{d}_{rpqt} \Big( \frac{\lambda^2}{2\omega} (\eta_r +\eta_q -\eta_p - \eta_t)^2 - 1 \Big) \Bigg) \\
& - \frac{\lambda^2}{\omega}\sum_{qt}{\tilde{g}}^a_{rsqt}\Tilde{d}_{rsqt} \Big( \frac{\lambda^2}{2\omega} (\eta_r +\eta_q -\eta_s - \eta_t)^2 - 1 \Big) \\
& + \frac{\lambda^2}{\omega}\sum_{qt}{\tilde{g}}^a_{rqst}\Tilde{d}_{rqst} \Big( \frac{\lambda^2}{2\omega} (\eta_r +\eta_s -\eta_q - \eta_t)^2 - 1 \Big) \\
& - \frac{\lambda^2}{\omega}\sum_{qt}{\tilde{g}}^a_{rqts}\Tilde{d}_{rqts} \Big( \frac{\lambda^2}{2\omega} (\eta_r +\eta_t -\eta_q - \eta_s)^2 - 1 \Big)  \\
& + \lambda^2 \Tilde{d}_{rrss}  - \frac{\lambda^2}{\omega}{\tilde{h}}^a_{rs}\Tilde{D}_{rs} \Big( \frac{\lambda^2}{2\omega} (\eta_r - \eta_s)^2 - 1 \Big) .
\end{split}
\end{equation}
The computational cost of building the $\eta$-$\eta$ Hessian matrix is $N^4$ and the matrix inversion is $N^3$, in this way recalculation of the integrals is avoided.

\section{3. Results and discussion}
\begin{figure}[H]
    \centering
    \subfigure{%
        \includegraphics[width=0.15\textwidth]{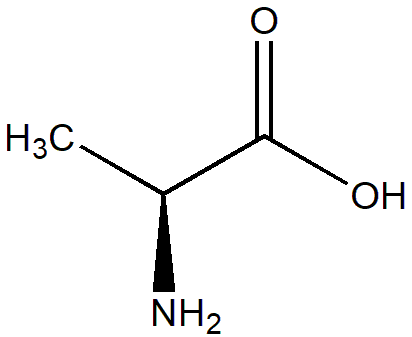}
    }
    \subfigure{%
        \includegraphics[width=0.10\textwidth]{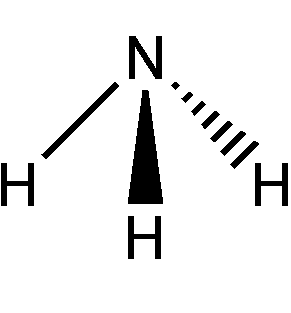}
    }
    \subfigure{%
        \includegraphics[width=0.125\textwidth]{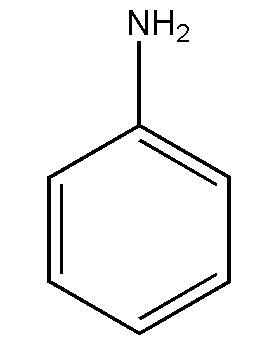}
    }
    \subfigure{%
        \includegraphics[width=0.2\textwidth]{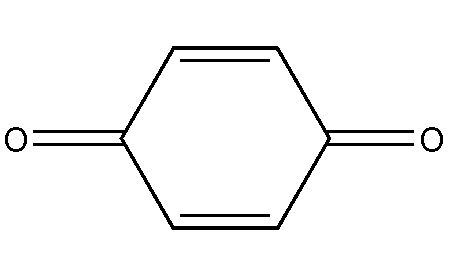}
    }
    \subfigure{%
        \includegraphics[width=0.12\textwidth]{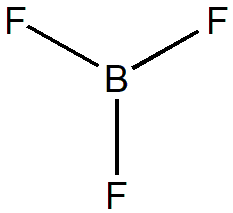}
    }
    \\
    \subfigure{%
        \includegraphics[width=0.14\textwidth]{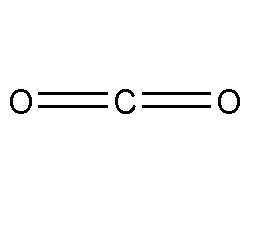}
    }
    \subfigure{%
        \includegraphics[width=0.13\textwidth]{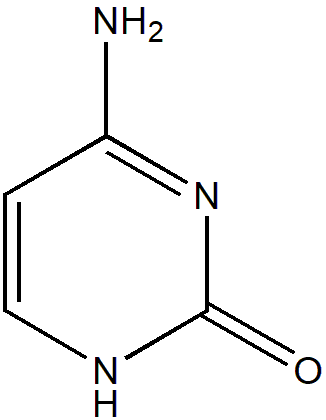}
    }
    \subfigure{%
        \includegraphics[width=0.14\textwidth]{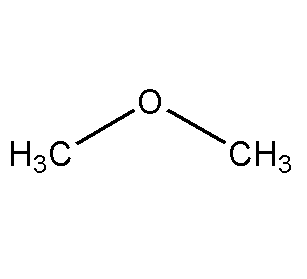}
    }
    \subfigure{%
        \includegraphics[width=0.13\textwidth]{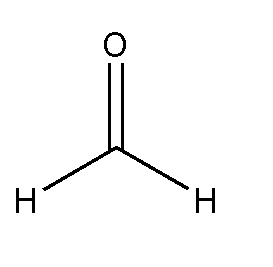}
    }
    \subfigure{%
        \includegraphics[width=0.14\textwidth]{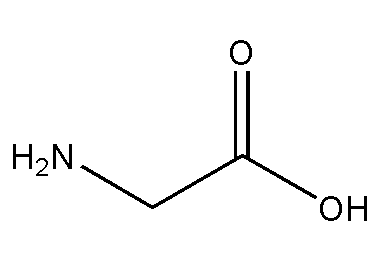}
    }
    \\
    \subfigure{%
        \includegraphics[width=0.1\textwidth]{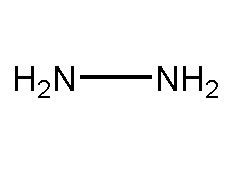}
    }
    \subfigure{%
        \includegraphics[width=0.14\textwidth]{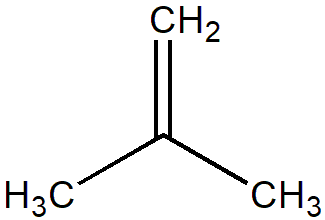}
    }
    \subfigure{%
        \includegraphics[width=0.18\textwidth]{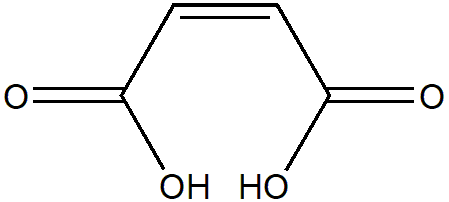}
    }
    \subfigure{%
        \includegraphics[width=0.12\textwidth]{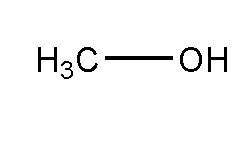}
    }
    \subfigure{%
        \includegraphics[width=0.13\textwidth]{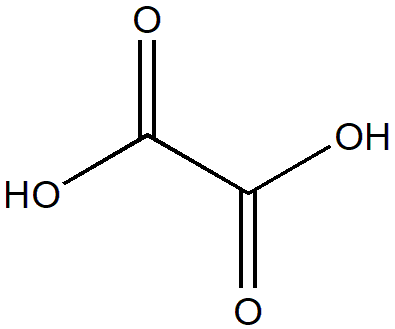}
    }
    \\
    \subfigure{%
        \includegraphics[width=0.14\textwidth]{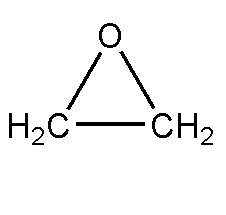}
    }
    \subfigure{%
        \includegraphics[width=0.10\textwidth]{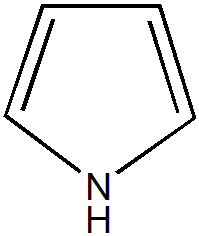}
    }
    \subfigure{%
        \includegraphics[width=0.19\textwidth]{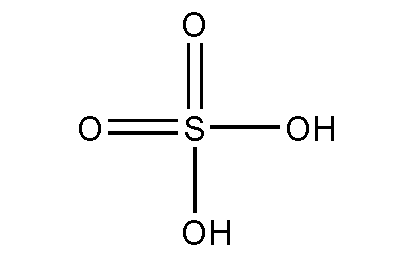}
    }
    \subfigure{%
        \includegraphics[width=0.1\textwidth]{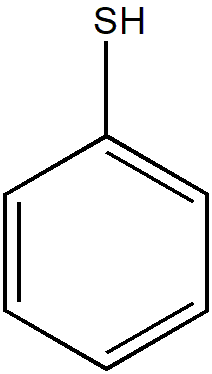}
    }
    \subfigure{%
        \includegraphics[width=0.16\textwidth]{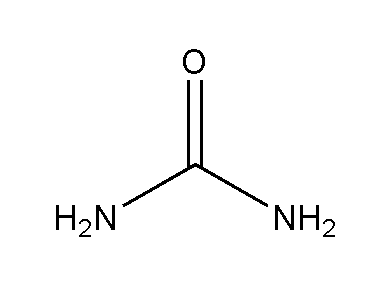}
    }
    \caption{Benchmark molecules.}
    \label{bench_molecules}
\end{figure}

In order to demonstrate the computational efficiency of the developed algorithms, we compare the performance for the set of 20 molecules shown in \Cref{bench_molecules}.
We also present a few calculations for larger molecular systems using the batching algorithm of the Cholesky decomposed two-electron integrals. All the calculations have been performed with a development version of the e$^\mathcal{T}$ program\cite{eT} using a dual-socket Intel(R) Xeon(R) Platinum 8380 system with 2 TB of memory. In the benchmark study we employed 20 cores, while for the largest systems 80 cores were used.
\subsection{Benchmark of the methods} 
\begin{figure}[H]
    \centering
    \includegraphics[scale = 0.31]{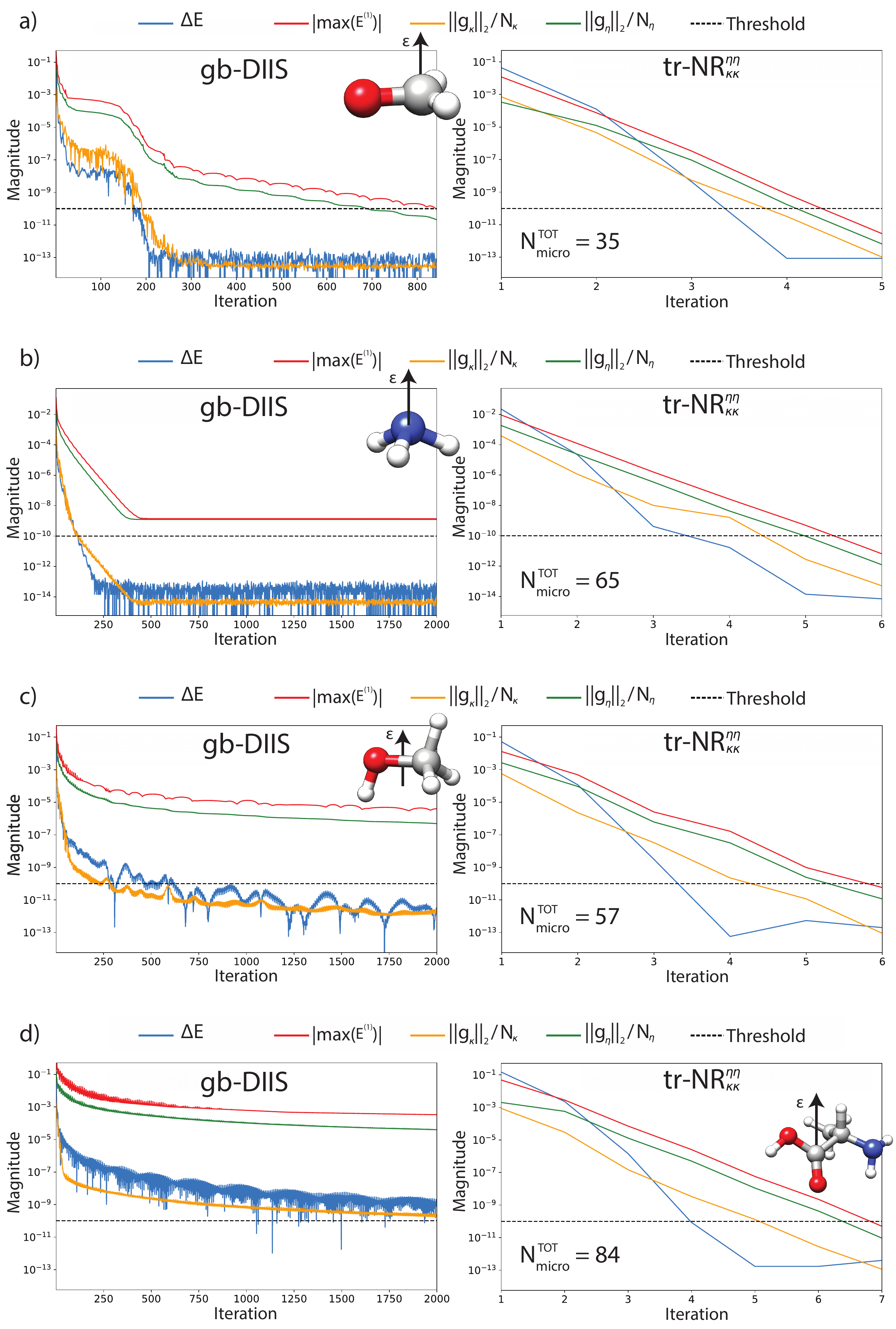}
    \caption{Convergence comparison between the gb-DIIS and the tr-NR$_{\kappa\kappa}^{\eta\eta}$ algorithms for a) formaldehyde, b) ammonia, c) methanol and d) alanine. For the tr-NR$_{\kappa\kappa}^{\eta\eta}$ algorithm we also report the total number of micro-iterations. See text for the definition of the other quantities.}
    \label{sc_vs_full2nd}
\end{figure}

In \Cref{sc_vs_full2nd}, we report the comparison of the convergence patterns for the DIIS accelerated gradient-based implementation (gb-DIIS) and the trust region Newton-Raphson algorithm using only the $\kappa$-$\kappa$ and $\eta$-$\eta$ blocks of the Hessian (tr-NR$_{\kappa\kappa}^{\eta\eta}$). For conciseness, we illustrate the convergence only for formaldehyde, ammonia, methanol, and alanine. All molecular geometries and the results for the remaining 16 molecules are reported in the Supporting information.
We used an aug-cc-pVDZ basis set,\cite{dunning1989a,pritchard2019a} light-matter coupling $\lambda=0.005$ a.u., vacuum cavity frequency $\omega=2.71$ eV, and a field polarization along the $z$-axis. 
The quantities plotted for each iteration are the absolute energy difference from the previous iteration
\begin{equation}
    \Delta E_n = \abs{E_{n}-E_{n-1}} ,
\end{equation}
the absolute maximum value of the total gradient vector $|\mathrm{max}\left(\mathbf{E}^{\left(1\right)}\right)|$, and the $L^2$-norms of the $\kappa$ and $\eta$-gradients (defined in \cref{kappa gradient,eta gradient}) scaled by the number of non-redundant parameters within each class: $\norm{\mathbf{g}_\kappa}_2 / N_\kappa$ and $\norm{\mathbf{g}_\eta}_2/ N_\eta$.
For all calculations, the convergence threshold is set to $10^{-10}$ a.u..
The results obtained with the gb-DIIS implementation are shown to the left in \Cref{sc_vs_full2nd} and clearly indicate that the convergence pattern of the $\eta$-parameters is not optimal. This is also corroborated by the rapid convergence observed when optimizing the orbitals while keeping the $\eta$-parameters frozen to the eigenvalues of the dipole operator.

For the gb-DIIS algorithm, only formaldehyde converges within 2000 iterations as can be seen in \Cref{sc_vs_full2nd}a.
For ammonia and methanol, in Figures \ref{sc_vs_full2nd}b and \ref{sc_vs_full2nd}c, the scaled $L^2$-norms of the $\eta$-gradient reach a plateau of $10^{-9}$ a.u. and $10^{-6}$ a.u., respectively. The $\kappa$ parameters keep oscillating around their stationary value, with the energy slowly decreasing by $10^{-11}$-$10^{-14}$ Hartree in each step. 
On the other hand, for alanine in \Cref{sc_vs_full2nd}d, we observe that not even the $\kappa$-parameters are converged within $2000$ iterations, while the $\eta$-parameters reach a plateau much higher than the convergence threshold.
In \Cref{sc_vs_full2nd}, to the right, we show the results obtained with the tr-NR$_{\kappa\kappa}^{\eta\eta}$ algorithm. In all four cases, convergence is reached in less than $10$ iterations, where each iteration requires on average less than $10$ micro-iterations to solve the linear equations in \cref{davidson}.

\begin{figure*}[ht]
    \centering
    \subfigure[sulfuric acid]{%
        \label{h2so4 hessian}
        \includegraphics[width=0.46\textwidth]{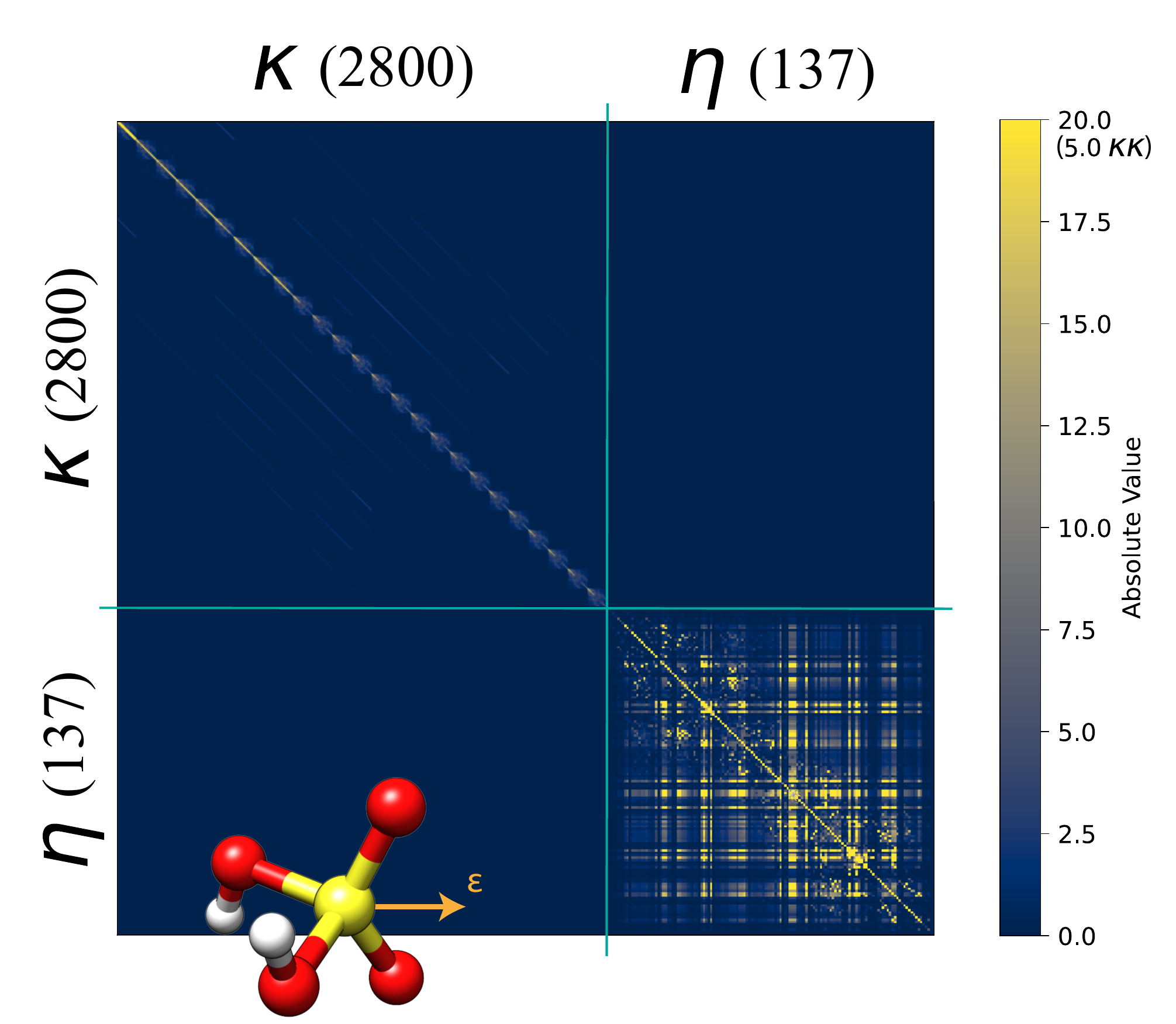}
    }%
    \hspace{0.02\textwidth}
    \subfigure[oxalic acid]{%
        \label{oxalic acid hessian}
        \includegraphics[width=0.46\textwidth]{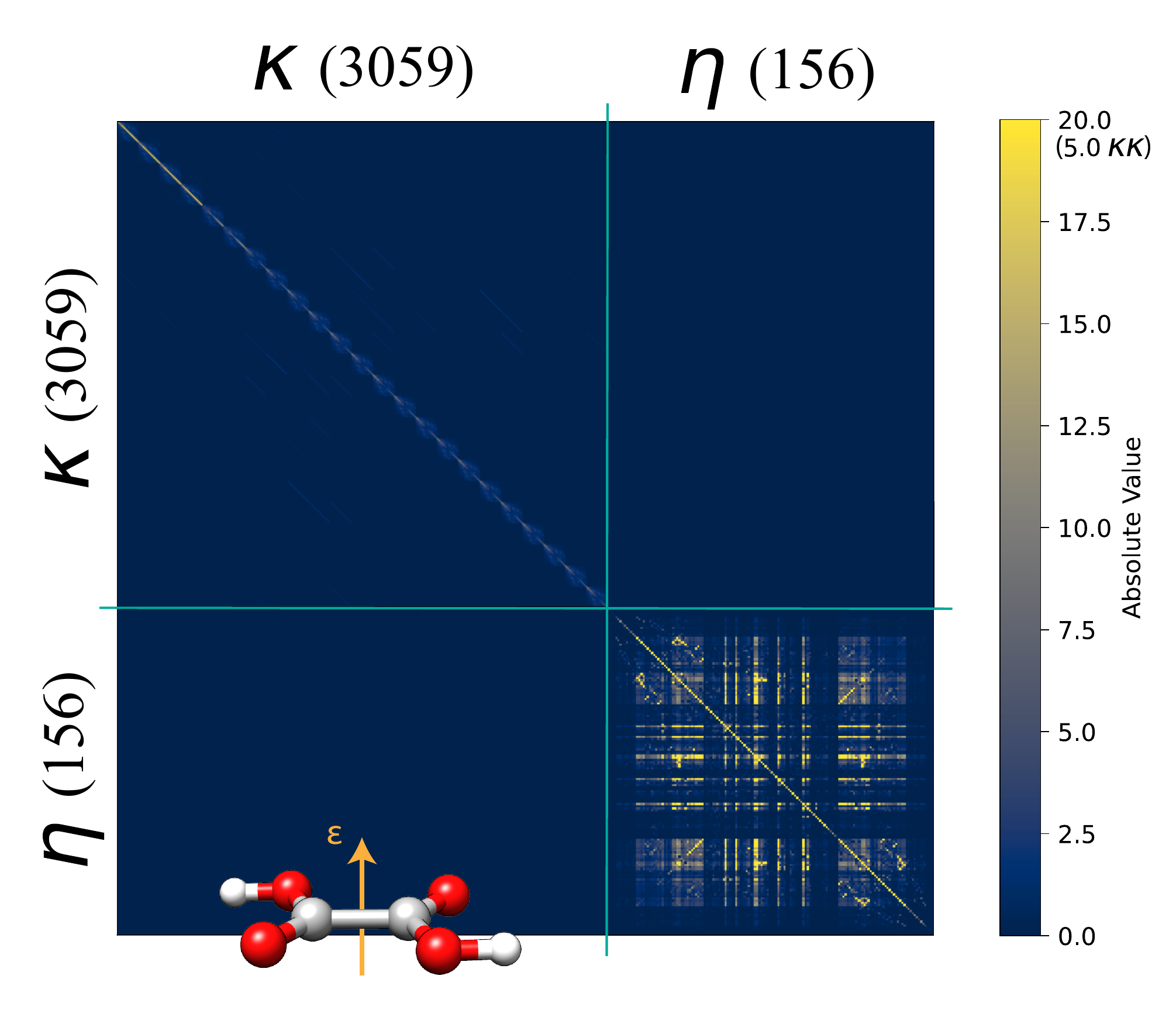}
    }\\
    \vspace{1em}
    \subfigure[glycine]{%
        \label{glycine hessian}
        \includegraphics[width=0.46\textwidth]{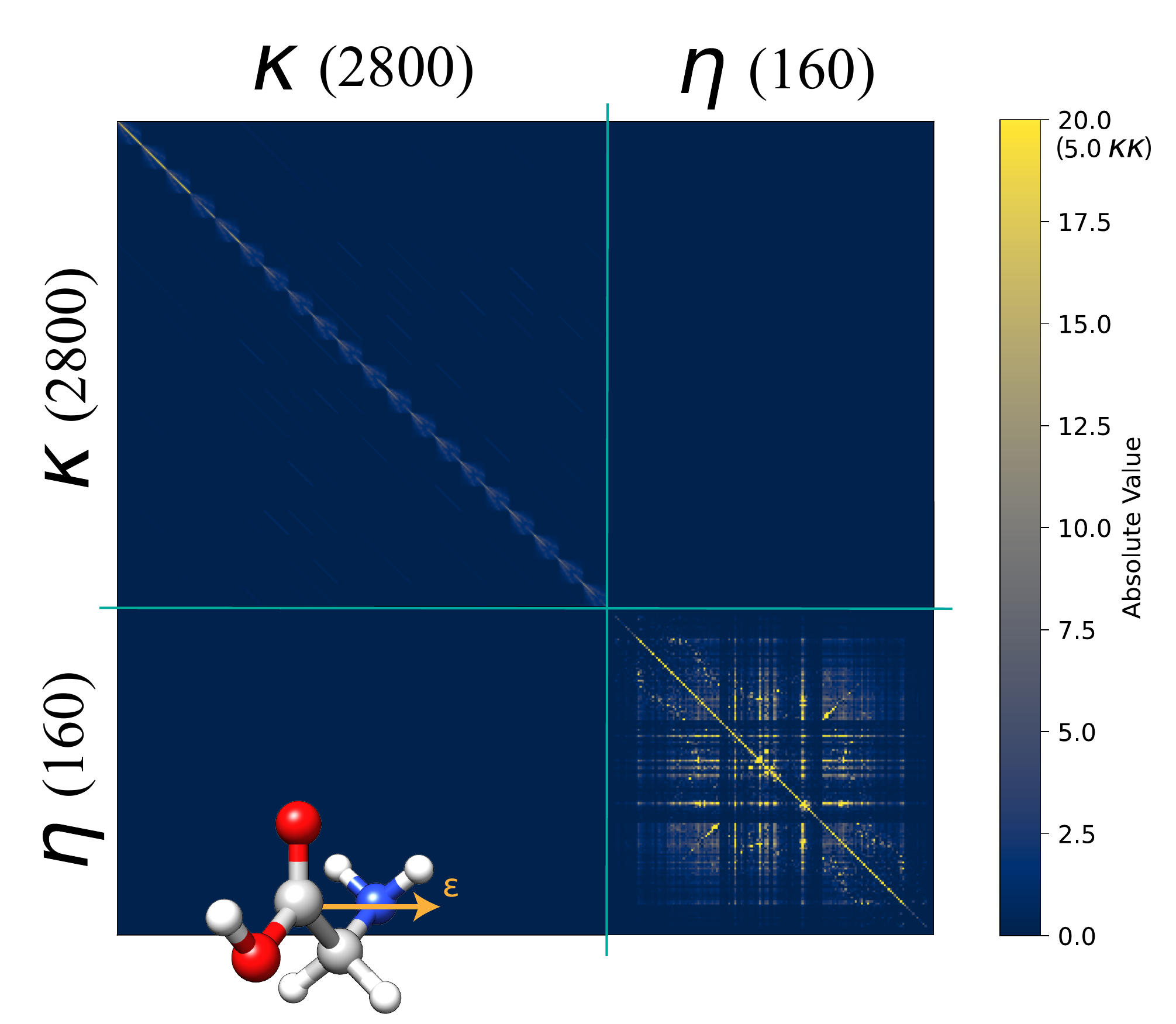}
    }%
    \hspace{0.02\textwidth}
    \subfigure[isobutylene]{%
        \label{isobulylene hessian}
        \includegraphics[width=0.46\textwidth]{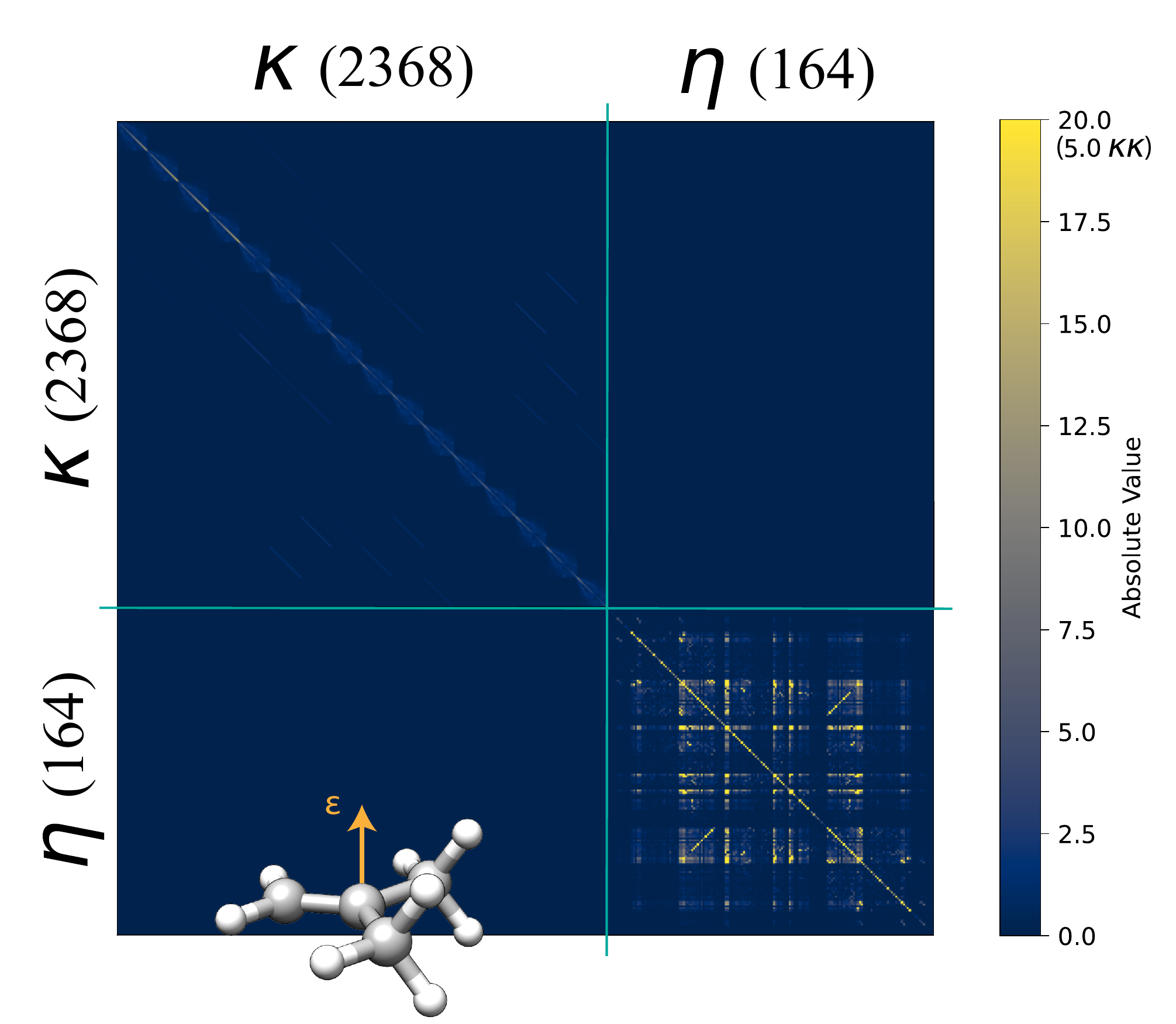}
    }%
    \caption{Hessian matrices at the first iteration for a) sulfuric acid, b) oxalic acid, c) glycine, and d) isobutylene. The $\eta$-$\eta$ and mixed parameters blocks are resized to provide better visualization of these critical terms. A cutoff of $20.0$ in the color scale is used to better illustrate the importance of the off-diagonal elements in the $\eta$-$\eta$ blocks. The cutoff of the $\kappa$-$\kappa$ blocks is placed at 5 to appreciate the diagonal dominance.}
    \label{hessian analysis}
\end{figure*}

We notice a fast and robust convergence when neglecting the mixed parameters blocks of the Hessian matrix.
To validate this, we analyzed the Hessian matrix, in the first iteration, for all the molecules in \Cref{bench_molecules}.
In \Cref{hessian analysis}, we show the heat map representations of the non-redundant Hessians for sulfuric acid, oxalic acid, glycine, and isobutyene. See the Supporting information for the heat map representations for the remaining 16 benchmark molecules.

As expected from Hartree-Fock theory, the $\kappa$-$\kappa$ blocks are diagonally dominant with the contribution from Fock matrix elements being the dominating part (see Supporting information for derivation):
\begin{equation}\label{k-k Hessian 4}
    \left(\frac{\partial^2E}{\partial \kappa_{ai}\partial\kappa_{bj}}\right) = 4\Big[F_{ab}\delta_{ij}-F_{ij}\delta_{ab} + 2({g}^b_{aibj}+{g}^b_{aijb} ) - {g}^b_{abji} - {g}^b_{ajbi} \Big] .
\end{equation}
In \Cref{hessian analysis} we observe that the $\eta$-$\eta$ blocks are highly non-diagonal indicating the parameters are strongly coupled. Although the diagonal elements are larger than the off-diagonal ones (Supporting information), the structure of the $\eta$-$\eta$ block still leads to convergence difficulties of the gradient-based optimization. This also explains why considering these couplings in the tr-NR$_{\kappa\kappa}^{\eta\eta}$ algorithm benefits the procedure.
The iterations saved by including the mixed blocks in the Hessian do not make up for the computational requirement (see the Supporting information for a detailed wall time comparison).

The observations made from the Hessian analysis suggest the development of a third algorithm.
To this end, the direct inversion of the $\eta$-$\eta$ block is performed concurrently with the orbital optimization process of the original implementation (gb-DBI$_{\eta\eta}$).
The plots with this algorithm are similar to the tr-NR$_{\kappa\kappa}^{\eta\eta}$ ones and are reported in the Supporting information. 
In \Cref{times table}, we show the wall time for the three algorithms. 
We stress that the timings reported for the gb-DIIS optimization refer to achieving 2000 iterations while still being orders of magnitude far from convergence. 
Only formaldehyde converged in 843 iterations.
The gb-DBI$_{\eta\eta}$ algorithm turns out to be faster in terms of wall time and number of iterations. 
These savings are obtained because the micro-iterations are no longer needed in favor of the direct inversion of the small $\eta$-$\eta$ block. 
\begin{table}[H]
  \centering
  \begin{tabular}{lccc}
    \toprule
    \textbf{Molecule} & \textbf{gb-DIIS} & \textbf{tr-NR}$_{\pmb{\kappa\kappa}}^{\pmb{\eta\eta}}$ \textbf{(micro-Iter.)} & \textbf{gb-DBI}$_{\pmb{\eta\eta}}$ \textbf{(Iter.)}\\
    \midrule
    alanine           & 7.71 h                          & 22.0 m  \ \ (84)                                               & 14.7 m  \ \ \ (21)                                               \\
    ammonia           & 87.3 s                       & 5.91 s \ \ \ (65)                                                  & 1.81 s \ \ \ \  (14)                                                 \\
    aniline           & 11.6 h                      & 23.7 m  \ \ (57)                                               & 21.9 m  \ \ \ (20)                                              \\
    benzoquinone      & 10.2 h                      & 25.1 m  \ \ (69)                                               & 18.4 m \ \ \  (19)                                              \\
    boron trifluoride & 19.4 m                       & \ 69.1 s \ \ \ (76)                                            & 24.7 s \ \ \ \  (14)                                                \\
    carbon dioxide    & 6.38 m                       & 20.2 s  \ \ \ (73)                                                 & 7.77 s  \ \ \ \ (13)                                                 \\
    cytosine          & 12.5 h                      & \ 47.8 m  \ \ (106)                                              & 27.4 m \ \ \  (23)                                              \\
    dimethyl ether    & 1.11 h                       & 2.09 m \ \  (49)                                                & 87.3 s \ \ \  \ (15)                                                \\
    formaldehyde      & 2.40 m                       & 10.2 s \ \ \  (35)                                                 & 7.88 s \ \ \  \ (14)                                                 \\
    glycine           & 2.87 h                       & 6.94 m \ \  (71)                                                & 5.29 m \ \ \  (20)                                               \\
    hydrazine         & 15.4 h                       & 26.8 s \ \ \  (44)                                                 & 19.1 s \ \ \ \ (16)                                                \\
    isobutylene       & 3.41 h                       & 6.37 s \ \ \  (51)                                                &  5.31 m \ \ \  (17)                                               \\
    maleic acid       & 10.2 h                      & \ 36.2 m \ \  (104)                                              & 22.2 m \ \ \  (23)                                              \\
    methanol          & 12.1 m                       & 31.2 s \ \ \  (57)                                                 & 17.4 s \ \ \ \ (16)                                                 \\
    oxalic acid       & 2.58 h                       & 5.37 m \ \  (56)                                                & 4.58 m \ \ \  (19)                                               \\
    oxirane           & 35.1 m                        & 83.3 s \ \ \  (69)                                                 & 46.6 s \ \ \  \ (16)                                                \\
    pyrrole           & 2.83 h                       & 5.23 m \ \  (47)                                                & 4.82 m \ \ \  (18)                                               \\
    sulfuric acid     & 1.68 h                       & \ 8.76 m  \ \ (145)                                               & 3.02 m \ \ \  (20)                                               \\
    thiophenol        & 10.0 h                      & 26.0 m \ \  (72)                                               & 22.2 m \ \ \  (23)                                              \\
    urea              & 1.24 h                       & 2.68 m \ \  (54)                                                & 2.16 m \ \ \  (18)                                               \\
    \bottomrule
  \end{tabular}
  \caption{Wall times and iterations comparison between the algorithms.}
  \label{times table}
\end{table}

\subsection{Polaritons for large molecular systems}

To investigate the polaritonic properties of larger molecular systems, we implemented a batching algorithm for the two-electron integrals in the dipole basis. While these integrals can be comfortably stored for smaller systems without significant memory requirements, larger systems need a more efficient handling. In our approach, the Cholesky vectors in the dipole basis are stored in memory, and the two-electron integrals are calculated on-the-fly in blocks that maximize the use of the  total available memory\cite{folkestad2022implementation,folkestad2019efficient,koch2003reduced,aquilante2011cholesky,nottoli2021black}:
\begin{equation}
    \Tilde{g}_{pqrs} \approx \sum_J\Tilde{L}_{pq}^J \Tilde{L}_{rs}^J \ .
\end{equation}

\begin{figure}[H]
    \centering
    \includegraphics[width=0.90\textwidth]{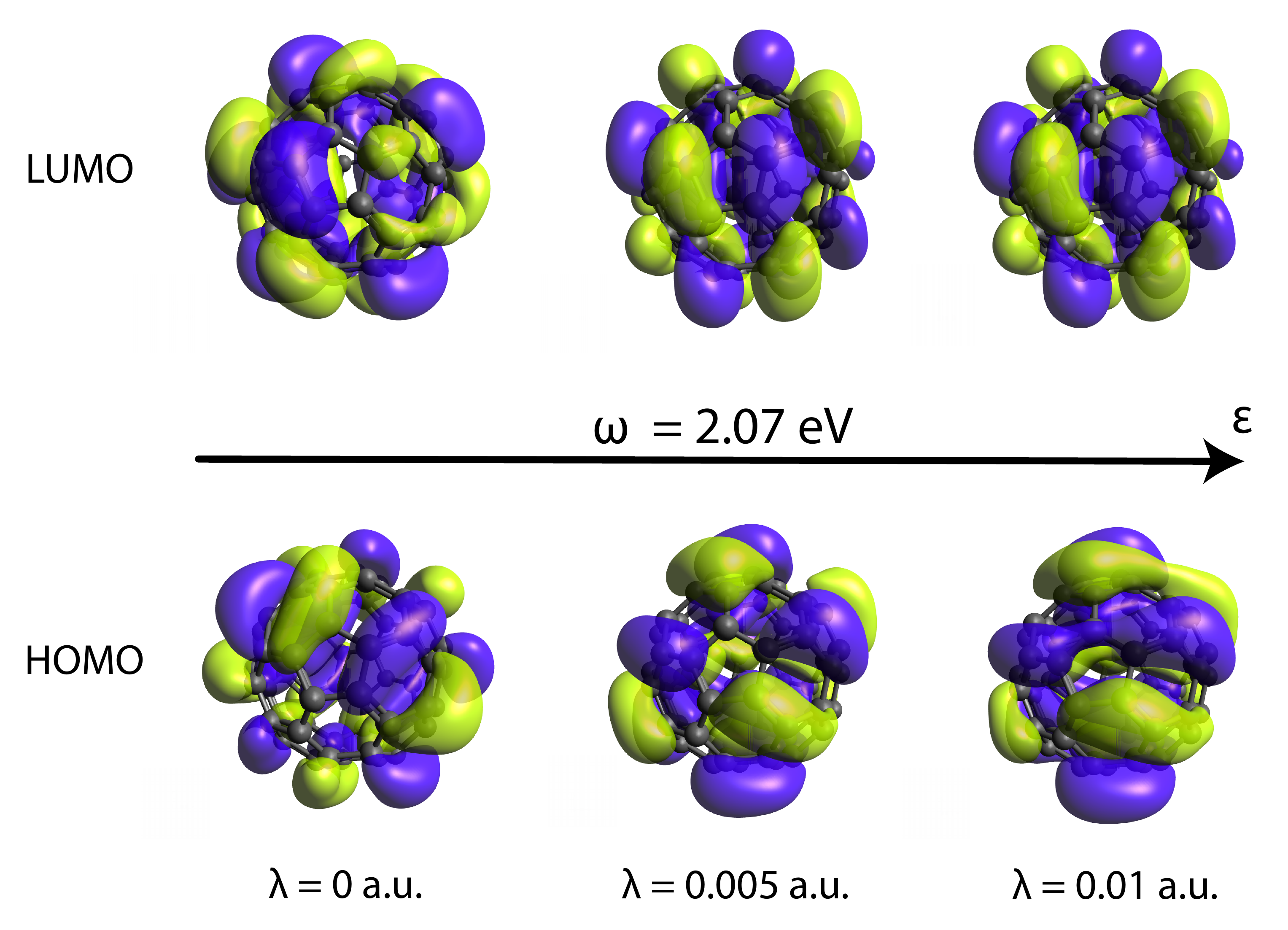}
    \caption{Fullerene (C$_{60}$) HOMO and LUMO orbital reshaping at various couplings and cavity frequency set to $\omega = 2.07$ eV. The arrow refers to the polarization vector. The surfaces are plotted using a 0.012 a.u. isosurface value.}
    \label{C60}
\end{figure}

\begin{figure}[H]
    \centering
    \includegraphics[width=\textwidth]{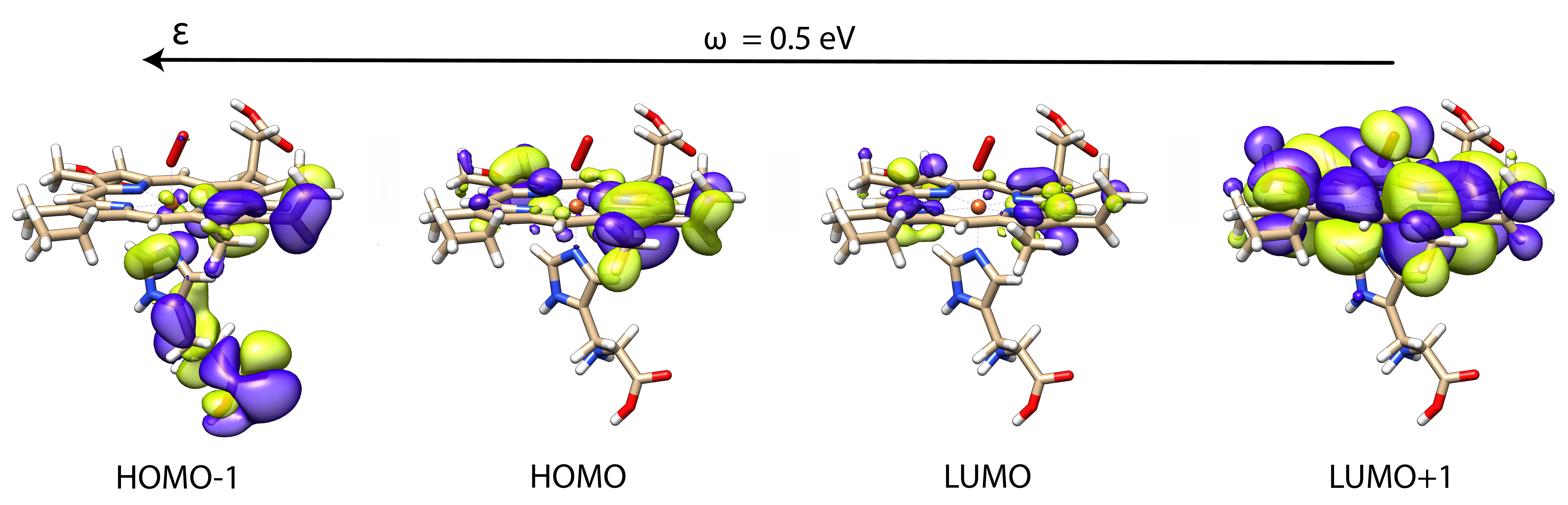}
    \caption{{Surface differences between the $\lambda=0.01$ a.u. and  $\lambda=0$ a.u. four frontier orbitals of a heme group coordinated to a proximal histidin amino acid and an oxygen molecule at $\omega=0.5$ eV. The arrow refers to the polarization vector. The surfaces are plotted using a 0.0002 a.u. isosurface value.}
    \label{HEME}}
\end{figure}

In \Cref{C60}, we show the reshaping of the HOMO and LUMO orbitals for fullerene using a cc-pVDZ basis set\cite{dunning1989a,pritchard2019a}. In the Supporting information we also show HOMO-1 and LUMO+1 orbitals as well as the molecular geometry. Interestingly we notice how the vacuum field polarization breaks the I$_h$ point group symmetry and how this is reflected in the orbital shapes at various coupling strengths. For the selected cavity frequency these effects are more pronounced for the HOMO, while no significant changes are observed in the LUMO passing from $\lambda = 0.005$ a.u. to $\lambda = 0.01$ a.u.
In \Cref{HEME} we analyse the four frontier orbitals of the heme group with the Fe$^{2+}$ ion coordinated to a proximal histidin amino acid and an oxygen molecule. We show $\lambda=0$ a.u. and  $\lambda=0.01$ a.u. differences at cavity frequency $\omega=0.5$ eV. The molecular geometry is given in the Supporting information. The calculations were performed using the tr-NR$_{\kappa\kappa}^{\eta\eta}$ algorithm for the first few iterations then followed by the faster gb-DBI$_{\eta\eta}$, due to the non-positive definite Hessian in the early stages of the optimization. 
We used a 6-31G basis set\cite{hehre1972self} without batching of the two-electron integrals. The observed differences in the orbitals are small due to the absence of cavity induced symmetry breaking. 

Our results show that the improvements in the convergence will allow us to study large molecular systems and address the electron-photon correlation using a properly dressed set of orbitals that can be used in post-mean-field approaches. 

\section{4. Conclusions}
In this work, we have reported a new and improved implementation of the strong coupling quantum electrodynamics Hartree-Fock model. Our new algorithms rely on the use of the second derivatives of the energy with respect to the wave function parameters. This provides faster convergence of the orbital-specific coherent state $\eta$-parameters. 
While a full implementation of the trust region Newton-Raphson scheme has been reported, our investigations reveal that only using the $\eta$-$\eta$ Hessian block is enough to provide robust and fast convergence in a memory-efficient manner. 
Our work provides new insight into the complex interplay between electrons and photons showing that, at the mean-field level, orbital rotations and electron-photon parameters are almost completely decoupled. In addition, our algorithms pave the way for developing computationally efficient post-mean-field methods. Specifically, coupled cluster and active space extensions would improve the description of electron-photon correlation while capturing the electron-electron correlation as well. 
Additionally, our improvements open new avenues for the development of multi-level methodologies to tackle the inclusion of solvent effects in QED environments. To this end, efficient screening of the photon-dressed two-electron integrals is necessary in order to reduce the computational scaling for large molecular systems. Future works will focus on response theory\cite{castagnola2024polaritonic} as well as the use of molecular orbitals to understand the cavity-induced modifications of molecular properties. Moreover, the generalization to a multi-mode Hamiltonian able to describe higher-order optical phenomena and the extension of the method to chiral cavities are currently in development.
\begin{acknowledgement}
We thank Sarai Dery Folkestad for insightful discussions. We acknowledge funding from the Research Council of Norway through FRINATEK Project No. 275506. This work has received funding from the European Research Council (ERC) under the European Union’s Horizon 2020 Research and Innovation Programme (Grant Agreement No. 101020016). 
\end{acknowledgement}
\begin{suppinfo}
The geometries used for the reported calculations as well as the results obtained for all the other molecules in \Cref{bench_molecules} can be found in the Supporting information. Detailed derivation of the Hessian equations and linear transformations are also presented.
\end{suppinfo}
\subsection*{Code availability} 
The e$^\mathcal{T}$ program\cite{eT} used to perform the calculations shown in this work is available from the corresponding author upon reasonable request.
\bibliography{references}

\begin{figure}[H]\label{For Table of Contents Only}
    \centering
    \includegraphics[width=\textwidth]{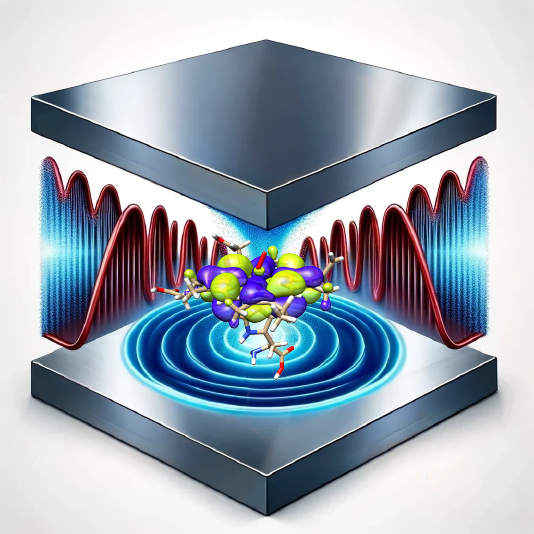}
    \caption*{For Table of Contents Only}
\end{figure}



\section*{Supplementary material}

while **d** is the molecular dipole operator defined as

$$\mathbf{d} = \sum_{pq} \mathbf{d}_{pq} E_{pq} = \sum_{pq} \left( \mathbf{d}^e_{pq} + \frac{\mathbf{d}^{nuc}}{N_e} \delta_{pq} \right) E_{pq}, \tag{3}$$

with $\mathbf{d}^e$ being the electronic dipole and $\mathbf{d}^{nuc}$ the nuclear dipole of a system of $N_e$ electrons. The electronic operators $E_{pq}$ and $e_{pqrs}$ are

$$E_{pq} = \sum_{\sigma} a^\dagger_{p\sigma} a_{q\sigma}$$
$$e_{pqrs} = E_{pq} E_{rs} - \delta_{rq} E_{ps}, \tag{4}$$

where $a^\dagger_{p\sigma}$ and $a_{p\sigma}$ are the creation and annihilation operators for an electron in orbital $p$ and spin $\sigma$. Finally, $h_{pq}$ and $g_{pqrs}$ are the one and two electron integrals associated to the electronic Hamiltonian in the Born-Oppenheimer approximation.

In the infinite coupling strength limit, the photonic part dominates and we thus employ the approximate Hamiltonian

$$H_\infty = \omega b^\dagger b - \lambda \sqrt{\frac{\omega}{2}} (\mathbf{d} \cdot \boldsymbol{\epsilon})(b^\dagger + b) + \frac{\lambda^2}{2} (\mathbf{d} \cdot \boldsymbol{\epsilon})^2. \tag{5}$$

The electronic part of the exact eigenfunctions are Slater determinants in the basis that diagonalizes the dipole operator **d**, referred to as the dipole basis. In this basis, the infinite coupling Hamiltonian reads:

$$H_\infty = \omega b^\dagger b - \lambda \sqrt{\frac{\omega}{2}} \sum_p (\tilde{\mathbf{d}} \cdot \boldsymbol{\epsilon})_{pp} \tilde{E}_{pp} (b^\dagger + b) + \sum_{pq} \frac{\lambda^2}{2} (\tilde{\mathbf{d}} \cdot \boldsymbol{\epsilon})_{pp} (\tilde{\mathbf{d}} \cdot \boldsymbol{\epsilon})_{qq} \tilde{E}_{pp} \tilde{E}_{qq}. \tag{6}$$

This expression suggests to introduce a coherent state transformation to reabsorb the bilinear term and move quantum picture:

$$H_\infty = \omega b^\dagger b \tag{7}$$

The transformation able to do so by mixing the photonic and electronic degrees of freedom is

$$U_\infty = \exp\left(-\frac{\lambda}{\sqrt{2\omega}} \sum_p (\tilde{\mathbf{d}} \cdot \boldsymbol{\epsilon})_{pp} \tilde{E}_{pp}(b - b^\dagger)\right). \tag{8}$$

Now, we can relax the infinite coupling limit to a finite coupling strength recovering the electronic Hamiltonian

$$H = \tilde{H}_e + \omega\left(b^\dagger - \frac{\lambda}{\sqrt{2\omega}} \sum_p (\tilde{\mathbf{d}} \cdot \boldsymbol{\epsilon})_{pp} \tilde{E}_{pp}\right)\left(b - \frac{\lambda}{\sqrt{2\omega}} \sum_p (\tilde{\mathbf{d}} \cdot \boldsymbol{\epsilon})_{pp} \tilde{E}_{pp}\right) \tag{9}$$

and introducing in place of the dipole integrals the $\eta$-parameters: a novel set of orbital specific parameters

$$U_{\text{SC}} = \exp\left(-\frac{\lambda}{\sqrt{2\omega}} \sum_p \eta_p \tilde{E}_{pp}(b - b^\dagger)\right). \tag{10}$$

The SC-QED-HF parametrization of the wave function is then

$$|\psi_{\text{SC}}\rangle = \exp\left(-\frac{\lambda}{\sqrt{2\omega}} \sum_p \eta_p \tilde{E}_{pp}\left(b - b^\dagger\right)\right) e^\kappa |\text{HF}\rangle \otimes |0\rangle, \tag{11}$$

where

$$\kappa = \sum_{p>q} \kappa_{pq}(E_{pq} - E_{qp}) \tag{12}$$

is the anti-hermitian operator used for the SCF orbital rotations.

The SC-QED-HF wave function parametrization is constituted by two set of parameters, the $\eta$-parameters and the $\kappa$-parameters, and therefore both of them need to be optimized during the SCF procedure.

# S2 SC-QED-HF Hessian matrix

The SC-parametrization in eq. (11) is composed by two different classes of parameters: the $\kappa$ and the $\eta$ parameters. For this reason the Hessian Matrix is be composed by 4 different blocks:

$$\mathbf{E}^{(2)} = \begin{pmatrix} \mathbf{E}^{\kappa\kappa} & \mathbf{E}^{\kappa\eta} \\ \mathbf{E}^{\eta\kappa} & \mathbf{E}^{\eta\eta} \end{pmatrix} = \begin{pmatrix} \frac{\partial^2 E}{\partial \kappa_{ai} \partial \kappa_{bj}} & \frac{\partial^2 E}{\partial \kappa_{ai} \partial \eta_r} \\ \frac{\partial^2 E}{\partial \eta_r \partial \kappa_{ai}} & \frac{\partial^2 E}{\partial \eta_r \partial \eta_s} \end{pmatrix}_{\boldsymbol{\kappa}=\mathbf{0}} . \tag{13}$$

where in top left we recognise the purely $\kappa$-$\kappa$ block, in bottom right the $\eta$-$\eta$ one and lastly the mixed parameters blocks in top right and bottom left. For each block, the derivatives can be taken at $\boldsymbol{\kappa} = \mathbf{0}$ if the Hessian is computed in the updated MO-basis. Instead, for the $\eta$-parameters, the derivatives are implicitly evaluated at the SCF converged values coming from the previous iteration.

In the next sections, each block contributing to the Hessian matrix $\mathbf{E}^{(2)}$ is explicitly derived for generic indices by calculating the proper commutators averaged on the reference wave function $|\text{HF}, 0\rangle$.

## S2.1 The $\kappa$-$\kappa$ block

The Hessian elements for the $\kappa$-$\kappa$ block are

$$\left(\frac{\partial^2 E}{\partial \kappa_{pq}\partial \kappa_{rs}}\right)_{\boldsymbol{\kappa}=\mathbf{0}} = (1 + P_{pq,rs}) \langle \text{HF}, 0| \left[[U_{\text{SC}}^\dagger H U_{\text{SC}}, E_{pq}], E_{rs}^-\right] |\text{HF}, 0\rangle \tag{14}$$

where $P_{mn,lo}$ permutes the $m$ and $n$ indices with the $l$ and $o$ ones.

By applying the SC-transformation (10) to the Hamiltonian (9), we obtain the SC-transformed dipole Hamiltonian, which reads

$$\begin{aligned}H_{\text{SC}} &= U_{\text{SC}}^\dagger H U_{\text{SC}} \\ &= H_e^{\text{SC}} + \omega\left(b^\dagger - \frac{\lambda}{\sqrt{2\omega}}\sum_p ((\tilde{\mathbf{d}}\cdot\boldsymbol{\epsilon})_{pp} - \eta_p)\tilde{E}_{pp}\right)\left(b - \frac{\lambda}{\sqrt{2\omega}}\sum_p ((\tilde{\mathbf{d}}\cdot\boldsymbol{\epsilon})_{pp} - \eta_p)\tilde{E}_{pp}\right)\end{aligned} \tag{15}$$

where the transformed electronic Hamiltonian is defined as:

$$\begin{aligned}H_e^{\text{SC}} &= \sum_{pq} \tilde{h}_{pq}\tilde{E}_{pq} \exp\left(\frac{\lambda}{\sqrt{2\omega}}(\eta_p - \eta_q)(b - b^\dagger)\right) \\ &+ \frac{1}{2}\sum_{pqrs} \tilde{g}_{pqrs}\tilde{e}_{pqrs} \exp\left(\frac{\lambda}{\sqrt{2\omega}}(\eta_p + \eta_r - \eta_q - \eta_s)(b - b^\dagger)\right).\end{aligned} \tag{16}$$

Notice that in the electronic Hessian of Eq. 14, the operators $E_{mn}$ and $E_{lo}^-$ act exclusively on the electronic degrees of freedom. Therefore, the average over the photon vacuum $|0\rangle$ involves only the Hamiltonian $H_{\text{SC}}$, giving an effective electronic operator. In particular, there are contributions only from the (transformed) electronic Hamiltonian $H_e^{\text{SC}}$ and the dipole self-energy.

Taking then the vacuum average on the SC-tranformed Hamiltonian (15) we obtain

$$\langle H_{\text{SC}} \rangle_0 = \sum_{pq} \tilde{h}_{pq} \exp\left(-\frac{\lambda^2}{4\omega}(\eta_p - \eta_q)^2\right) \tilde{E}_{pq}$$
$$+ \frac{1}{2} \sum_{pqrs} \tilde{g}_{pqrs} \exp\left(-\frac{\lambda^2}{4\omega}(\eta_p + \eta_r - \eta_q - \eta_s)^2\right) \tilde{e}_{pqrs} \qquad (17)$$
$$+ \frac{\lambda^2}{2} \sum_{pq} ((\tilde{\mathbf{d}} \cdot \boldsymbol{\epsilon})_{pp} - \eta_p)((\tilde{\mathbf{d}} \cdot \boldsymbol{\epsilon})_{qq} - \eta_q)(\tilde{e}_{ppqq} + \tilde{E}_{qp}\delta_{pq})$$

It is convenient to rewrite the above equation in an electronic Hamiltonian fashion

$$\langle H_{\text{SC}} \rangle_0 = \sum_{pq} \tilde{h}^b_{pq} \tilde{E}_{pq} + \frac{1}{2} \sum_{pqrs} \tilde{g}^b_{pqrs} \tilde{e}_{pqrs} \qquad (18)$$

where we have redefined the one and two electrons integrals:

$$\tilde{h}^b_{pq} = \tilde{h}_{pq} \exp\left(-\frac{\lambda^2}{4\omega}(\eta_p - \eta_q)^2\right) + \frac{\lambda^2}{2}((\tilde{\mathbf{d}} \cdot \boldsymbol{\epsilon})_{pp} - \eta_p)^2 \delta_{pq}, \qquad (19)$$

$$\tilde{g}^b_{pqrs} = \tilde{g}_{pqrs} \exp\left(-\frac{\lambda^2}{4\omega}(\eta_p + \eta_r - \eta_q - \eta_s)^2\right) + \lambda^2((\tilde{\mathbf{d}} \cdot \boldsymbol{\epsilon})_{pp} - \eta_p)((\tilde{\mathbf{d}} \cdot \boldsymbol{\epsilon})_{rr} - \eta_r)\delta_{pq}\delta_{rs}. \qquad (20)$$

The $\kappa$-$\kappa$ block elements of the SC-QED-HF Hessian read

$$\left(\frac{\partial^2 E}{\partial \kappa_{pq} \partial \kappa_{rs}}\right)_{\boldsymbol{\kappa}=\mathbf{0}} = (1 + P_{pq,rs}) \langle \text{HF} | \left[\left[\langle H_{\text{SC}} \rangle_0, E_{pq}\right], E^-_{rs}\right] | \text{HF} \rangle, \qquad (21)$$

where the vacuum averaged SC-transformed Hamiltonian $\langle H_{\text{SC}} \rangle_0$ involved in the commutators assumes the shape of a purely electronic Hamiltonian with redefined integrals. This electron-like redefinition is also useful because, once we have explicitly derived the equations for the Hessian elements, we can compare them with the ones coming from the Hartree Fock theory:

$$E^{(2)}_{\text{HF},aibj} = 4\left[\delta_{ab}\delta_{ij}(\epsilon_a - \epsilon_i) + 4g_{aibj} - g_{abji} - g_{ajbi}\right]. \qquad (22)$$

The Hessian elements in eq. (21) can be rewritten as

$$\left(\frac{\partial^2 E}{\partial \kappa_{pq} \partial \kappa_{rs}}\right)_{\boldsymbol{\kappa}=\mathbf{0}} = (1 + P_{pq,rs})(G_{pqrs} - G_{pqsr}) \quad (23)$$

where

$$G_{pqrs} = \langle \mathrm{HF}| \left[\left[\langle H_{\mathrm{SC}}\rangle_0, E_{pq}\right], E_{rs}\right] |\mathrm{HF}\rangle. \quad (24)$$

The first common commutator for both $G_{pqrs}$ and $G_{pqrs}$ is

$$\left[\langle H_{\mathrm{SC}}\rangle_0, E_{pq}\right] = \sum_t \left(\sum_{rs} U_{rt}\tilde{h}^b_{rs}U^*_{sp}\right)E_{tq} - \sum_u \left(\sum_{rs} U_{rq}\tilde{h}^b_{rs}U^*_{su}\right)E_{pu}$$
$$+ \sum_{txy}\left(\sum_{rsvz} U_{rt}U_{vx}\tilde{g}^b_{rsvz}U^*_{sp}U^*_{zy}\right)e_{tqxy} - \sum_{uxy}\left(\sum_{rsvz} U_{rq}U_{vy}\tilde{g}^b_{rsvz}U^*_{su}U^*_{zx}\right)e_{puxy}$$
$$\quad (25)$$

where $\mathbf{U}$ is the canonical to dipole basis transformation.

Transforming the $\tilde{h}^b_{pq}$ and $\tilde{g}^b_{pqrs}$ integrals back to the canonical basis and calculating the second commutator with $E_{rs}$ in $G_{pqrs}$, we obtain

$$G_{pqrs} = \langle \mathrm{HF}| \Big(\sum_t h^b_{tp}(E_{ts}\delta_{rq} - E_{rq}\delta_{ts}) - \sum_u h^b_{qu}(E_{ps}\delta_{ru} - E_{ru}\delta_{ps})$$
$$+ \sum_{tvz} g^b_{tpvz}(e_{tsvz}\delta_{rq} + e_{tqvs}\delta_{rz} - e_{rqvz}\delta_{ts} - e_{tqrz}\delta_{vs}) \quad (26)$$
$$- \sum_{uvz} g^b_{quvz}(e_{psvz}\delta_{ru} + e_{puvs}\delta_{rz} - e_{ruvz}\delta_{ps} - e_{purz}\delta_{vs})\Big)|\mathrm{HF}\rangle.$$

Now we can average the electronic operators on the Hartree Fock determinant and obtain the density matrices elements which, in the canonical basis, read

$$D_{pq} = \langle \text{HF} | E_{pq} | \text{HF} \rangle = 2\delta_{pq}\delta_{p,occ}, \tag{27}$$

$$d_{pqrs} = \langle \text{HF} | e_{pqrs} | \text{HF} \rangle = 4\delta_{pq}\delta_{rs}\delta_{p,occ}\delta_{r,occ} - 2\delta_{ps}\delta_{qr}\delta_{p,occ}\delta_{q,occ}. \tag{28}$$

Then, $G_{pqrs}$ becomes

$$G_{pqrs} = 2\Big(\delta_{rq}(h^b_{sp} + \sum_j^{occ}(2g^b_{sjj} - g^b_{sjjp})) - \delta_{ps}(h^b_{qr} + \sum_j^{occ}(2g^b_{qrjj} - g^b_{qjjr}))\Big)(\delta_{s,occ} - \delta_{r,occ}) \\ + (\delta_{q,occ} - \delta_{p,occ})(\delta_{s,occ} - \delta_{r,occ})\Big(g^b_{pqrs} - \frac{1}{2}g^b_{psrq}\Big), \tag{29}$$

while $G_{pqsr}$ is easily obtained from the previous equation by simply swapping the $r$ and $s$ indices. We can recognise from the above eq. (29) the presence of the Fock matrix elements for our electronic-like Hamiltonian (18), transformed back to the canonical basis

$$F_{pq} = h^b_{pq} + \sum_j^{occ}(2g^b_{pqjj} - g^b_{pjjq}) \tag{30}$$

which is related to the non-redundant Hartree-Fock gradient

$$f^\kappa_{pq} = \left(\frac{\partial E}{\partial \kappa_{pq}}\right)_{\boldsymbol{\kappa}=\boldsymbol{0}} = \langle \text{HF} | \left[\langle H_{\text{SC}}\rangle_0, E^-_{pq}\right] | \text{HF}\rangle = 2(F_{pq} - F_{qp})(\delta_{q,occ} - \delta_{p,occ}), \tag{31}$$

where

$$F_{pq} = \sum_{rs} V_{pr}\tilde{F}_{rs}V_{qs}. \tag{32}$$

and

$$\tilde{F}_{pq} = \frac{\partial E}{\partial \tilde{D}_{pq}} = \tilde{h}_{pq} \exp\left(-\frac{\lambda^2}{4\omega}(\eta_p - \eta_q)^2\right) + \frac{\lambda^2}{2}\delta_{pq}((\tilde{\mathbf{d}} \cdot \boldsymbol{\epsilon})_{pp} - \eta_p)^2$$
$$+ \frac{1}{2}\sum_{rs}(2\tilde{g}_{pqrs} - \tilde{g}_{psrq})\tilde{D}_{rs}\exp\left(-\frac{\lambda^2}{4\omega}(\eta_p + \eta_r - \eta_q - \eta_s)^2\right)$$
$$- \frac{\lambda^2}{2}\tilde{D}_{qp}((\tilde{\mathbf{d}} \cdot \boldsymbol{\epsilon})_{pp} - \eta_p)((\tilde{\mathbf{d}} \cdot \boldsymbol{\epsilon})_{qq} - \eta_q)$$
$$+ \lambda^2 \delta_{pq}((\tilde{\mathbf{d}} \cdot \boldsymbol{\epsilon})_{pp} - \eta_p)\sum_r \tilde{D}_{rr}((\tilde{\mathbf{d}} \cdot \boldsymbol{\epsilon})_{rr} - \eta_r). \quad (33)$$

By introducing the Fock matrix elements in $G_{pqrs}$ and $G_{pqsr}$ and substituting the last two in eq. (23), we obtain

$$\left(\frac{\partial^2 E}{\partial \kappa_{pq} \partial \kappa_{rs}}\right)_{\boldsymbol{\kappa}=\mathbf{0}} = (1 + P_{pq,rs})\Big(2(\delta_{qr}F_{ps} - \delta_{ps}F_{qr} + \delta_{qs}F_{pr} - \delta_{pr}F_{qs})(\delta_{s,occ} - \delta_{r,occ}) \\ + \Big[4(g^b_{pqrs} + g^b_{pqsr}) - 2(g^b_{psrq} + g^b_{prsq})\Big](\delta_{q,occ} - \delta_{p,occ})(\delta_{s,occ} - \delta_{r,occ})\Big). \quad (34)$$

Finally, by the application of the permutator $P_{pq,rs}$ we obtain

$$\left(\frac{\partial^2 E}{\partial \kappa_{pq} \partial \kappa_{rs}}\right)_{\boldsymbol{\kappa}=\mathbf{0}} = \Big[(\delta_{s,occ} - \delta_{r,occ}) - (\delta_{q,occ} - \delta_{p,occ})\Big]2(\delta_{qr}F_{ps} - \delta_{ps}F_{qr}) \\ + \Big[(\delta_{s,occ} - \delta_{r,occ}) + (\delta_{q,occ} - \delta_{p,occ})\Big]2(\delta_{qs}F_{pr} - \delta_{pr}F_{qs}) \\ + (\delta_{q,occ} - \delta_{p,occ})(\delta_{s,occ} - \delta_{r,occ})\Big[8(\mathring{g}_{pqrs} + \mathring{g}_{pqsr}) - 4(\mathring{g}_{psrq} + \mathring{g}_{prsq})\Big]. \quad (35)$$

From this last equation we can recognise that the only non-redundant derivatives are those arising from two $\kappa$s both respectively constituted by one occupied and one virtual indices. Considering that in eq. (12) we have only $\kappa_{pq}$ parameters where $p > q$, the non-vanishing elements of the $\kappa$-$\kappa$ block of the Hessian matrix are

$$\left(\frac{\partial^2 E}{\partial \kappa_{ai} \partial \kappa_{bj}}\right) = 4\Big[F_{ab}\delta_{ij} - F_{ij}\delta_{ab} + 2(g^b_{aibj} + g^b_{aijb}) - g^b_{abji} - g^b_{ajbi}\Big] \quad (36)$$

which can be further simplified recognising that the Fock matrix is diagonal in the canonical basis:

$$\left(\frac{\partial^2 E}{\partial \kappa_{ai} \partial \kappa_{bj}}\right)_{\boldsymbol{\kappa=0}} = 4\left[\delta_{ab}\delta_{ij}(\epsilon_a - \epsilon_i) + 2(g^b_{aibj} + g^b_{aijb}) - g^b_{abji} - \mathring{g}_{ajbi}\right]. \tag{37}$$

Equation (37) represents the $\kappa$-$\kappa$ block elements of the Hessian and by comparing it with the Hartree Fock Hessian in eq. (22) we can observe that the $g^b_{aibj}$ and $g^b_{aijb}$ terms don't sum up giving a factor of 4 like in the Hartree Fock case. This happens because, according to eq. (20), the redefined two electrons integrals pass from an eigth-fold symmetry to a fourth-fold one: the symmetry of interchanging two real orbitals belonging to the same electron and leaving unchanged the ones for the other electron is lost. Only interchanging both the couples is allowed due to the presence of the cavity Gaussian factors.

Lastly, we can observe how the $\kappa$-$\kappa$ block of the Hessian reduces to the Hartree Fock one at zero coupling restoring also the lost symmetry of the integrals:

$$g^b_{pqrs} \xrightarrow{\lambda=0} g_{pqrs}. \tag{38}$$

## S2.2 The η-η block

The Hessian elements for the $\eta$-$\eta$ block are

$$\left(\frac{\partial^2 E}{\partial \eta_r \partial \eta_s}\right)_{\boldsymbol{\kappa}=\mathbf{0}} = \frac{\lambda^2}{2\omega} \langle \text{HF}, 0 | \left[\tilde{E}_{ss}(b - b^\dagger), \left[\tilde{E}_{rr}(b - b^\dagger), H_{\text{SC}}\right]\right] | \text{HF}, 0 \rangle \tag{39}$$

where $H_{\text{SC}}$ is defined in eq. (15).

Unlike for the $\kappa$-$\kappa$ block of the Hessian, we cannot average $H_{\text{SC}}$ on the photonic vacuum and work with a simpler electronic-like Hamiltonian with redefined integrals. This happens because in the commutators in eq. (39) we have several photonic operators ($b$ and $b^\dagger$) so that the effect of the vacuum average is not trivial to evaluate.

However, we can notice which terms of $H_{\text{SC}}$ have a non-zero contribution by expanding the commutators to separate the photonic and electronic operators. Using

$$[AB, C] = A[B, C] + [A, C]B, \tag{40}$$

the first commutator can be written as

$$\left[\tilde{E}_{rr}(b - b^\dagger), \tilde{H}_{SC}\right] = \tilde{E}_{rr}\left[(b - b^\dagger), \tilde{H}_{SC}\right] + \left[\tilde{E}_{rr}, \tilde{H}_{SC}\right](b - b^\dagger) \tag{41}$$

and by substituting it into the second commutator we obtain:

$$\left[\tilde{E}_{ss}(b - b^\dagger), \tilde{E}_{rr}\left[(b - b^\dagger), \tilde{H}_{SC}\right] + \left[\tilde{E}_{rr}, \tilde{H}_{SC}\right](b - b^\dagger)\right]. \tag{42}$$

Using again eq. (40) we finally end up with the sum of the following 4 terms:

$$\tilde{E}_{ss}\tilde{E}_{rr}\big[(b-b^\dagger), \big[(b-b^\dagger), \tilde{H}_{SC}\big]\big], \tag{43}$$

$$\tilde{E}_{ss}\big[(b-b^\dagger), \big[\tilde{E}_{rr}, \tilde{H}_{SC}\big]\big](b-b^\dagger), \tag{44}$$

$$\tilde{E}_{rr}\big[\tilde{E}_{rr}, \big[(b-b^\dagger), \tilde{H}_{SC}\big]\big](b-b^\dagger), \tag{45}$$

$$\big[\tilde{E}_{ss}, \big[\tilde{E}_{rr}, \tilde{H}_{SC}\big]\big](b-b^\dagger)(b-b^\dagger). \tag{46}$$

Now, it's easy to verify that, once the vacuum average is taken, the addends in eqs. (44) and (45) vanish while for those in eqs. (43) and (46) only the purely photonic $\omega b^\dagger b$ and SC-electronic $H_e^{SC}$ in eq. (16) respectively contribute.

Equation 39 for the $\eta$-$\eta$ block elements of the Hessian reduces to

$$\left(\frac{\partial^2 E}{\partial \eta_r \partial \eta_s}\right)_{\boldsymbol{\kappa}=\mathbf{0}} = \frac{\lambda^2}{2\omega}\Big(\langle \text{HF}|\tilde{E}_{ss}\tilde{E}_{rr}|\text{HF}\rangle\langle 0|\big[(b-b^\dagger), \big[(b-b^\dagger), \omega b^\dagger b\big]\big]|0\rangle \\ + \langle \text{HF}, 0|\big[\tilde{E}_{ss}, \big[\tilde{E}_{rr}, H_e^{SC}\big]\big](b-b^\dagger)^2|\text{HF}, 0\rangle\Big). \tag{47}$$

The first term in the parenthesis is simply

$$\langle \text{HF}|\tilde{E}_{ss}\tilde{E}_{rr}|\text{HF}\rangle\langle 0|\big[(b-b^\dagger), \big[(b-b^\dagger), \omega b^\dagger b\big]\big]|0\rangle = 2\omega(\tilde{D}_{rr}\delta_{rs} + \tilde{d}_{rrss}) \tag{48}$$

where $\tilde{D}_{pq}$ and $\tilde{d}_{pqrs}$ are the one and two body density matrices in the dipole basis.

For the second more complex term we have:

12

$$\langle \text{HF}, 0 | \left[ \tilde{E}_{ss}, \left[ \tilde{E}_{rr}, H_e^{\text{SC}} \right] \right] (b - b^\dagger)^2 | \text{HF}, 0 \rangle =$$

$$= \sum_{pq} \tilde{h}_{pq} \langle \text{HF} | \left[ \tilde{E}_{ss}, \left[ \tilde{E}_{rr}, \tilde{E}_{pq} \right] \right] | \text{HF} \rangle \langle 0 | \exp\left( \frac{\lambda}{\sqrt{2\omega}} (\eta_p - \eta_q) (b - b^\dagger) \right) (b - b^\dagger)^2 | 0 \rangle$$

$$+ \frac{1}{2} \sum_{pqtu} \tilde{g}_{pqtu} \langle \text{HF} | \left[ \tilde{E}_{ss}, \left[ \tilde{E}_{rr}, \tilde{e}_{pqtu} \right] \right] | \text{HF} \rangle \langle 0 | \exp\left( \frac{\lambda}{\sqrt{2\omega}} (\eta_p + \eta_r - \eta_q - \eta_s) (b - b^\dagger) \right) (b - b^\dagger)^2 | 0 \rangle.$$

(49)

Using the following relation

$$\langle 0 | \exp(\alpha(b - b^\dagger))(b - b^\dagger)^2 | 0 \rangle = (\alpha^2 - 1) \exp\left( -\frac{\alpha^2}{2} \right)$$

(50)

and expliciting all the commutators in eq. (49), we end up with

$$\langle \text{HF}, 0 | \left[ \tilde{E}_{ss}, \left[ \tilde{E}_{rr}, H_{el}^{\text{SC}} \right] \right] (b - b^\dagger)^2 | \text{HF}, 0 \rangle =$$

$$= 2 \Bigg( \delta_{rs} \sum_q \tilde{h}_{rq}^a \tilde{D}_{rq} \Big( \frac{\lambda^2}{2\omega} (\eta_r - \eta_q)^2 - 1 \Big) - \tilde{h}_{rs}^a \tilde{D}_{rs} \Big( \frac{\lambda^2}{2\omega} (\eta_r - \eta_s)^2 - 1 \Big)$$

$$+ \delta_{rs} \sum_{pqt} \tilde{g}_{rpqt}^a \tilde{d}_{rpqt} \Big( \frac{\lambda^2}{2\omega} (\eta_r + \eta_q - \eta_p - \eta_t)^2 - 1 \Big)$$

$$- \sum_{qt} \tilde{g}_{rsqt}^a \tilde{d}_{rsqt} \Big( \frac{\lambda^2}{2\omega} (\eta_r + \eta_q - \eta_s - \eta_t)^2 - 1 \Big)$$

$$+ \sum_{qt} \tilde{g}_{rqst}^a \tilde{d}_{rqst} \Big( \frac{\lambda^2}{2\omega} (\eta_r + \eta_s - \eta_q - \eta_t)^2 - 1 \Big)$$

$$- \sum_{qt} \tilde{g}_{rqts}^a \tilde{d}_{rqts} \Big( \frac{\lambda^2}{2\omega} (\eta_r + \eta_t - \eta_q - \eta_s^2 - 1) \Big) \Bigg)$$

(51)

where we introduced the Gaussian scaled integrals $\bar{\tilde{h}}_{pq}$ and $\bar{\tilde{g}}_{pqrs}$ :

$$\tilde{h}_{pq}^a = \tilde{h}_{pq} \exp\left( -\frac{\lambda^2}{4\omega} (\eta_p - \eta_q)^2 \right),$$

(52)

$$\tilde{g}_{pqrs}^a = \tilde{g}_{pqrs} \exp\left( -\frac{\lambda^2}{4\omega} (\eta_p + \eta_r - \eta_q - \eta_s)^2 \right).$$

(53)

13

Finally, by substituting the terms in eqs. (48) and (51) into eq. (47) we end up with

$$\left(\frac{\partial^2 E}{\partial \eta_r \partial \eta_s}\right)_{\boldsymbol{\kappa}=0} = \delta_{rs}\left(\frac{\lambda^2}{\omega}\sum_q \tilde{h}^a_{rq}\tilde{D}_{rq}\left(\frac{\lambda^2}{2\omega}(\eta_r - \eta_q)^2 - 1\right) + \lambda^2\tilde{D}_{rr}\right.$$
$$\left. + \frac{\lambda^2}{\omega}\sum_{pqt}\tilde{g}^a_{rpqt}\tilde{d}_{rpqt}\left(\frac{\lambda^2}{2\omega}(\eta_r + \eta_q - \eta_p - \eta_t)^2 - 1\right)\right)$$
$$- \frac{\lambda^2}{\omega}\sum_{qt}\tilde{g}^a_{rsqt}\tilde{d}_{rsqt}\left(\frac{\lambda^2}{2\omega}(\eta_r + \eta_q - \eta_s - \eta_t)^2 - 1\right)$$
$$+ \frac{\lambda^2}{\omega}\sum_{qt}\tilde{g}^a_{rqst}\tilde{d}_{rqst}\left(\frac{\lambda^2}{2\omega}(\eta_r + \eta_s - \eta_q - \eta_t)^2 - 1\right)$$
$$- \frac{\lambda^2}{\omega}\sum_{qt}\tilde{g}^a_{rqts}\tilde{d}_{rqts}\left(\frac{\lambda^2}{2\omega}(\eta_r + \eta_t - \eta_q - \eta_s)^2 - 1\right)$$
$$+ \lambda^2\tilde{d}_{rrss} - \frac{\lambda^2}{\omega}\tilde{h}^a_{rs}\tilde{D}_{rs}\left(\frac{\lambda^2}{2\omega}(\eta_r - \eta_s)^2 - 1\right).$$

(54)

for the $\eta$-$\eta$ block elements of the SC-Hessian matrix.

Note that this block was derived in the dipole basis contrary to the $\kappa$-$\kappa$ block which was derived in the canonical one. Note also that this result is in totally agreement with the result we would have more tediously obtained by directly deriving from the $\eta$ gradient equation written in the dipole basis

$$\tilde{f}^\eta_r = \frac{\partial E}{\partial \eta_r} = \frac{\lambda^2}{\omega}\sum_q \tilde{h}^a_{rq}\tilde{D}_{rq}(\eta_q - \eta_r) - \lambda^2\tilde{D}_{rr}((\tilde{\mathbf{d}}\cdot\boldsymbol{\epsilon})_{rr} - \eta_r)$$
$$+ \frac{\lambda^2}{\omega}\sum_{pqt}\tilde{g}^a_{rpqt}\tilde{d}_{rpqt}(\eta_p + \eta_t - \eta_r - \eta_q)$$
$$- \lambda^2\sum_q \tilde{d}_{qqrr}((\tilde{\mathbf{d}}\cdot\boldsymbol{\epsilon})_{qq} - \eta_q).$$

(55)

To conclude, we point out that no redundancies are obtained for this block of the Hessian contrary to the $\kappa$-$\kappa$ block derivide in the canonical basis.

## S2.3 The $\kappa$-$\eta$ and $\eta$-$\kappa$ blocks

The Hessian elements for the $\kappa$-$\eta$ and $\eta$-$\kappa$ blocks are

$$\left(\frac{\partial^2 E}{\partial \kappa_{pq} \partial \eta_r}\right)_{\boldsymbol{\kappa}=\mathbf{0}} = \left(\frac{\partial^2 E}{\partial \eta_r \partial \kappa_{pq}}\right)_{\boldsymbol{\kappa}=\mathbf{0}} = \lambda \sqrt{\frac{2}{\omega}} \langle \text{HF}, 0| \left[\left[\tilde{E}_{rr}(b - b^\dagger), H_{\text{SC}}\right], E_{pq}\right] |\text{HF}, 0\rangle \quad (56)$$

where $H_{\text{SC}}$ is defined in eq. (15).

We point out that these blocks turn to be naturally identical and in the Hessian matrix they would simply be one the transpose of the other and vice versa.

Like for the $\eta$-$\eta$ block case, we cannot directly take the vacuum average on $H_{\text{SC}}$ and we should split the first commutator in order to separate the photonic and the electronic operators. Then, using eq. (41) and making observations on the resulting commutators looking forward on what would be the effect of taking the vacuum average, we obtain

$$\left(\frac{\partial^2 E}{\partial \kappa_{pq} \partial \eta_r}\right)_{\boldsymbol{\kappa}=\mathbf{0}} = \left(\frac{\partial^2 E}{\partial \eta_r \partial \kappa_{pq}}\right)_{\boldsymbol{\kappa}=\mathbf{0}} = \sum_{tu} U^*_{tp} \tilde{H}_{r,tu} U_{uq} \quad (57)$$

where $\tilde{H}_{r,tu}$ has the following definition

$$\tilde{H}_{r,tu} = 2\lambda^2 \sum_p ((\tilde{\mathbf{d}} \cdot \boldsymbol{\epsilon})_{pp} - \eta_p) \langle \text{HF}| \left[\tilde{E}_{tu}, \tilde{e}_{rrpp} + \tilde{E}_{pr}\delta_{rp}\right] |\text{HF}\rangle$$
$$+ \lambda \sqrt{\frac{2}{\omega}} \langle \text{HF}, 0| \left[\left[\tilde{E}_{rr}, H_e^{SC}\right](b - b^\dagger), \tilde{E}_{tu}\right] |\text{HF}, 0\rangle, \quad (58)$$

$\mathbf{U}$ is the canonical to dipole basis transformation and $H_e^{\text{SC}}$ is defined in eq. (16).

15

By expliciting the commutator in the first term we obtain

$$2\lambda^2 \sum_p ((\mathbf{d}\cdot\boldsymbol{\epsilon})_{pp} - \eta_p) \langle \mathrm{HF}| \left[\tilde{E}_{tu}, \tilde{e}_{rrpp} + \tilde{E}_{pr}\delta_{rp}\right] |\mathrm{HF}\rangle =$$

$$= (1 - P_{t,u})2\lambda^2 \left(\delta_{ru}\left(\tilde{D}_{tr}((\tilde{\mathbf{d}}\cdot\boldsymbol{\epsilon})_{rr} - \eta_r) + \sum_p ((\tilde{\mathbf{d}}\cdot\boldsymbol{\epsilon})_{pp} - \eta_p)\tilde{d}_{trpp}\right) + \tilde{d}_{rrtu}((\tilde{\mathbf{d}}\cdot\boldsymbol{\epsilon})_{uu} - \eta_u)\right) \quad (59)$$

where $P_{t,u}$ swaps the $t$ and $u$ indices.

For the second more complex term, by expliciting $H_e^{\mathrm{SC}}$ given eq. (16), we have

$$\lambda\sqrt{\frac{2}{\omega}} \langle \mathrm{HF}, 0| \left[\left[\tilde{E}_{rr}, H_e^{\mathrm{SC}}\right](b - b^\dagger), \tilde{E}_{tu}\right] |\mathrm{HF}, 0\rangle =$$

$$= \lambda\sqrt{\frac{2}{\omega}} \sum_{pq} \tilde{h}_{pq} \langle \mathrm{HF}| \left[\left[\tilde{E}_{rr}, \tilde{E}_{pq}\right], \tilde{E}_{tu}\right] |\mathrm{HF}\rangle \langle 0| \exp\left(\frac{\lambda}{\sqrt{2\omega}}(\eta_p - \eta_q)(b - b^\dagger)\right) b^\dagger |0\rangle$$

$$+ \frac{\lambda}{\sqrt{2\omega}} \sum_{pqvz} \tilde{g}_{pqvz} \langle \mathrm{HF}| \left[\left[\tilde{E}_{rr}, \tilde{e}_{pqvz}\right], \tilde{E}_{tu}\right] |\mathrm{HF}\rangle \langle 0| \exp\left(\frac{\lambda}{\sqrt{2\omega}}(\eta_p + \eta_v - \eta_q - \eta_z)(b - b^\dagger)\right) b^\dagger |0\rangle. \quad (60)$$

Using the following relation

$$\langle 0| \exp(\alpha(b - b^\dagger))b^\dagger |0\rangle = \alpha \exp\left(-\frac{\alpha^2}{2}\right) \quad (61)$$

and expanding all the commutators, we obtain

$$\lambda\sqrt{\frac{2}{\omega}} \langle \mathrm{HF}, 0| \left[\left[\tilde{E}_{rr}, H_e^{\mathrm{SC}}\right](b - b^\dagger), \tilde{E}_{tu}\right] |\mathrm{HF}, 0\rangle =$$

$$= (1 - P_{t,u})\frac{\lambda^2}{\omega}\left(\delta_{ru}\left(\sum_q \tilde{h}^c_{qr}\tilde{D}_{qt} + \sum_{qvz} \tilde{g}^c_{qrvz}\tilde{d}_{qtvz}\right) + \tilde{h}^c_{rt}\tilde{D}_{ru} \right.$$

$$\left. + \sum_{vz} \tilde{g}^c_{rtvz}\tilde{d}_{ruvz} - \sum_{qz} \tilde{g}^c_{rquz}\tilde{d}_{rqtz} + \sum_{qv} \tilde{g}^c_{rqvt}\tilde{d}_{rqvu}\right) \quad (62)$$

where the $\tilde{h}^c_{pq}$ and $\tilde{g}^c_{pqrs}$ integrals are defined as follows:

$$\tilde{h}^c_{pq} = \tilde{h}_{pq} \exp\left(-\frac{\lambda^2}{4\omega}(\eta_p - \eta_q)^2\right)(\eta_q - \eta_p) , \tag{63}$$

$$\tilde{g}^c_{pqrs} = \tilde{g}_{pqrs} \exp\left(-\frac{\lambda^2}{4\omega}(\eta_p + \eta_r - \eta_q - \eta_s)^2\right)(\eta_q + \eta_s - \eta_p - \eta_r) . \tag{64}$$

Finally, by adding together the terms in eqs. (59) and (62) we end up with the following formula for the $\kappa$-$\eta$ and $\eta$-$\kappa$ blocks elements of the SC-Hessian:

$$\left(\frac{\partial^2 E}{\partial \kappa_{pq} \partial \eta_r}\right)_{\boldsymbol{\kappa}=\mathbf{0}} = \left(\frac{\partial^2 E}{\partial \eta_r \partial \kappa_{pq}}\right)_{\boldsymbol{\kappa}=\mathbf{0}} = \sum_{tu} U^*_{tp} \tilde{H}_{r,tu} U_{uq} \tag{65}$$

where $\tilde{H}_{r,tu}$ is defined as:

$$\tilde{H}_{r,tu} = (1 - P_{t,u})\left(\frac{\lambda^2}{\omega}\left(\delta_{ru}\left(\sum_q \tilde{h}^c_{qr}\tilde{D}_{qt} + \sum_{qvz}\tilde{g}^c_{qrvz}\tilde{d}_{qtvz}\right) + \tilde{h}^c_{rt}\tilde{D}_{ru} \right.\right.$$
$$\left.+ \sum_{vz}\tilde{g}^c_{rtvz}\tilde{d}_{ruvz} - \sum_{qz}\tilde{g}^c_{rquz}\tilde{d}_{rqtz} + \sum_{qv}\tilde{g}^c_{rqvt}\tilde{d}_{rqvu}\right)$$
$$\left.+ 2\lambda^2\left(\delta_{ru}\left(\tilde{D}_{tr}((\tilde{\mathbf{d}}\cdot\boldsymbol{\epsilon})_{rr} - \eta_r) + \sum_p((\tilde{\mathbf{d}}\cdot\boldsymbol{\epsilon})_{pp} - \eta_p)\tilde{d}_{trpp}\right) + \tilde{d}_{rrtu}((\tilde{\mathbf{d}}\cdot\boldsymbol{\epsilon})_{uu} - \eta_u)\right)\right). \tag{66}$$

We point out that no apparent redundancies emerged from the derivation of the mixed parameter blocks because we derived $\tilde{H}_{r,tu}$ in the dipole basis without explicitly transforming back the $t$ and $u$ indices to the canonical one. However, by doing so we would obtain the usual occupied-occupied and virtual-virtual redundancies brought by the $\kappa$-parameters:

$$\left(\frac{\partial^2 E}{\partial \kappa_{ai} \partial \eta_r}\right)_{\boldsymbol{\kappa}=\mathbf{0}} = \left(\frac{\partial^2 E}{\partial \eta_r \partial \kappa_{ai}}\right)_{\boldsymbol{\kappa}=\mathbf{0}} = \sum_{tu} U^*_{ta} \tilde{H}_{r,tu} U_{ui} . \tag{67}$$

# S3 Trust region Newton-Raphson algorithm

If we consider an Hamiltonian $H$ and a parametrized wave function $|\psi(\mathbf{z})\rangle$ where $\mathbf{z}$ is the parameters vector, the energy hypersurface reads

$$E(\mathbf{z}) = \langle \psi(\mathbf{z}) | H | \psi(\mathbf{z}) \rangle. \tag{68}$$

By approximating the above equation to a quadratic model we obtain

$$E(\mathbf{z} + \boldsymbol{\delta}) \approx Q(\boldsymbol{\delta}) = E(\mathbf{z}) + \mathbf{g}^T(\mathbf{z})\boldsymbol{\delta} + \frac{1}{2}\boldsymbol{\delta}^T \mathbf{G}(\mathbf{z})\boldsymbol{\delta} \tag{69}$$

where $\mathbf{g}$ is the gradient vector, $\mathbf{G}$ is the Hessian matrix and $\boldsymbol{\delta}$ is the step vector. If we impose the stationary conditions for the optimization

$$\frac{\partial Q(\boldsymbol{\delta})}{\partial \boldsymbol{\delta}} = 0, \tag{70}$$

we obtain the following set of linear equations:

$$\mathbf{G}(\mathbf{z})\boldsymbol{\delta} = -\mathbf{g}(\mathbf{z}). \tag{71}$$

We can show that this model is quadratically convergent by expanding the gradient vector

$$\mathbf{g}(\mathbf{z} + \boldsymbol{\delta}) = \mathbf{g}(\mathbf{z}) + \mathbf{G}(\mathbf{z})\boldsymbol{\delta} + \mathcal{O}(\boldsymbol{\delta}^2) \tag{72}$$

which turn to scale quadratically with the error.

Convergence is guaranteed if the Hessian is positive definite, but in the initial steps this is not always the case as the quadratic approximation holds only locally close to a minimum. To strengthen the robustness of the algorithm we can introduce a Lagrange constrain to the

quadratic model such that the step is accepted only if it lies within a trusted hypervolume around the current iterate. Then, we can define a Lagrangian

$$\mathcal{L}(\boldsymbol{\delta}, \mu) = Q(\boldsymbol{\delta}) - \frac{1}{2}\mu \left( \|\boldsymbol{\delta}\|^2 - R_t^2 \right) \tag{73}$$

where $\mu$ is the Lagrange multiplier and $R_t$ the trust radius. From this Lagrangian we can obtain the active constrain condition as well as the new set of linear equations for determine the step, the so called Karush-Kuhn-Tucker (KKT) conditions:

$$\frac{\partial \mathcal{L}}{\partial \mu} = \|\boldsymbol{\delta}\|^2 - R_t^2 = 0, \tag{74}$$

$$\frac{\partial \mathcal{L}}{\partial \boldsymbol{\delta}} = \mathbf{G}(\mathbf{z})\boldsymbol{\delta} + \mathbf{g}(\mathbf{z}) - \mu\boldsymbol{\delta} = 0 \rightarrow \left(\mathbf{G}(\mathbf{z}) - \mu\mathbf{I}\right)\boldsymbol{\delta} = -\mathbf{g}(\mathbf{z}). \tag{75}$$

The role of the Lagrange multiplier $\mu$ is to shift the eigenvalues of the Hessian matrix $\mathbf{G}$ such that it becomes positive definite and convergence to a stationary minima is ensured. To this purpose, a multiplier smaller than the smallest eigenvalue of $\mathbf{G}$ is the most suitable choice. The common way of achieving this is to define an augmented Hessian

$$\tilde{\mathbf{G}} = \begin{pmatrix} & & & 1 \\ & \mathbf{G} & & \vdots \\ & & & 1 \\ \hline 1 & \cdots & 1 & 0 \end{pmatrix} \tag{76}$$

whose first eigenvalue is smaller then all the eigenvalues of $\mathbf{G}$ thanks to the Hylleraas-Undheim-MacDonald theorem. A successful variant of this approach is the NEO algorithm.

Lastly, we mention that the trust radius is updated after each iteration step by using the Fletcher method based on the ratio between the actual energy change and the one predicted by the approximate quadratic model.

## S4 Hessian linear transformations

In the trust region Newton-Raphson algorithm the optimization is carried out by solving iteratively the following set of linear equations

$$\left(\mathbf{E}_n^{(2)} - \mu \mathbf{I}\right) \Delta \mathbf{z}_n = -\mathbf{E}_n^{(1)} \tag{77}$$

featuring the level-shifted Hessian at the n-th iteration $(\mathbf{E}_n^{(2)} - \mu \mathbf{I})$, the new step in the variables $(\Delta \mathbf{z}_n)$ and the local gradient $(\mathbf{E}_n^{(1)})$ defined as

$$\mathbf{E}^{(1)} = \begin{pmatrix} \partial E/\partial \boldsymbol{\kappa} \\ \partial E/\partial \boldsymbol{\eta} \end{pmatrix} \equiv \begin{pmatrix} \mathbf{f}^{\kappa} \\ \mathbf{f}^{\eta} \end{pmatrix}. \tag{78}$$

In principle, to resolve this set of equations, the Hessian matrix should be calculated at each n-th iteration step and then be inverted, but this is very computationally demanding due to the large number of the non-redudant $\kappa$-parameters. For this reason the common way for defining Newton steps is to calculate directly the Hessian linear transformations on a trial step vector

$$\boldsymbol{\sigma} = \mathbf{E}^{(2)} \Delta \mathbf{z} \tag{79}$$

and solve iteratively with a Davidson algorithm.

Considering our SC-parametrization in eq. (11) composed by two different classes of parameters, the linear transformations equations read as

$$\boldsymbol{\sigma} = \begin{pmatrix} \mathbf{E}^{\kappa\kappa} & \mathbf{E}^{\kappa\eta} \\ \mathbf{E}^{\eta\kappa} & \mathbf{E}^{\eta\eta} \end{pmatrix} \begin{pmatrix} \Delta \boldsymbol{\kappa} \\ \Delta \boldsymbol{\eta} \end{pmatrix} = \begin{pmatrix} \mathbf{E}^{\kappa\kappa} \Delta \boldsymbol{\kappa} + \mathbf{E}^{\kappa\eta} \Delta \boldsymbol{\eta} \\ \mathbf{E}^{\eta\kappa} \Delta \boldsymbol{\kappa} + \mathbf{E}^{\eta\eta} \Delta \boldsymbol{\eta} \end{pmatrix}. \tag{80}$$

In the following sections each term in eq. (80) is derived in terms of gradient elements.

## S4.1 Pure $\kappa$ linear transformations

Starting with the purely $\kappa$-transformations, we have

$$(\mathbf{E}^{\kappa\kappa}\Delta\boldsymbol{\kappa})_{pq} = \frac{1+P_{pq,rs}}{2} \sum_{r>s} \langle \text{HF}, 0 | \left[\left[H_{\text{SC}}, E_{pq}^-\right], E_{rs}^-\right] | \text{HF}, 0 \rangle \Delta\kappa_{rs} \quad (81)$$

which can be easily rearranged as

$$(\mathbf{E}^{\kappa\kappa}\Delta\boldsymbol{\kappa})_{pq} = 2 \langle \text{HF} | \left[\left[\langle H_{\text{SC}}\rangle_0, \Delta\hat{\kappa}\right], E_{pq}\right] | HF \rangle + \langle HF | \left[\left[\Delta\hat{\kappa}, E_{pq}\right], \langle H_{\text{SC}}\rangle_0\right] | HF \rangle \quad (82)$$

where $\langle H_{SC}\rangle_0$ is an electronic-like Hamiltonian defined in eq. (18), while $\Delta\hat{\kappa}$ is defined as

$$\Delta\hat{\kappa} = \sum_{pq} \Delta\kappa_{pq} E_{pq}. \quad (83)$$

For the first term in eq. (82) we can observe that the commutator between $\langle H_{\text{SC}}\rangle_0$ and $\Delta\hat{\kappa}$ remains an electronic-like Hamiltonian

$$\langle H_{\text{SC}}\rangle_0^{\Delta\kappa} = \left[\langle H_{\text{SC}}\rangle_0, \Delta\hat{\kappa}\right] = \sum_{pq} h_{pq}^{b,\Delta\kappa} E_{pq} + \frac{1}{2} \sum_{pqrs} g_{pqrs}^{b,\Delta\kappa} e_{pqrs} \quad (84)$$

where the integrals in eqs. (19) and (20) (transformed back to the canonical basis) are $\Delta\kappa$-one-index transformed:

$$h_{pq}^{b,\Delta\kappa} = \sum_t (\Delta\kappa_{tp} h_{tq}^b + \Delta\kappa_{tq} h_{pt}^b), \quad (85)$$

$$g_{pqrs}^{b,\Delta\kappa} = \sum_t (\Delta\kappa_{tp} g_{tqrs}^b + \Delta\kappa_{tq} g_{ptrs}^b + \Delta\kappa_{tr} g_{pqts}^b + \Delta\kappa_{ts} g_{pqrt}^b). \quad (86)$$

For the second term in eq. (82), instead, we have

$$\langle \text{HF} | \left[ \left[ \Delta \hat{\kappa}, E_{pq} \right], \langle H_{\text{SC}} \rangle_0 \right] | \text{HF} \rangle =$$
$$= \sum_r \left( \Delta \kappa_{pr} \langle \text{HF} | \left[ E_{rq}, \langle H_{\text{SC}} \rangle_0 \right] | \text{HF} \rangle - \langle \text{HF} | \left[ E_{pr}, \langle H_{\text{SC}} \rangle_0 \right] | \text{HF} \rangle \Delta \kappa_{rq} \right). \tag{87}$$

By substituting eqs. (84) and (87) in eq. (82) we end up with the following definition for the purely $\kappa$-transformations:

$$\left( \mathbf{E}^{\kappa\kappa} \Delta \boldsymbol{\kappa} \right)_{pq} = f_{pq}^{\kappa}(h_{pq}^{b,\Delta\kappa}, g_{pqrs}^{b,\Delta\kappa}) + \frac{1}{2} \left( \left[ \Delta \boldsymbol{\kappa}, \mathbf{f}^{\kappa} \right] \right)_{pq}. \tag{88}$$

This result is completely analogous with the one found in the Hartree Fock which is recovered by simply setting $\lambda = 0$.

## S4.2 Pure $\eta$ linear transformations

Moving to the purely $\eta$-transformations, we have

$$\left(\mathbf{E}_{\eta\eta}^{(2)}\Delta\boldsymbol{\eta}\right)_r = \frac{\lambda}{\sqrt{2\omega}} \langle\mathrm{HF}, 0| \left[\tilde{E}_{rr}(b - b^\dagger), [\Delta\hat{\eta}, H_{\mathrm{SC}}]\right] |\mathrm{HF}, 0\rangle \tag{89}$$

where $\Delta\hat{\eta}$ is defined as:

$$\Delta\hat{\eta} = \frac{\lambda}{\sqrt{2\omega}} \sum_s \Delta\eta_s \tilde{E}_{ss}(b - b^\dagger). \tag{90}$$

Calculating the first commutator we obtain the $\Delta\eta$-transformed SC-Hamiltonian:

$$\begin{aligned}
H_{\mathrm{SC}}^{\Delta\eta} &= \left[\Delta\hat{\eta}, H_{\mathrm{SC}}\right] \\
&= \frac{\lambda}{\sqrt{2\omega}} \sum_{pq} \tilde{h}_{pq}(\Delta\eta_p - \Delta\eta_q)\tilde{E}_{pq} \exp\left(\frac{\lambda}{\sqrt{2\omega}}(\eta_p - \eta_q)(b - b^\dagger)\right)(b - b^\dagger) \\
&+ \frac{\lambda}{\sqrt{2\omega}} \sum_{pqrs} \frac{\tilde{g}_{pqrs}}{2}(\Delta\eta_p - \Delta\eta_q + \Delta\eta_r - \Delta\eta_s)\tilde{e}_{pqrs} \exp\left(\frac{\lambda}{\sqrt{2\omega}}(\eta_p - \eta_q + \eta_r - \eta_s)(b - b^\dagger)\right)(b - b^\dagger) \\
&+ \lambda\sqrt{\frac{\omega}{2}} \sum_s \Delta\eta_s \tilde{E}_{ss}(b + b^\dagger) - \frac{\lambda^2}{\omega} \sum_{ps}((\tilde{\mathbf{d}} \cdot \boldsymbol{\epsilon})_{pp} - \eta_p)\Delta\eta_s(\tilde{e}_{sspp} + \tilde{E}_{ps}\delta_{ps}).
\end{aligned} \tag{91}$$

Substituting the $\Delta\eta$-transformed Hamiltonian just found inside the second commutator in eq. (54), we obtain:

$$\begin{aligned}
\left[\tilde{E}_{rr}(b - b^\dagger), H_{\mathrm{SC}}^{\Delta\eta}\right] &= \\
&= \frac{\lambda}{\sqrt{2\omega}} \sum_{pq} \tilde{h}_{pq}(\Delta\eta_p - \Delta\eta_q)\left[\tilde{E}_{rr}, \tilde{E}_{pq}\right] \exp\left(\frac{\lambda}{\sqrt{2\omega}}(\eta_p - \eta_q)(b - b^\dagger)\right)(b - b^\dagger)^2 \\
&+ \frac{\lambda}{\sqrt{2\omega}} \sum_{pqrs} \frac{\tilde{g}_{pqrs}}{2}(\Delta\eta_p - \Delta\eta_q + \Delta\eta_r - \Delta\eta_s)\left[\tilde{E}_{rr}, \tilde{e}_{pqrs}\right] \exp\left(\frac{\lambda}{\sqrt{2\omega}}(\eta_p - \eta_q + \eta_r - \eta_s)(b - b^\dagger)\right)(b - b^\dagger)^2 \\
&+ \lambda\sqrt{2\omega} \sum_s \Delta\eta_s(\tilde{e}_{rrss} + \tilde{E}_{sr}\delta_{rs}).
\end{aligned} \tag{92}$$

Expliciting the last commutators in the electronic part and averaging over the reference wave function we end up with

$$(\mathbf{E}^{\eta\eta}\Delta\boldsymbol{\eta})_r = \frac{\lambda^2}{\omega}\sum_q \tilde{h}^{a,\eta}_{rq}(\Delta\eta_q - \Delta\eta_r)\tilde{D}_{rq}$$
$$+ \frac{\lambda^2}{\omega}\sum_{qvz}\tilde{g}^{a,\eta}_{rqvz}(\Delta\eta_q - \Delta\eta_r + \Delta\eta_z - \Delta\eta_v)\tilde{d}_{rqvz} \quad (93)$$
$$+ \lambda^2\sum_s \Delta\eta_s \tilde{d}_{rrss} + \lambda^2 \Delta\eta_r \tilde{D}_{rr}$$

where the integrals $\tilde{h}^{a,\eta}_{pq}$ and $\tilde{g}^{a,\eta}_{pqrs}$ are

$$\tilde{h}^{a,\eta}_{pq} = \tilde{h}^a_{pq}\left(1 - \frac{\lambda^2}{2\omega}(\eta_p - \eta_q)^2\right), \quad (94)$$

$$\tilde{g}^{a,\eta}_{pqrs} = \tilde{g}^a_{pqrs}\left(1 - \frac{\lambda^2}{2\omega}(\eta_p - \eta_q + \eta_r - \eta_s)^2\right). \quad (95)$$

Comparing eq. (93) with the one for the $\eta$-gradient elements in eq. (55) we are suggested to rewrite the purely $\eta$-transformations as

$$(\mathbf{E}^{\eta\eta}\Delta\boldsymbol{\eta})_r = \tilde{f}^{\Delta\eta}_r(\tilde{h}^{a,\eta}_{pq}, \tilde{g}^{a,\eta}_{pqrs}) + \lambda^2\sum_q (\tilde{\mathbf{d}}\cdot\boldsymbol{\epsilon})_{qq}\tilde{d}_{qqrr} + \lambda^2(\tilde{\mathbf{d}}\cdot\boldsymbol{\epsilon})_{rr}\tilde{D}_{rr} \quad (96)$$

where we see the presence of the $\Delta\eta$-gradient calculated with $\eta$-transformed integrals. The same result can of course be obtained by directly contracting one index of the $\eta$-$\eta$ Hessian in eq. (54) with the $\eta$-parameters.

## S4.3 Mixed $\kappa$ linear transformations

Jumping to the mixed parameter block contribute to the linear transformations in eq. (80), we notice that this block is either contracted with the $\kappa$-parameters or the $\eta$-parameters.

Starting with the transformations of the $\kappa$-parameters, we have

$$(\mathbf{E}^{\eta\kappa}\Delta\boldsymbol{\kappa})_r = \sum_{pq} \Delta\kappa_{pq} \sum_{tu} U^*_{tp}\tilde{H}_{r,tu}U_{uq} = \sum_{tu}\Delta\tilde{\kappa}_{tu}\tilde{H}_{r,tu} \qquad (97)$$

where $\Delta\tilde{\kappa}_{pq}$ are the $\kappa$-steps transformed to the dipole basis. Proceeding with the calculation of the last equation directly through the definition of $\tilde{H}_{r,tu}$ in eq. (66), we obtain

$$\begin{aligned}(\mathbf{E}^{\eta\kappa}\Delta\boldsymbol{\kappa})_r &= 2\frac{\lambda^2}{\omega}\sum_q \tilde{h}^c_{rq}\sum_t(\Delta\tilde{\kappa}_{rt}\tilde{D}_{tq} + \Delta\tilde{\kappa}_{qt}\tilde{D}_{rt}) \\ &+ 2\frac{\lambda^2}{\omega}\sum_{qvz} \tilde{g}^c_{rqvz}\sum_t(\Delta\tilde{k}_{rt}\tilde{d}_{tqvz} + \Delta\tilde{k}_{qt}\tilde{d}_{rtvz} + \Delta\tilde{k}_{vt}\tilde{d}_{rqtz} + \Delta\tilde{k}_{st}\tilde{d}_{rqvt}) \\ &- 2\lambda^2\sum_p((\tilde{\mathbf{d}}\cdot\boldsymbol{\epsilon})_{pp} - \eta_p)\sum_t(\Delta\tilde{k}_{rt}\tilde{d}_{trpp} + \Delta\tilde{k}_{rt}\tilde{d}_{rtpp} + \Delta\tilde{k}_{pt}\tilde{d}_{rrtp} + \Delta\tilde{k}_{pt}\tilde{d}_{rrpt}) \\ &- 2\lambda^2((\tilde{\mathbf{d}}\cdot\boldsymbol{\epsilon})_{rr} - \eta_r)\sum_t(\Delta\tilde{\kappa}_{tr}\tilde{D}_{rt} + \Delta\tilde{\kappa}_{tr}\tilde{D}_{tr})\end{aligned} \qquad (98)$$

which, comparing it with the $\eta$-gradient equation in eq. (55), can be rewritten as two times the $\eta$-gradient calculated using $\Delta\kappa$-one-index transformed density matrices:

$$(\mathbf{E}^{\eta\kappa}\Delta\boldsymbol{\kappa})_r = 2\tilde{f}^{\Delta\eta}_r(\mathbf{D}^{\Delta\kappa},\ \mathbf{d}^{\Delta\kappa}) \qquad (99)$$

with

$$\tilde{D}^{\Delta\kappa}_{pq} = \sum_t(\Delta\tilde{\kappa}_{pt}\tilde{D}_{tq} + \Delta\tilde{\kappa}_{qt}\tilde{D}_{pt}), \qquad (100)$$

$$\tilde{d}^{\Delta\kappa}_{pqrs} = \sum_t(\Delta\tilde{\kappa}_{pt}\tilde{d}_{tqrs} + \Delta\tilde{\kappa}_{qt}\tilde{d}_{ptrs}\Delta\tilde{\kappa}_{rt}\tilde{d}_{pqts} + \Delta\tilde{\kappa}_{st}\tilde{d}_{pqrt}). \qquad (101)$$

## S4.4 Mixed $\eta$ linear transformations

Finally, moving to the transformation of the $\eta$-parameters through the mixed block, we have

$$(\mathbf{E}^{\kappa\eta}\Delta\boldsymbol{\eta})_{pq} = \lambda\sqrt{\frac{2}{\omega}}\sum_r \langle\text{HF}, 0|\left[\left[\tilde{E}_{rr}(b - b^\dagger), H_{\text{SC}}\right], E_{pq}\right]|\text{HF}, 0\rangle \Delta\eta_r. \qquad (102)$$

Using eq. (41) for the inner commutator we obtain

$$(\mathbf{E}^{\kappa\eta}\Delta\boldsymbol{\eta})_{pq} = \lambda\sqrt{\frac{2}{\omega}}\sum_r \Delta\eta_r \Big(\langle\text{HF}, 0|\left[\tilde{E}_{rr}\left[(b - b^\dagger), H_{\text{SC}}\right], E_{pq}\right]|\text{HF}, 0\rangle \\ + \langle\text{HF}, 0|\left[\left[\tilde{E}_{rr}, H_{\text{SC}}\right](b - b^\dagger), E_{pq}\right]|\text{HF}, 0\rangle\Big) \qquad (103)$$

where for the first term only the bilinear term of the SC-transformed Hamiltonian contributes, while for the second only the electronic one in eq. (16) does.

For the second addend in eq. (103): expliciting the inner commutator, taking the vacuum average and exploiting the particle antisymmetry of eqs. (63) and (64), we obtain

$$\sum_r \Delta\eta_r \langle\text{HF}, 0|\left[\left[\tilde{E}_{rr}, H_{\text{SC}}\right](b - b^\dagger), E_{pq}\right]|\text{HF}, 0\rangle = \frac{\lambda}{\sqrt{2\omega}}\langle\text{HF}|\left[H^{\#}_{tmp}, E_{pq}\right]|\text{HF}\rangle \qquad (104)$$

where

$$H^{\#}_{tmp} = \sum_{rq}\tilde{h}^c_{rq}(\Delta\eta_r - \Delta\eta_q)\tilde{E}_{rq} + \frac{1}{2}\sum_{rqvz}\tilde{g}^c_{rqvz}(\Delta\eta_r - \Delta\eta_q + \Delta\eta_v - \Delta\eta_z)\tilde{e}_{rqvz} \qquad (105)$$

is an electronic-like Hamiltonian inside the definition of an Hartree Fock $\kappa$-gradient.

For the first addend in eq. (103): proceeding as we did before we obtain

$$\sum_r \Delta\eta_r \langle \text{HF}, 0 | \left[ \tilde{E}_{rr}[(b - b^\dagger), \tilde{H}_{SC}], E_{pq} \right] | \text{HF}, 0 \rangle = \\ = -\lambda\sqrt{2\omega} \langle \text{HF} | \left[ \sum_{rs} ((\tilde{\mathbf{d}} \cdot \boldsymbol{\epsilon})_{rr} - \eta_r)\Delta\eta_s (\tilde{e}_{rrss} + \tilde{E}_{rs}\delta_{rs}), E_{pq} \right] | \text{HF} \rangle . \quad (106)$$

To further continue with the derivation is useful to give the particle symmetry carried by the electronic operators on left hand side of the commutator to the product $((\tilde{\mathbf{d}} \cdot \boldsymbol{\epsilon})_{rr} - \eta_r)\Delta\eta_s$:

$$\sum_{rs} ((\tilde{\mathbf{d}} \cdot \boldsymbol{\epsilon})_{rr} - \eta_r)\Delta\eta_s \, \tilde{e}_{rrss} = \frac{1}{2} \sum_{rs} \left( ((\tilde{\mathbf{d}} \cdot \boldsymbol{\epsilon})_{rr} - \eta_r)\Delta\eta_s + ((\tilde{\mathbf{d}} \cdot \boldsymbol{\epsilon})_{ss} - \eta_s)\Delta\eta_r \right) \tilde{e}_{rrss}. \quad (107)$$

Using this symmetrization is possible to reabsorb all the first addend inside the second in eq. (104) redefining the one and two electronic integrals as follows:

$$\sum_r \langle \text{HF}, 0 | \left[ [\tilde{E}_{rr}(b - b^\dagger), H_{SC}], E_{pq} \right] | \text{HF}, 0 \rangle \Delta\eta_r = \frac{\lambda}{\sqrt{2\omega}} \langle \text{HF} | \left[ H_e^{\#}, E_{pq} \right] | \text{HF} \rangle \quad (108)$$

where the new electronic-like Hamiltonian $H_{el}^{\#}$ is

$$H_e^{\#} = \sum_{pq} \tilde{h}_{pq}^{\#} \tilde{E}_{pq} + \frac{1}{2} \sum_{pqrs} \tilde{g}_{pqrs}^{\#} \tilde{e}_{pqrs} \quad (109)$$

with

$$\tilde{h}_{pq}^{\#} = \tilde{h}_{pq}^c (\Delta\eta_p - \Delta\eta_q) - 2\omega((\tilde{\mathbf{d}} \cdot \boldsymbol{\epsilon})_{pp} - \eta_p)\Delta\eta_p \delta_{pq}, \quad (110)$$

$$\tilde{g}_{pqrs}^{\#} = \tilde{g}_{pqrs}^c (\Delta\eta_p - \Delta\eta_q + \Delta\eta_r - \Delta\eta_s) - 2\omega \left( ((\tilde{\mathbf{d}} \cdot \boldsymbol{\epsilon})_{pp} - \eta_p)\Delta\eta_r + ((\tilde{\mathbf{d}} \cdot \boldsymbol{\epsilon})_{rr} - \eta_r)\Delta\eta_p \right) \delta_{pq}\delta_{rz}. \quad (111)$$

Finally, we can substitute eq. (108) in eq. (102) and obtain

27

$$\left(\mathbf{E}^{\kappa\eta}\Delta\boldsymbol{\eta}\right)_{pq} = \frac{\lambda^2}{\omega}\left\langle\mathrm{HF}\right|\left[H_e^{\#}, E_{pq}\right]\left|\mathrm{HF}\right\rangle \tag{112}$$

in which we recognise the $\kappa$-gradient calculated with the transformed integrals $\tilde{h}_{pq}^{\#}$ and $\tilde{g}_{pqrs}^{\#}$ transformed back to the canonical basis:

$$\left(\mathbf{E}^{\kappa\eta}\Delta\boldsymbol{\eta}\right)_{pq} = \frac{\lambda^2}{2\omega} f_{pq}^{\kappa}(h_{pq}^{\#}, g_{pqrs}^{\#}) \tag{113}$$

where

$$h_{pq}^{\#} = \sum_{tu} U_{tp}\tilde{h}_{tu}^{\#}U_{uq}^{*}, \tag{114}$$

$$g_{pqrs}^{\#} = \sum_{tuvz} U_{tp}U_{vr}\tilde{g}_{tuvz}^{\#}U_{uq}^{*}U_{zs}^{*}. \tag{115}$$

# S5 Algorithms convergence comparison

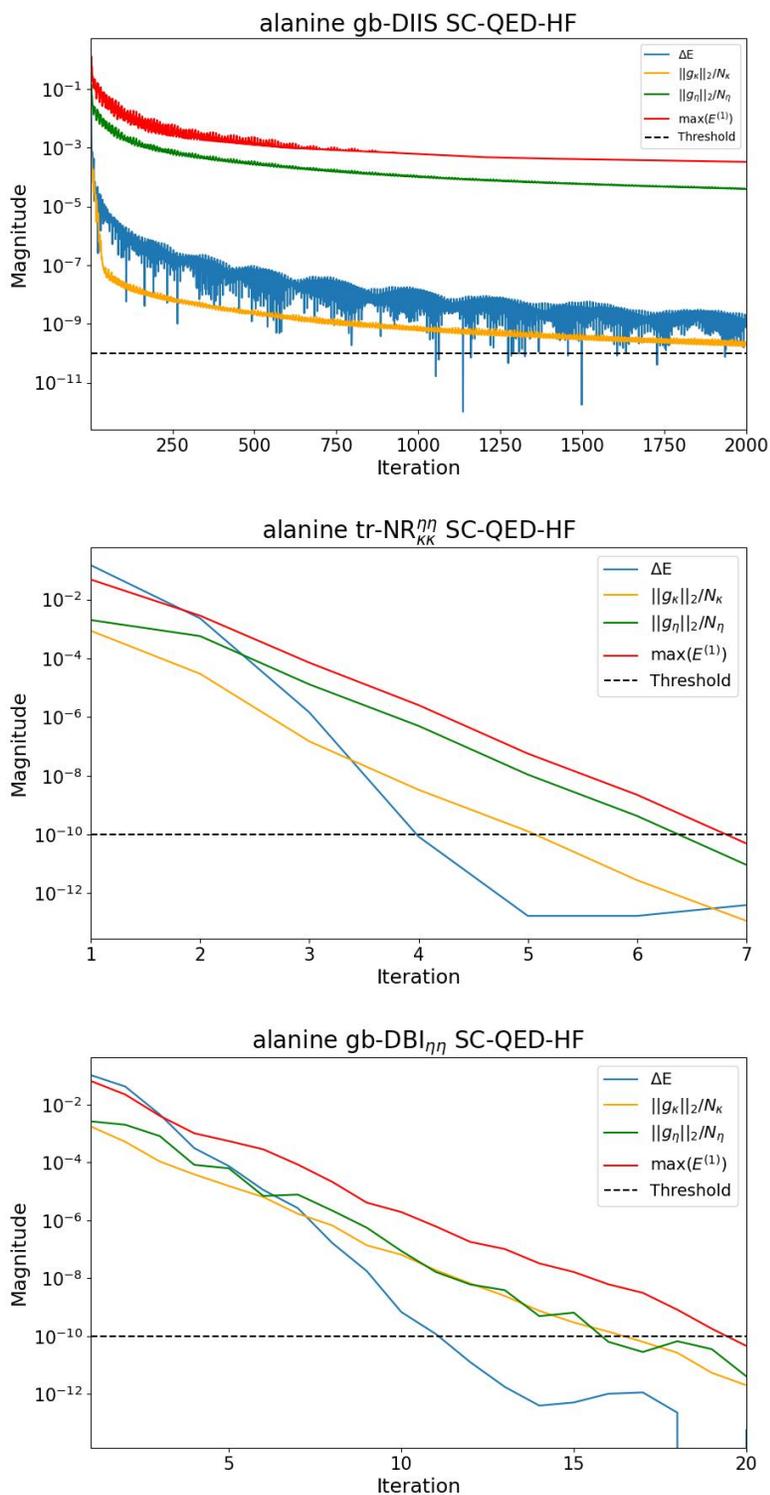

Figure S1: alanine convergence comparison

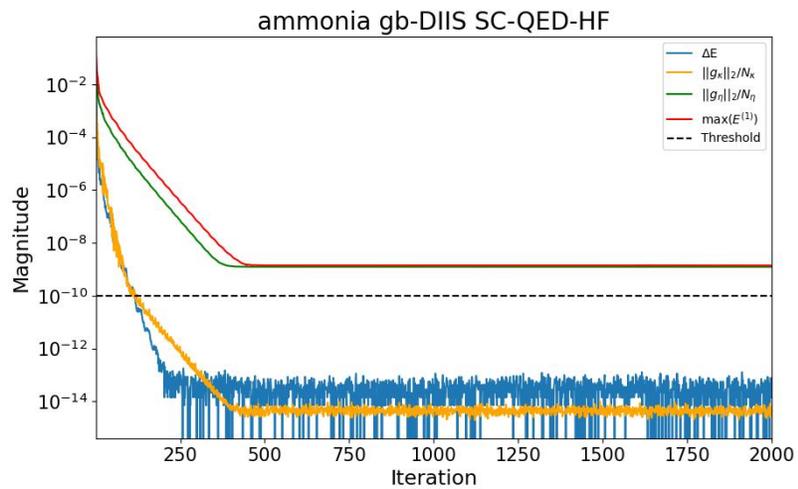

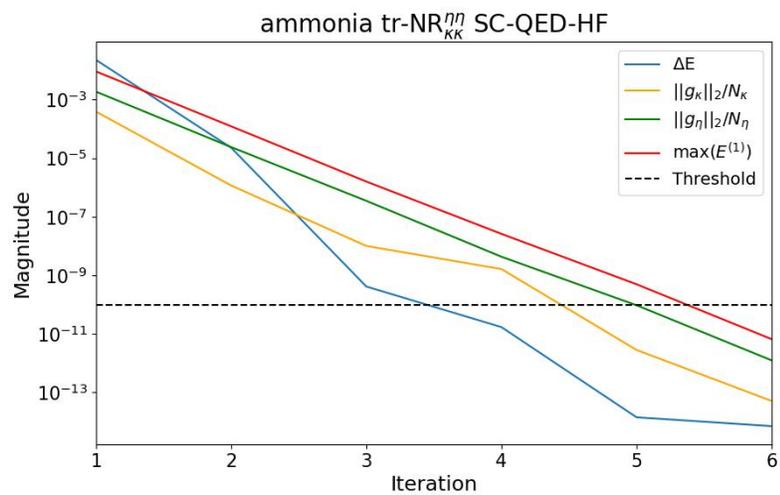

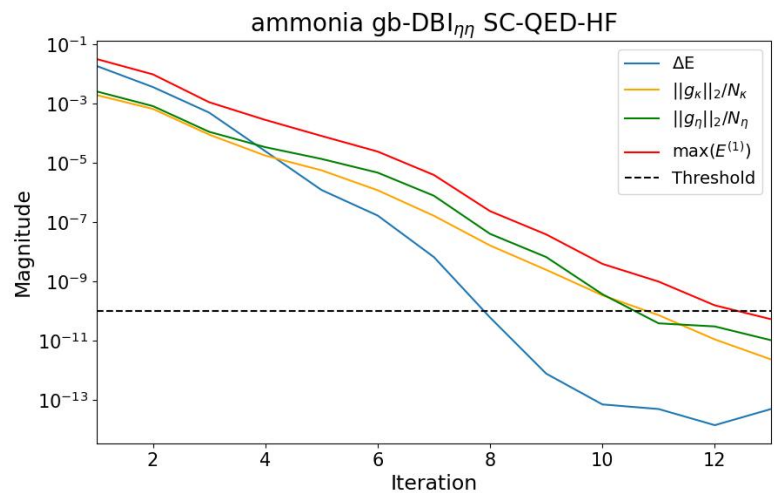

Figure S2: ammonia convergence comparison

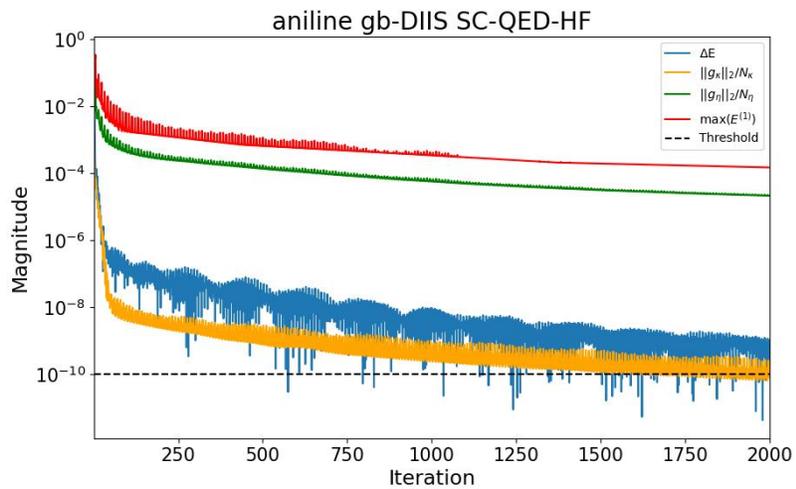

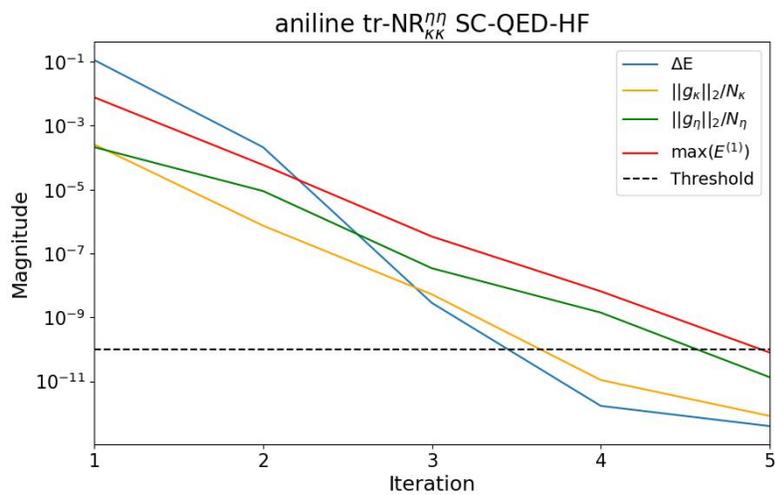

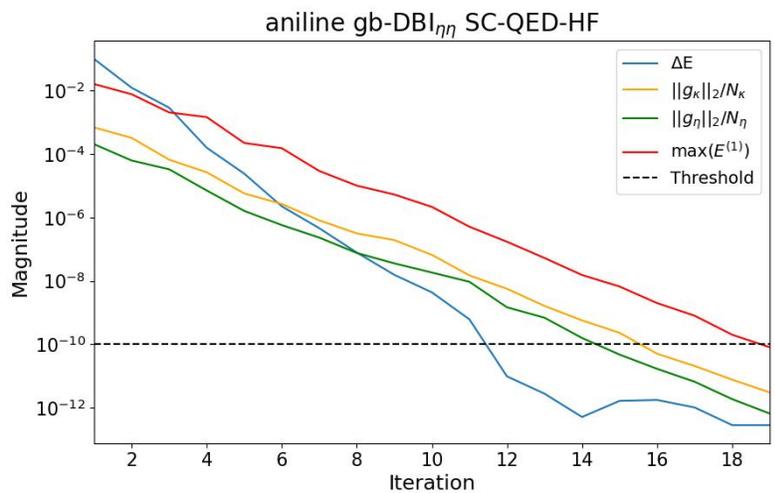

Figure S3: aniline convergence comparison

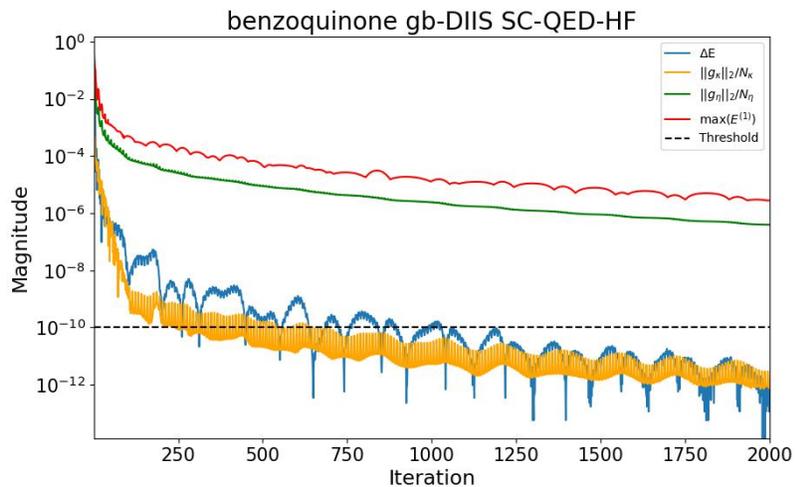
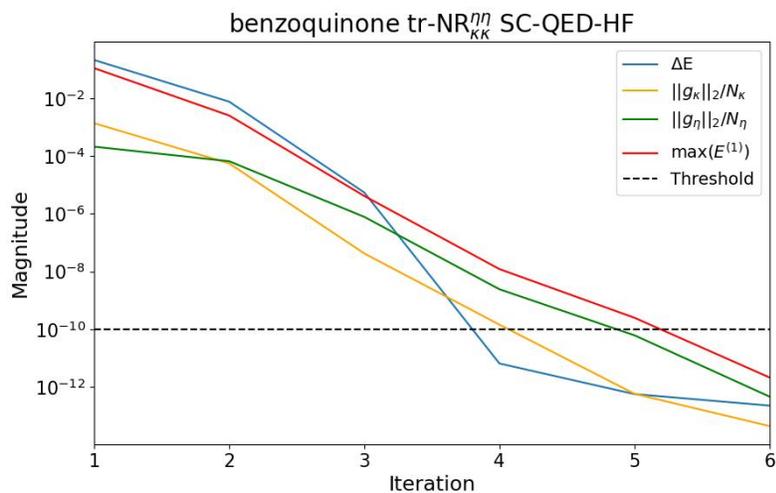
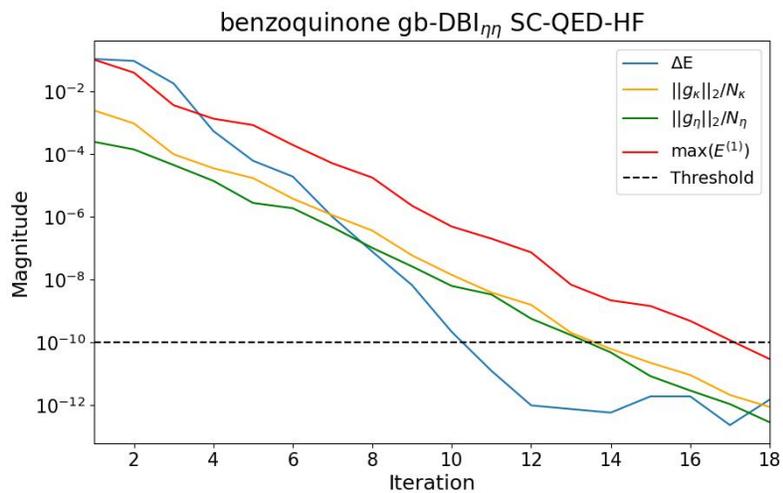

Figure S4: benzoquinone convergence comparison

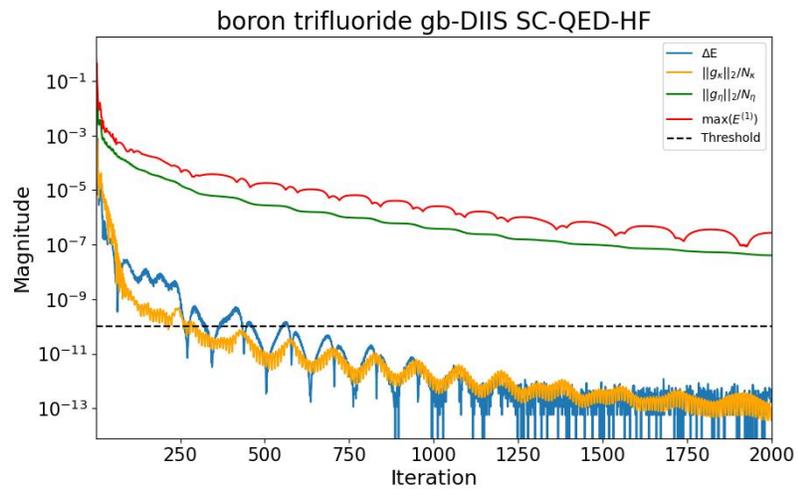

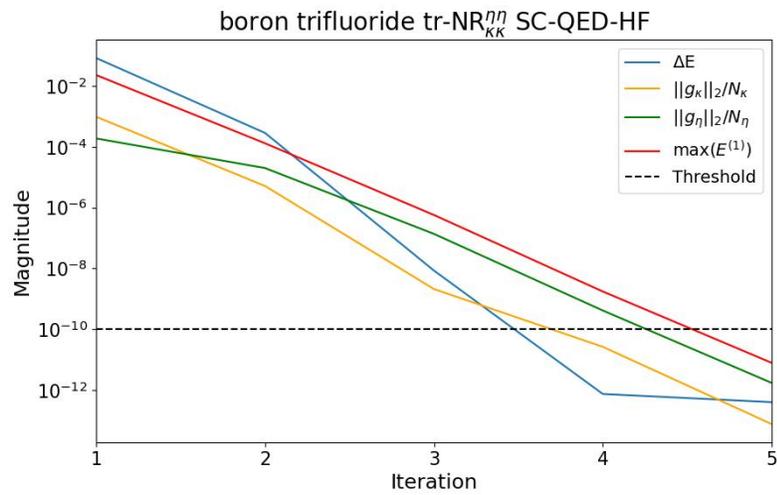

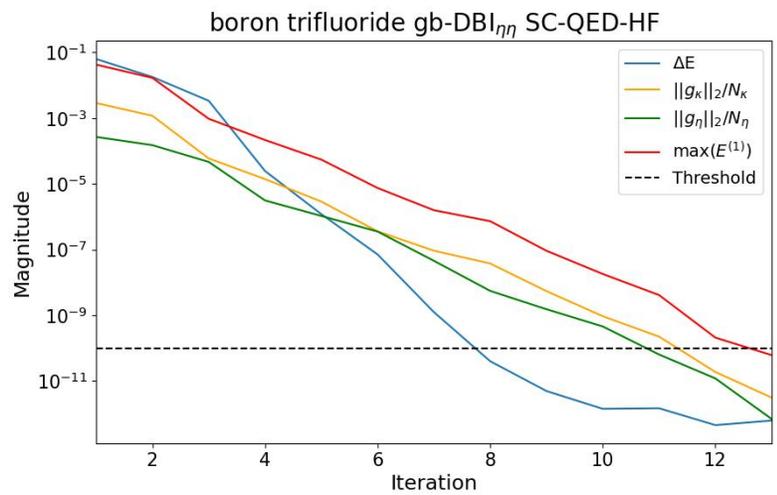

Figure S5: boron trifluoride convergence comparison

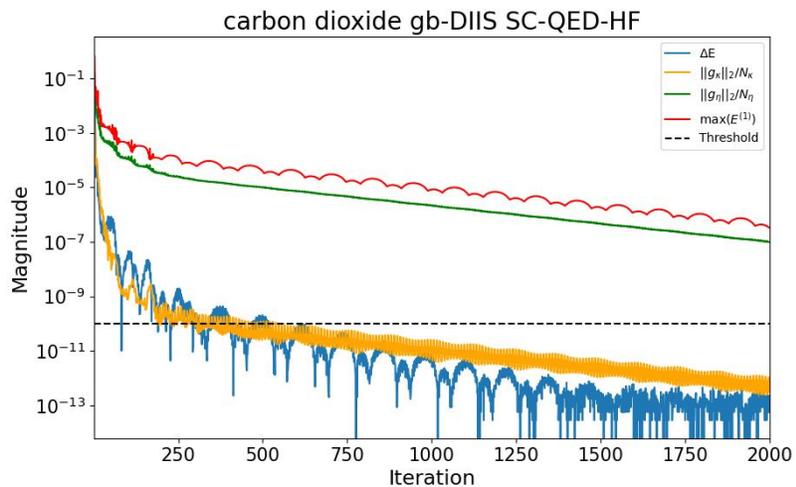

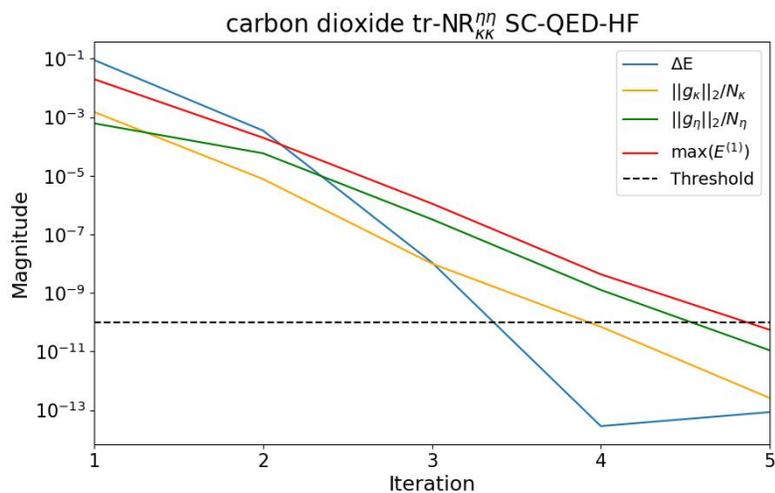

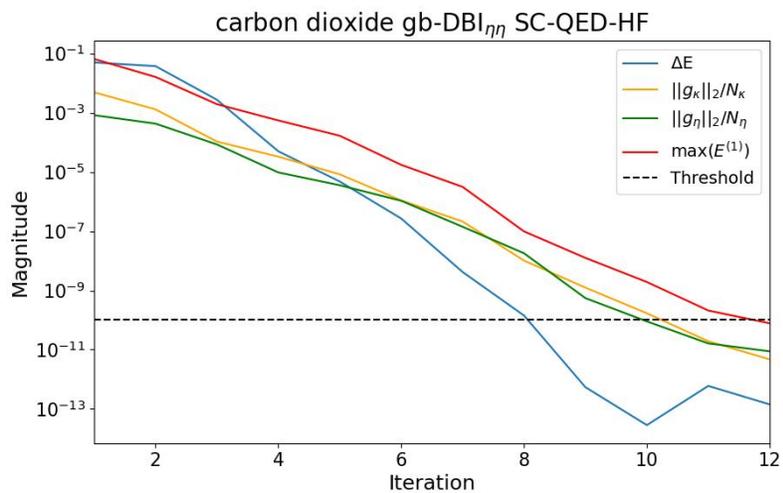

Figure S6: carbon dioxide convergence comparison

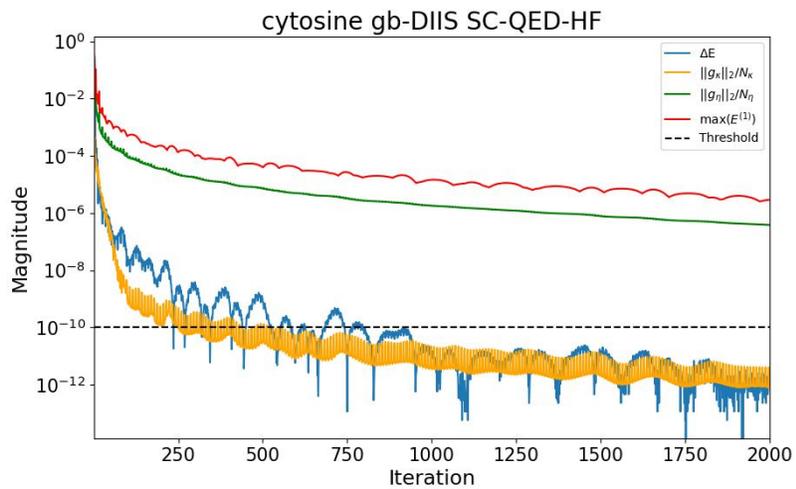
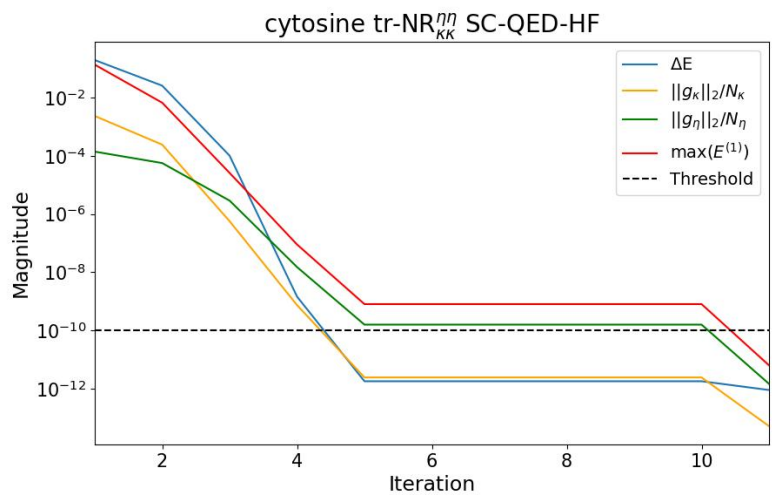
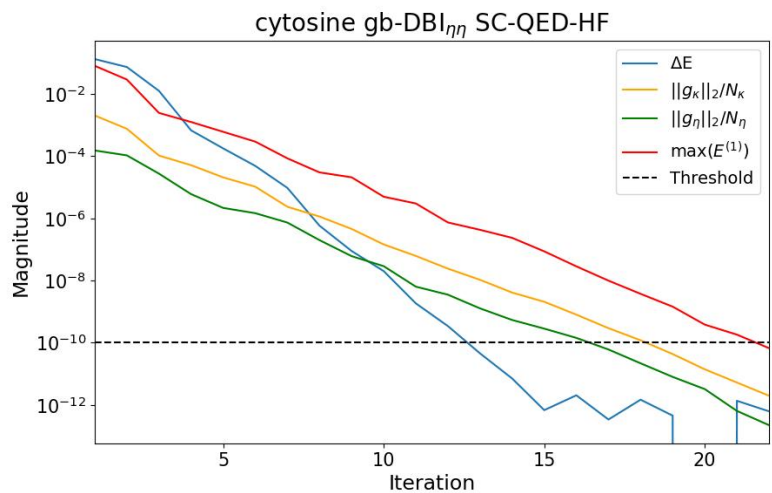

Figure S7: cytosine convergence comparison

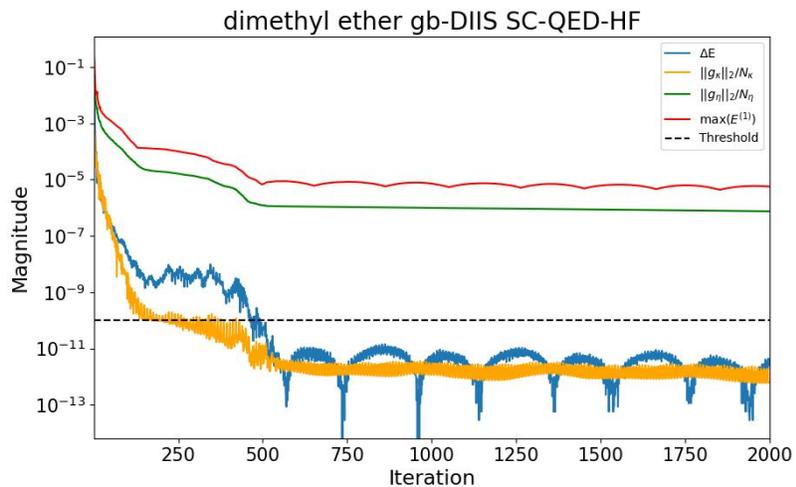
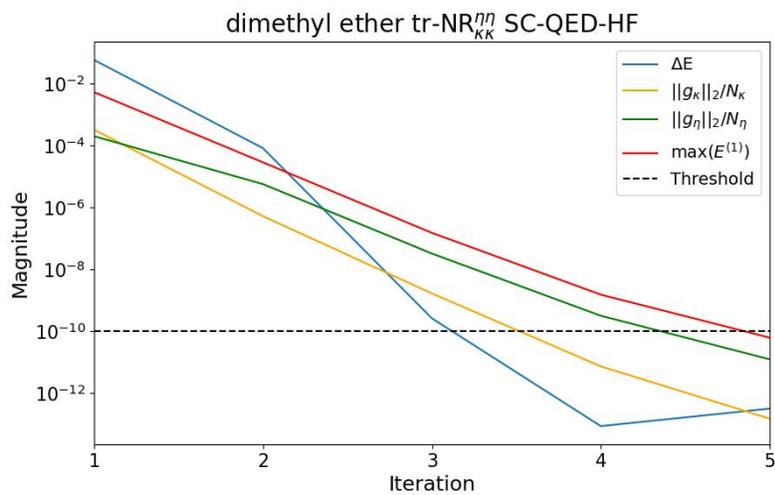
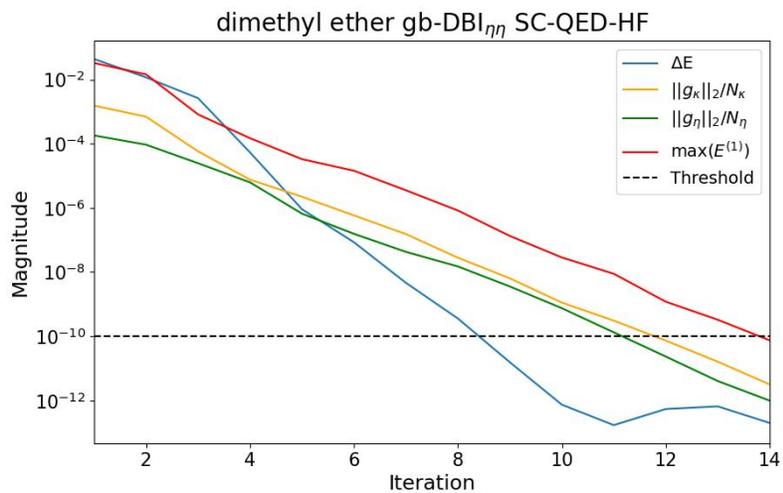

Figure S8: dimethyl ether convergence comparison

36

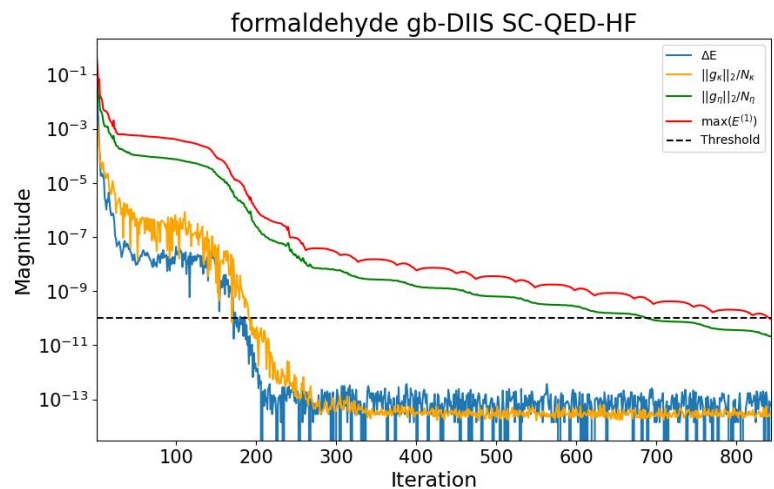
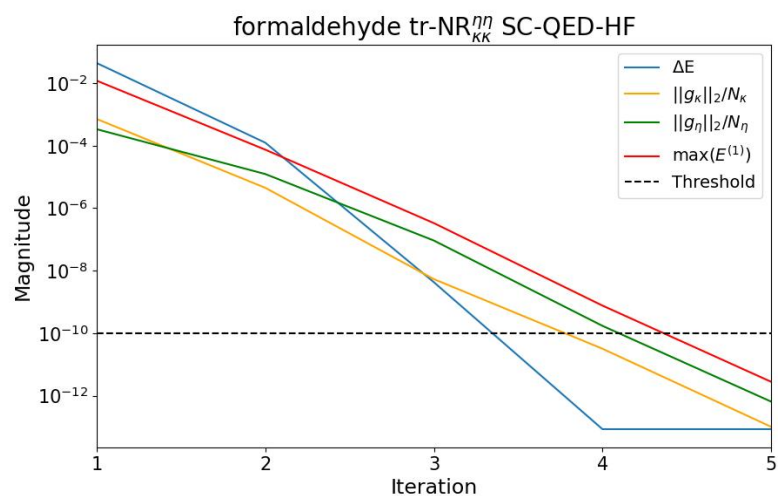
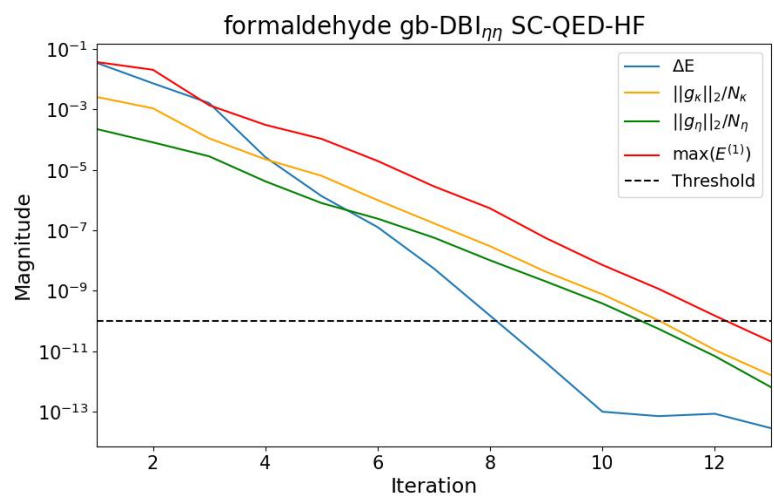

Figure S9: formaldehyde convergence comparison

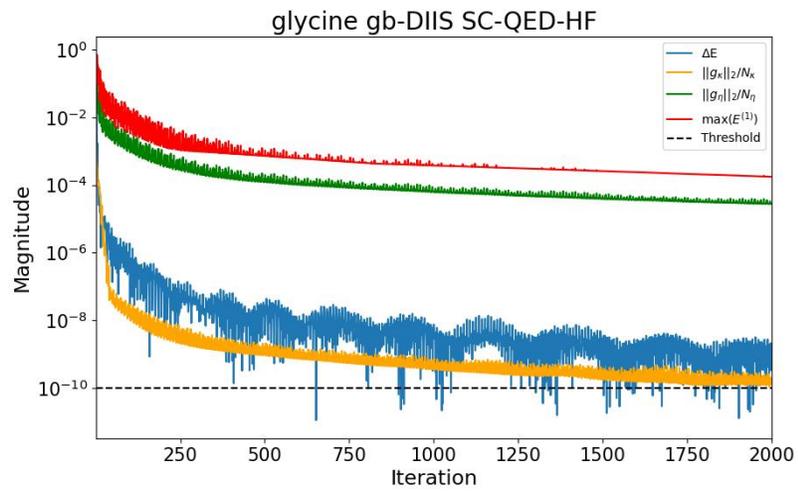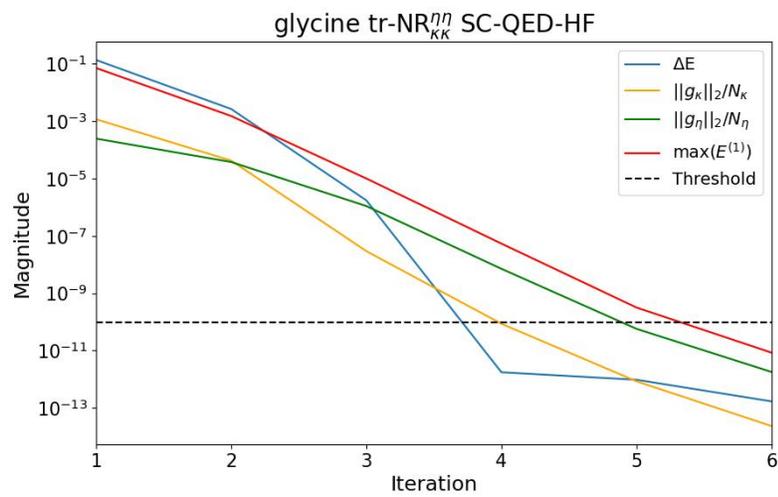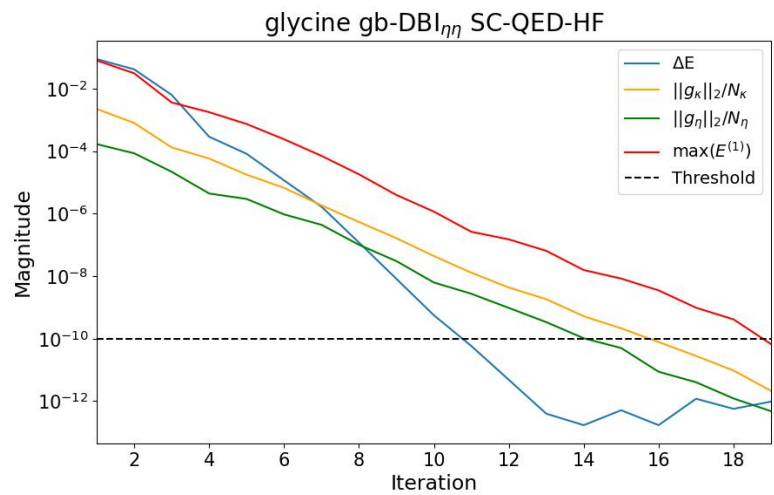

Figure S10: glycine convergence comparison

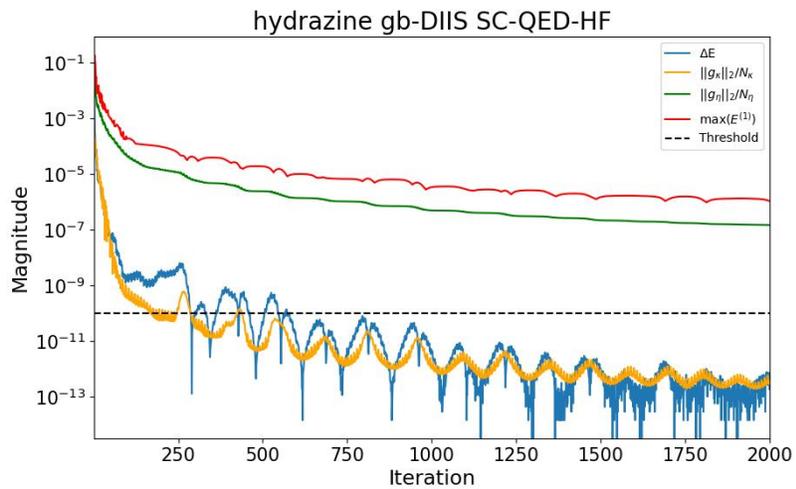

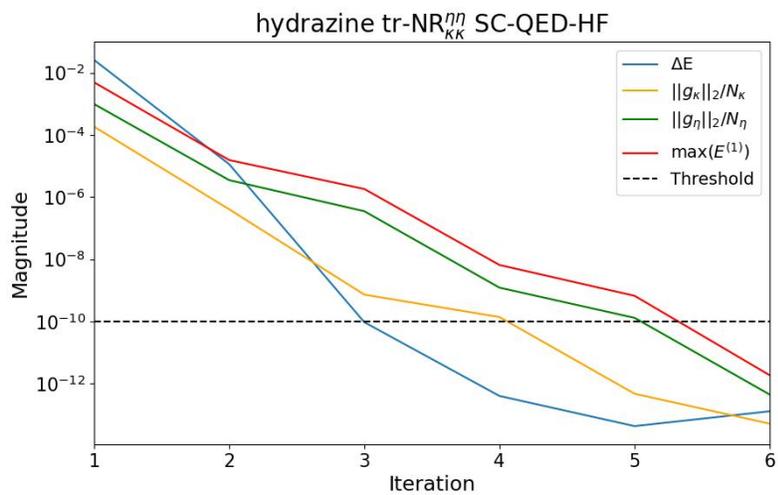

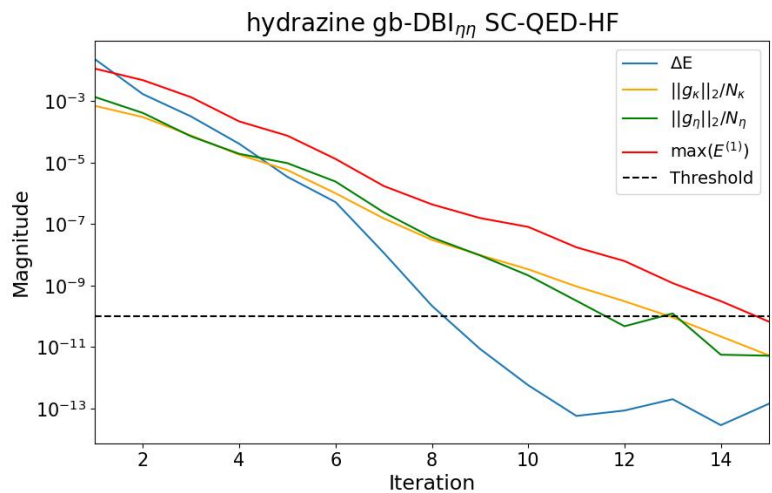

Figure S11: hydrazine convergence comparison

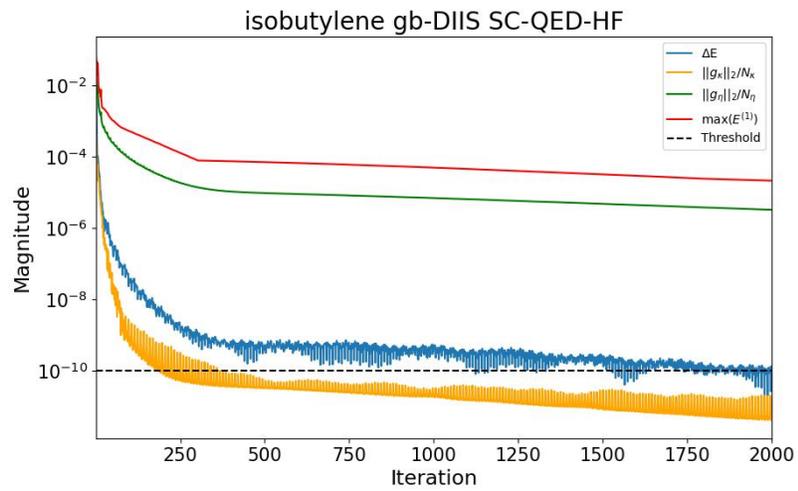
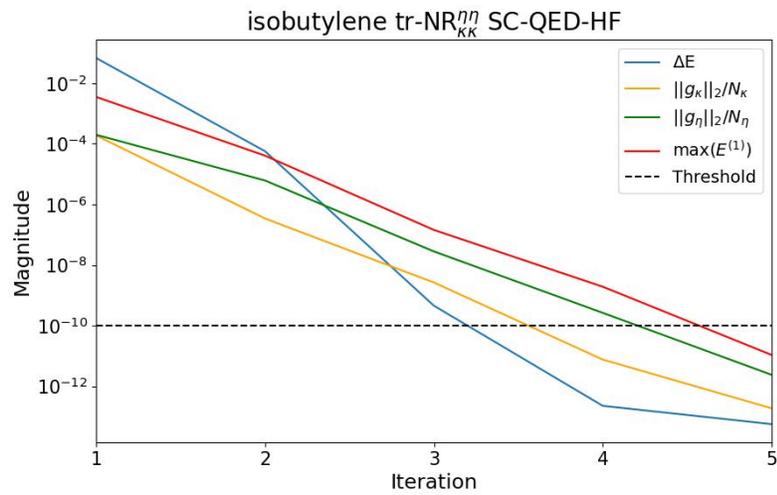
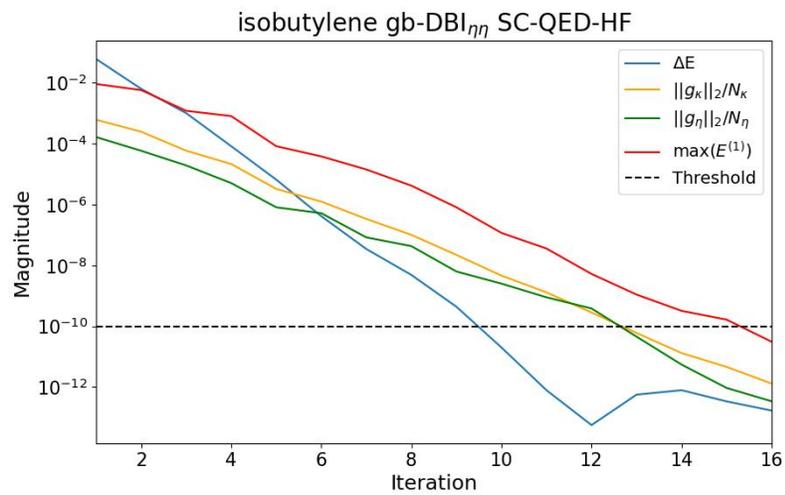

Figure S12: isobutylene convergence comparison

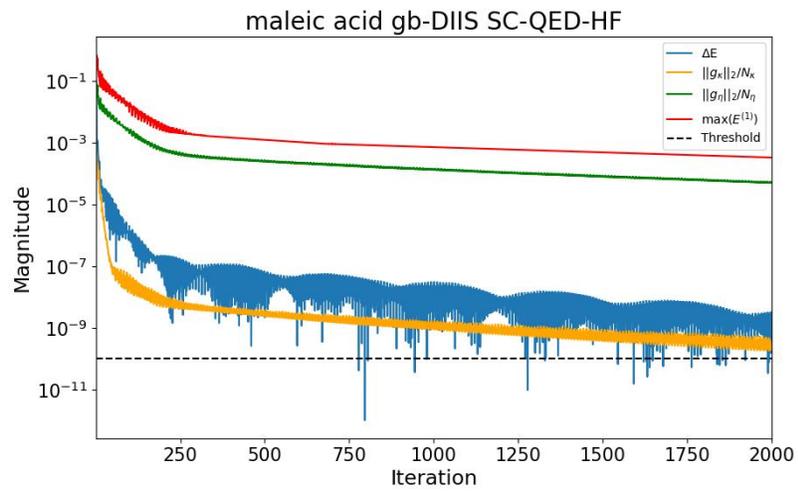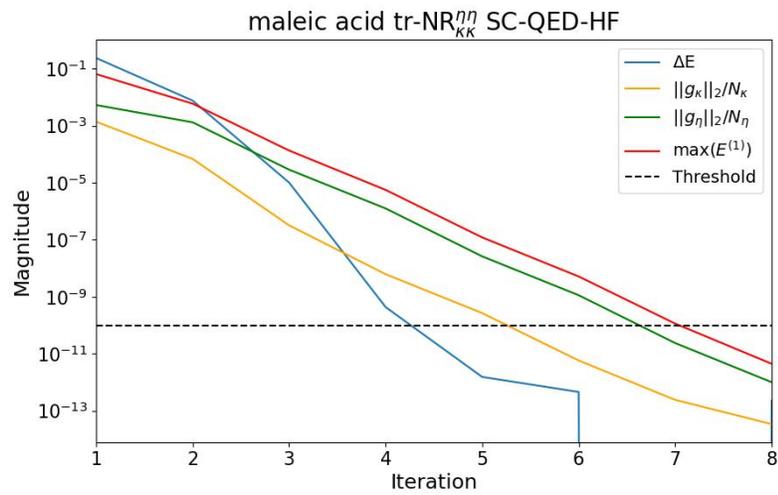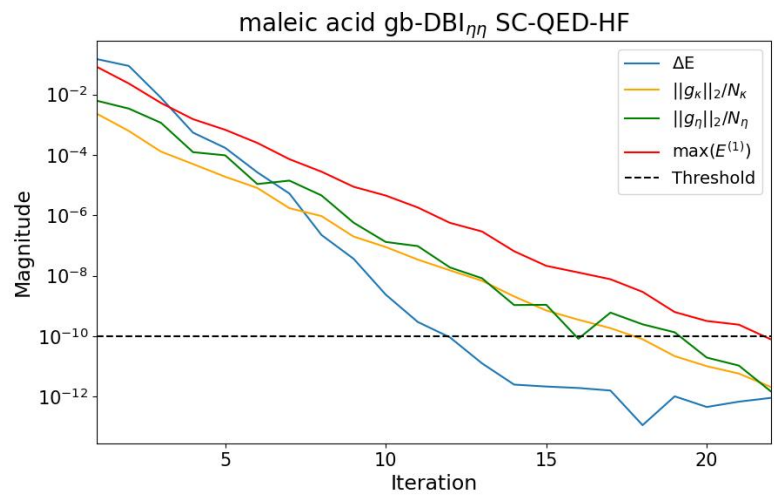

Figure S13: maleic acid convergence comparison

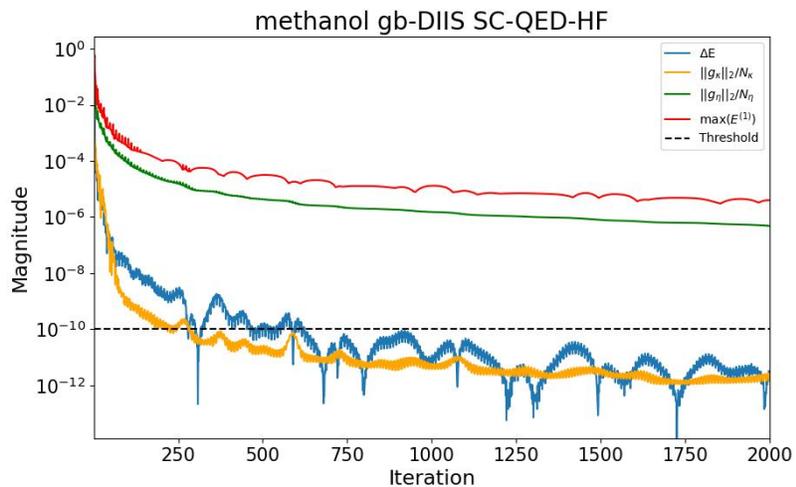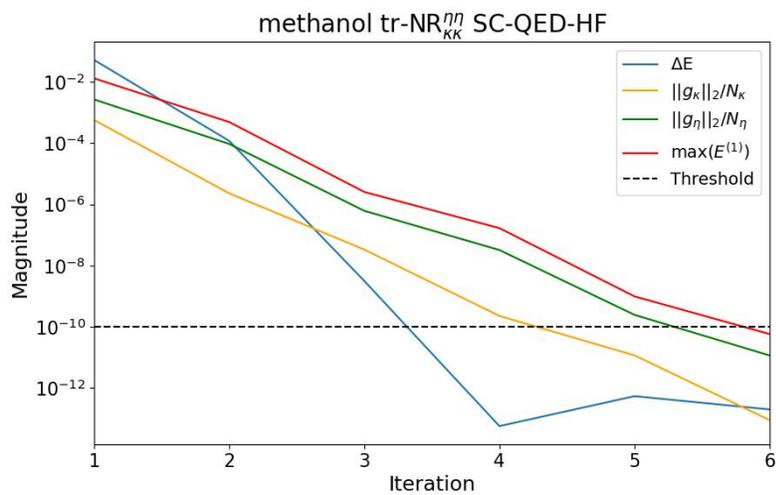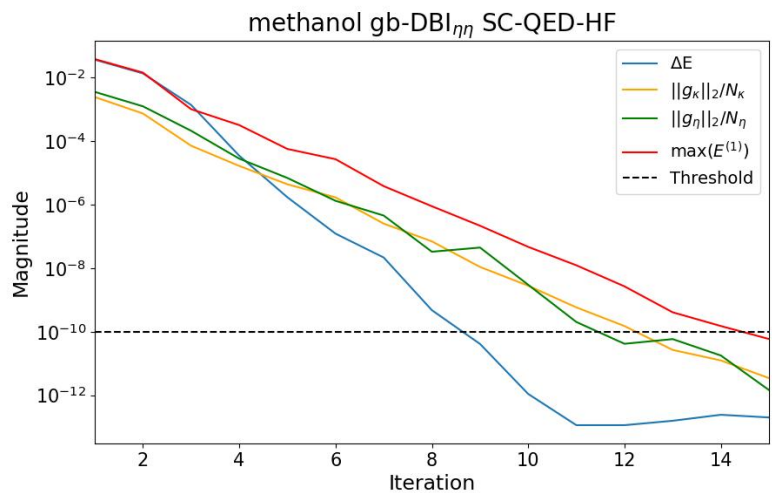

Figure S14: methanol convergence comparison

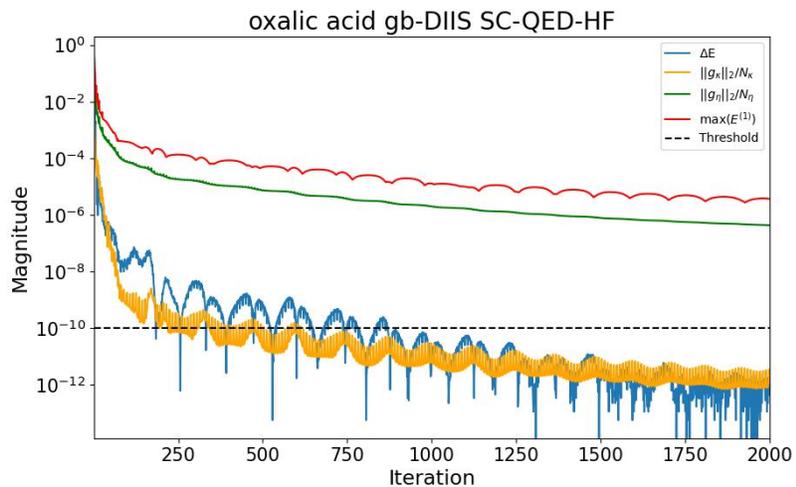
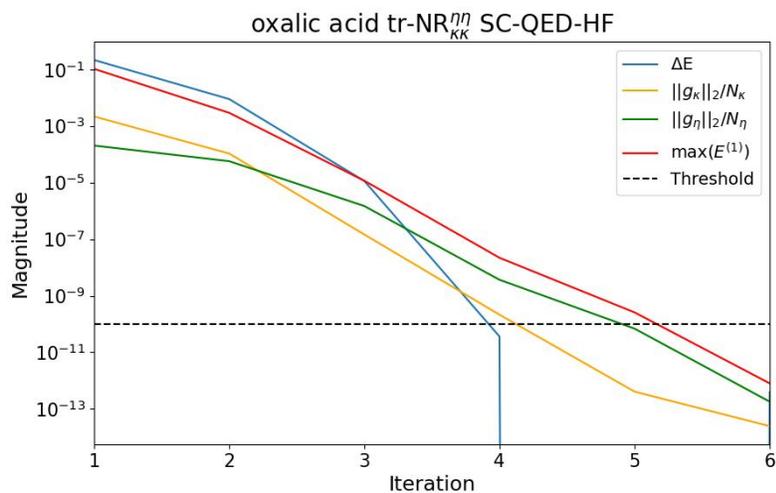
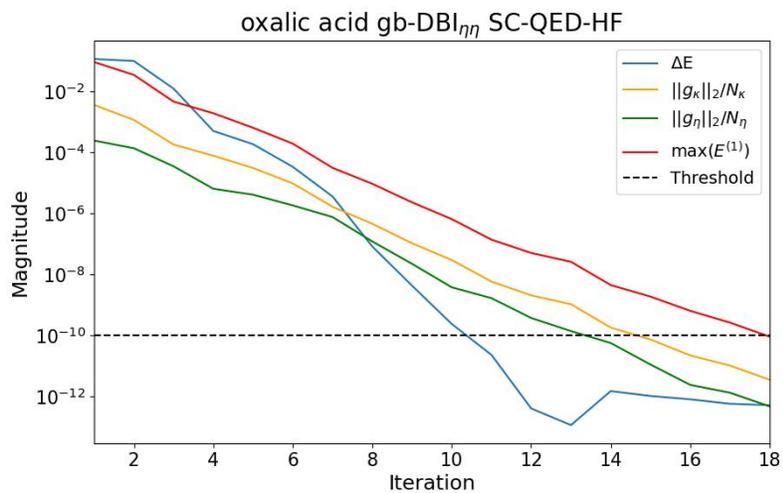

Figure S15: oxalic acid convergence comparison

43

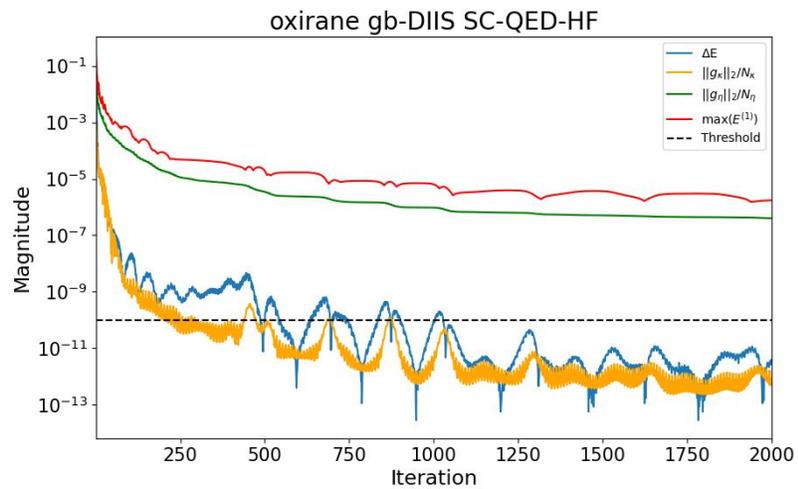
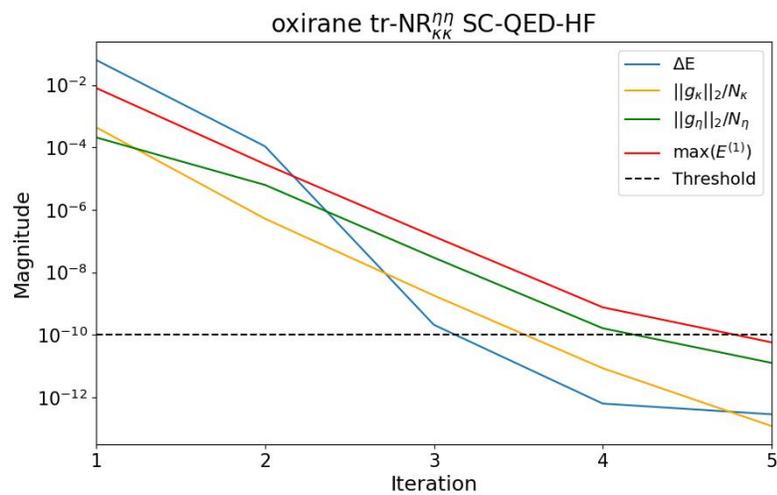
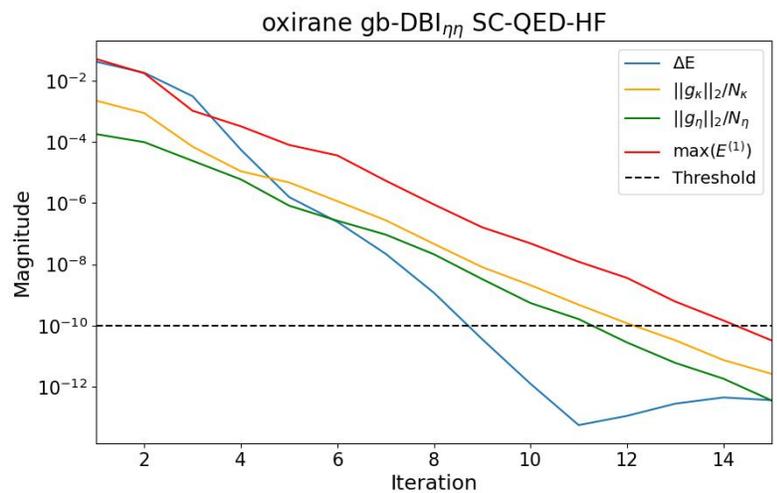

Figure S16: oxirane convergence comparison

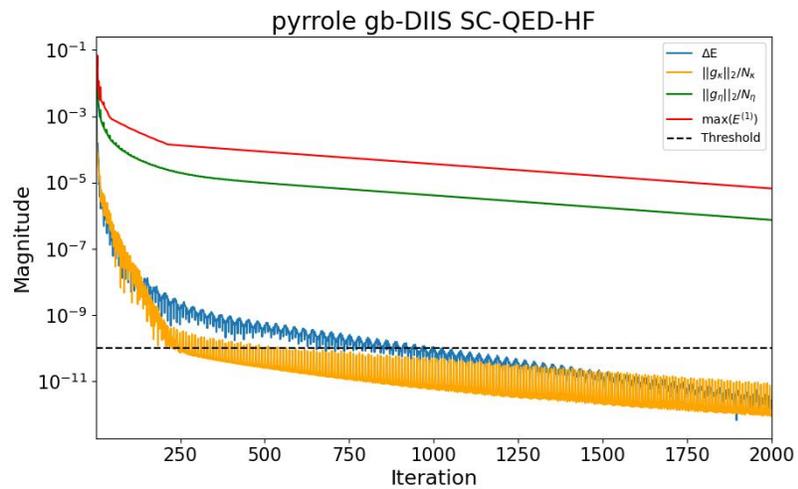
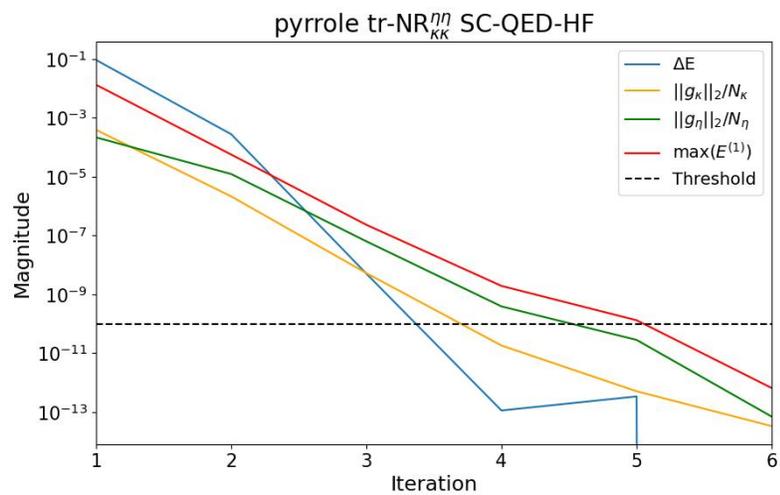
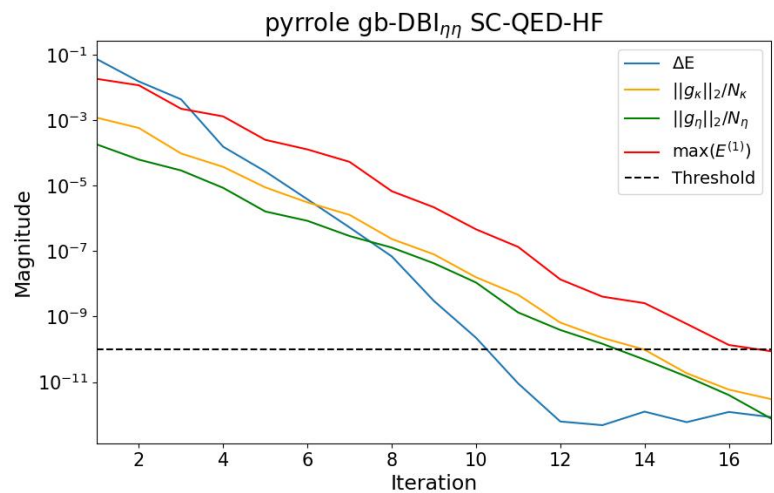

Figure S17: pyrrole convergence comparison

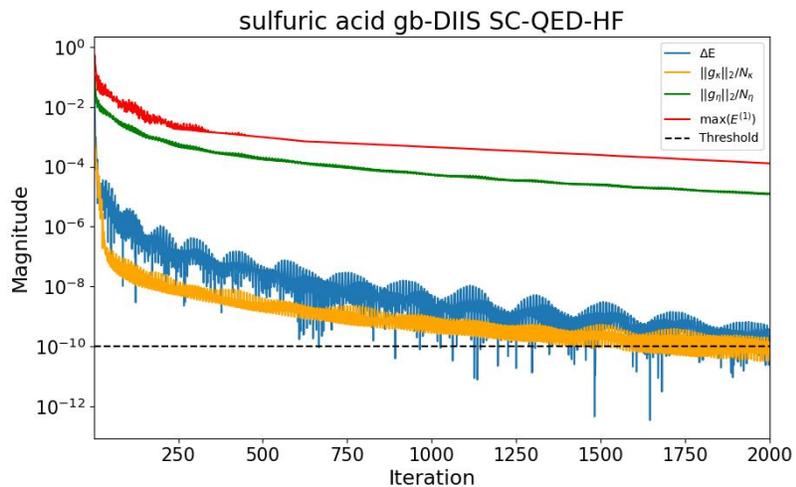

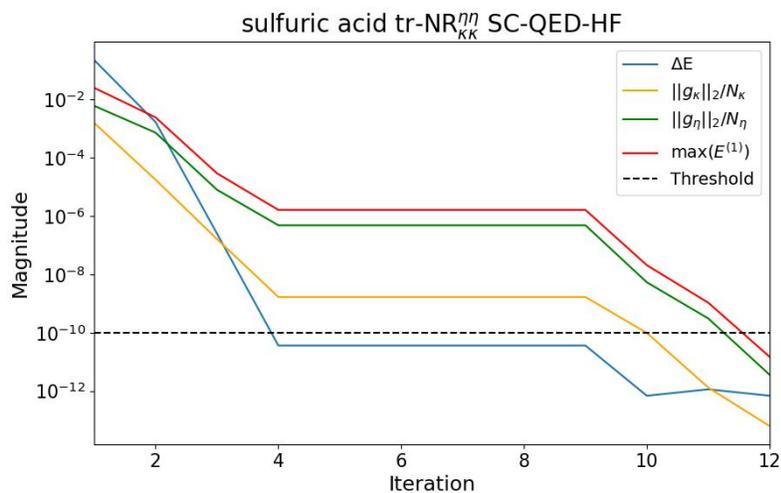

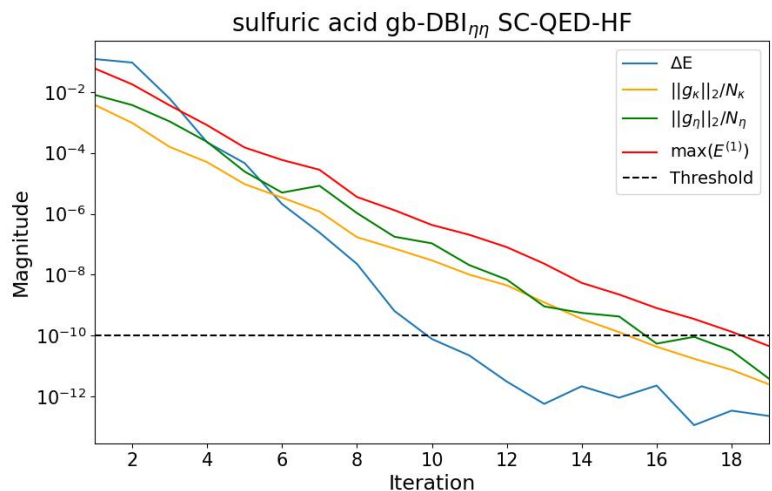

Figure S18: sulfuric acid convergence comparison

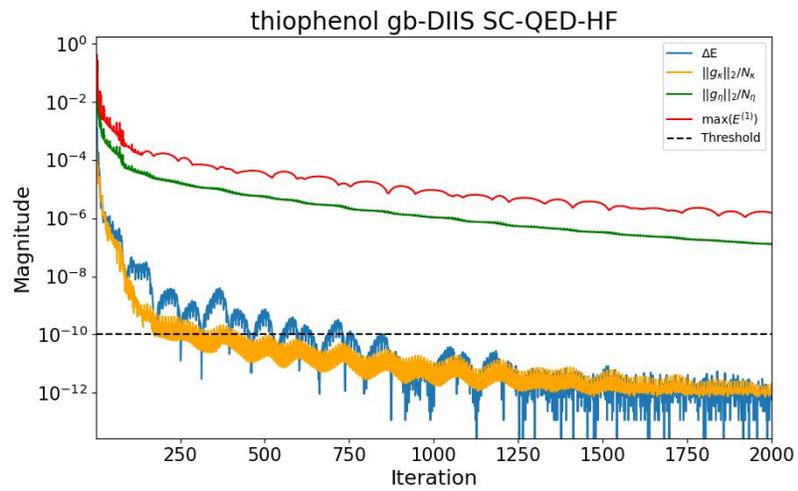
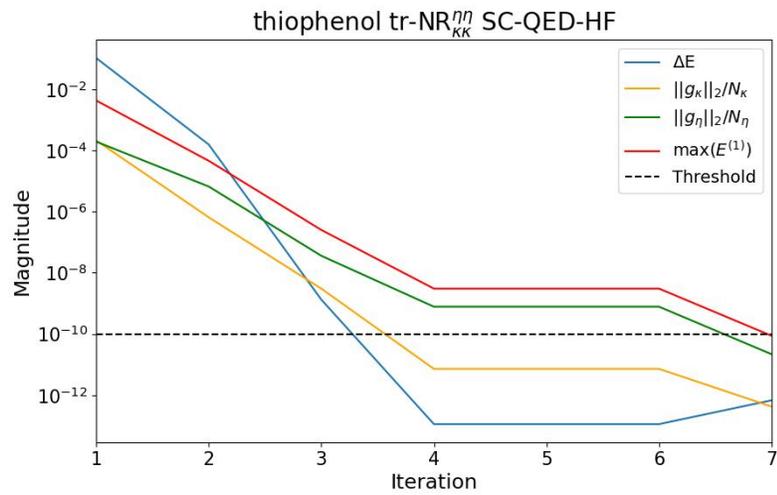
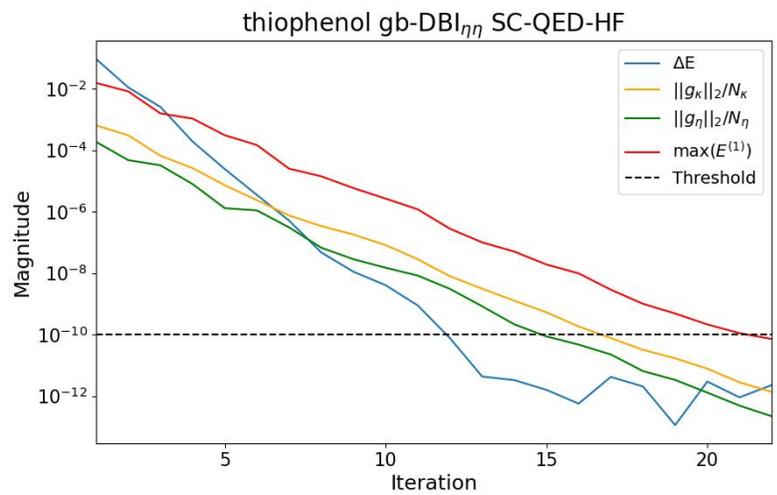

Figure S19: thiophenol convergence comparison

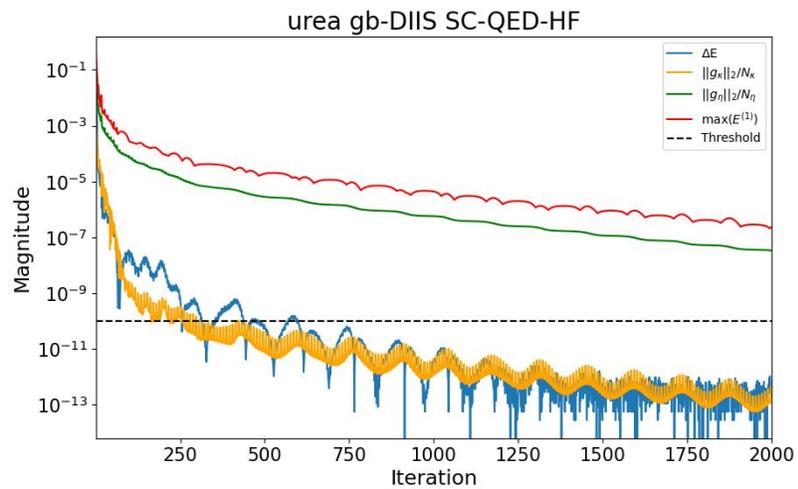
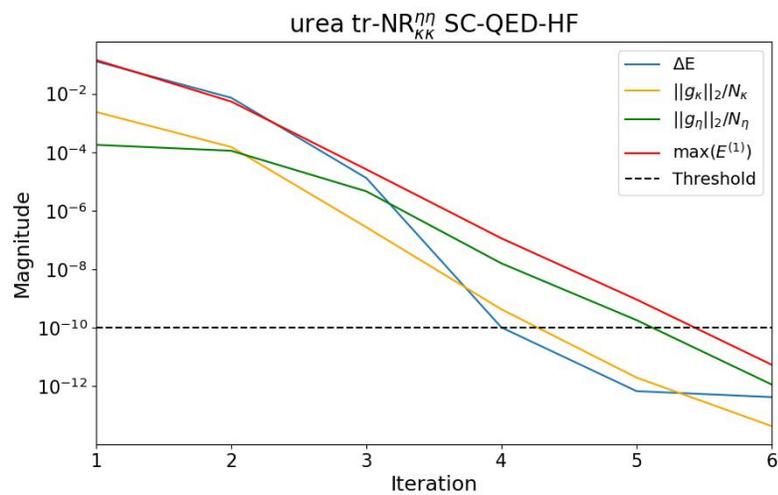
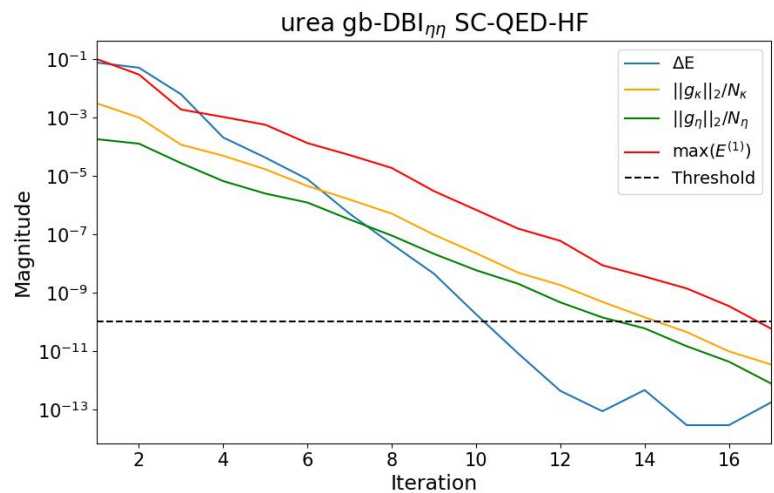

Figure S20: urea convergence comparison

# S6 Hessian blocks analysis

The heat map representations of the $\kappa$-$\kappa$ and $\eta$-$\eta$ Hessian blocks are reported. The $\eta$-$\eta$ Hessian blocks turn to be highly non-diagonal.

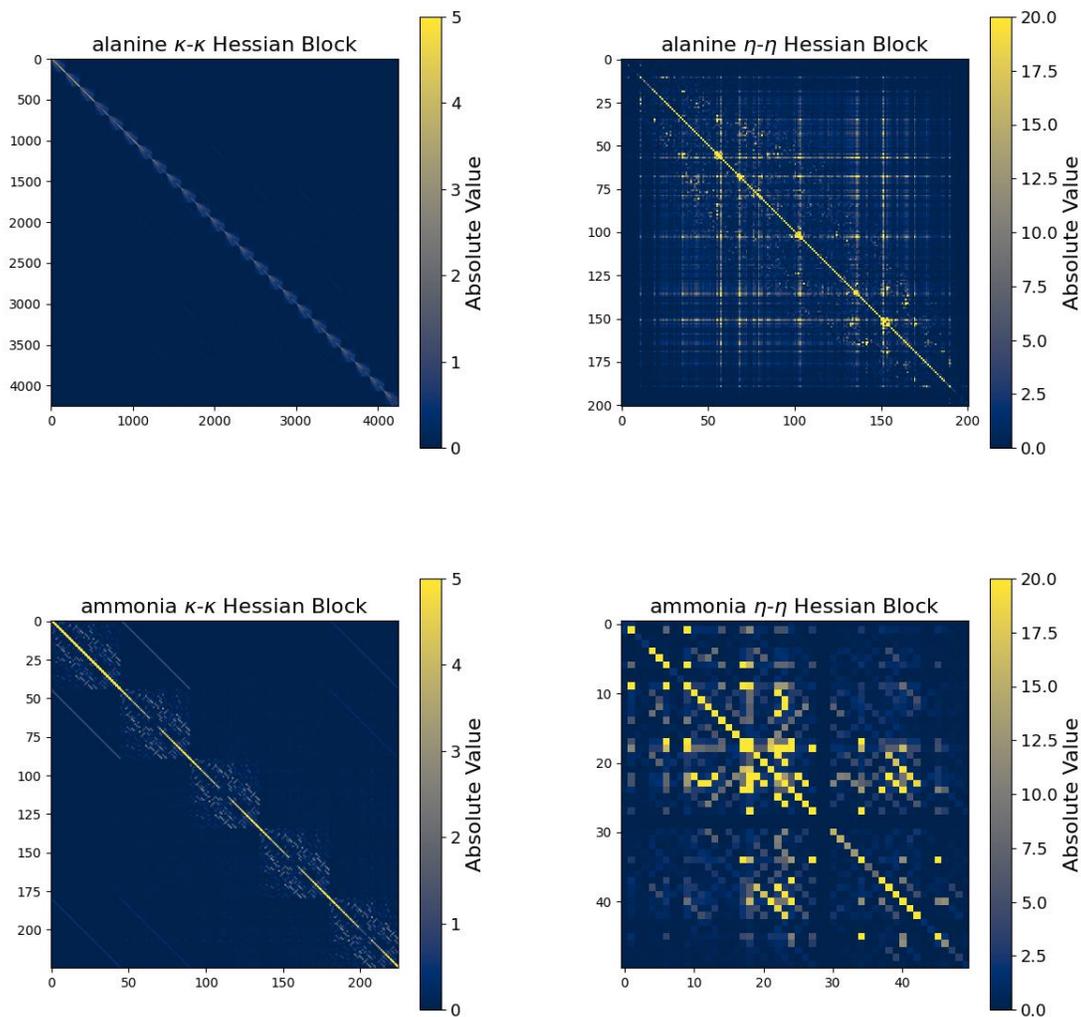

Figure S21: $\kappa$-$\kappa$ and $\eta$-$\eta$ Hessian heat map representations (part 1)

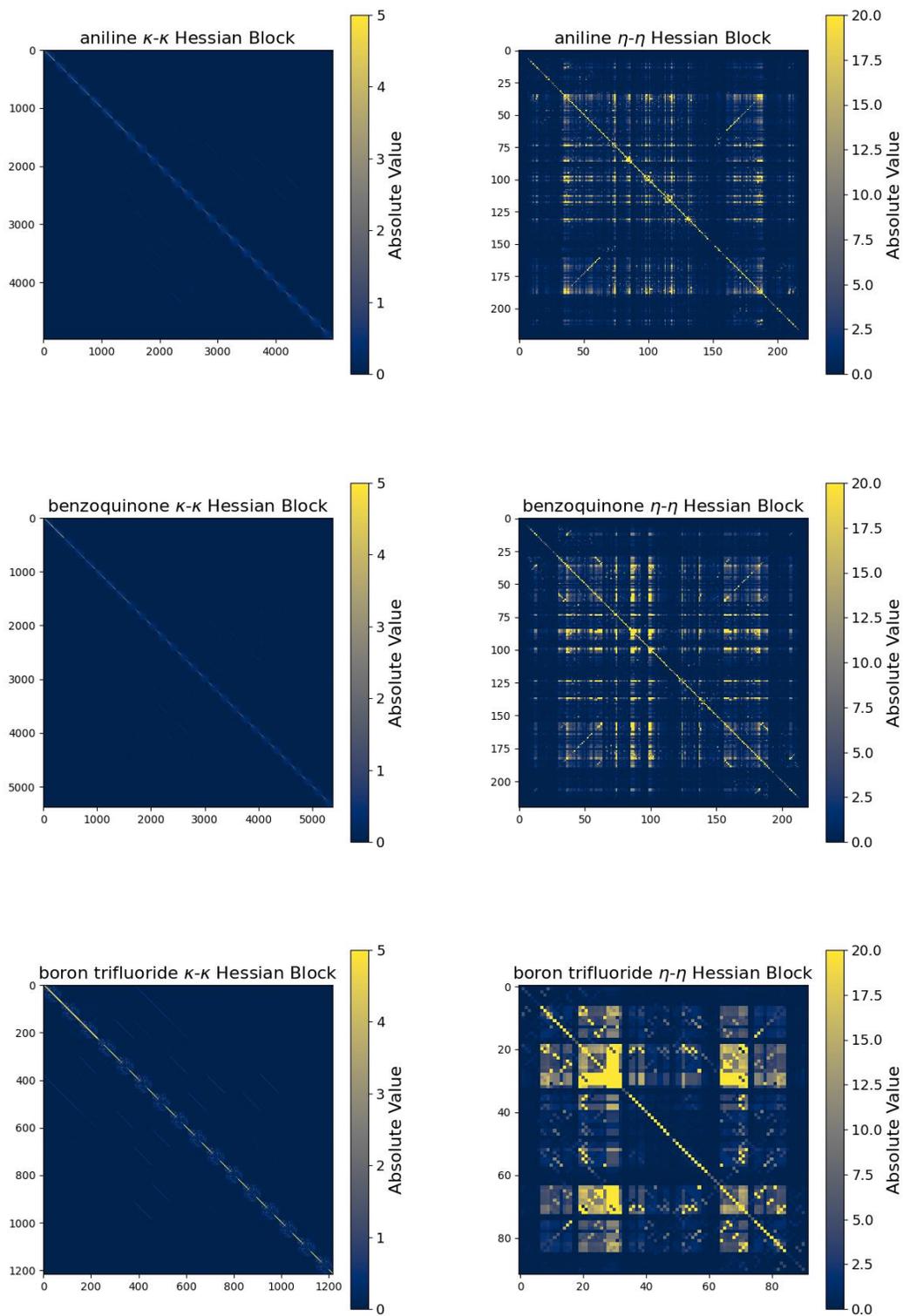

Figure S22: $\kappa$-$\kappa$ and $\eta$-$\eta$ Hessian heat map representations (part 2)

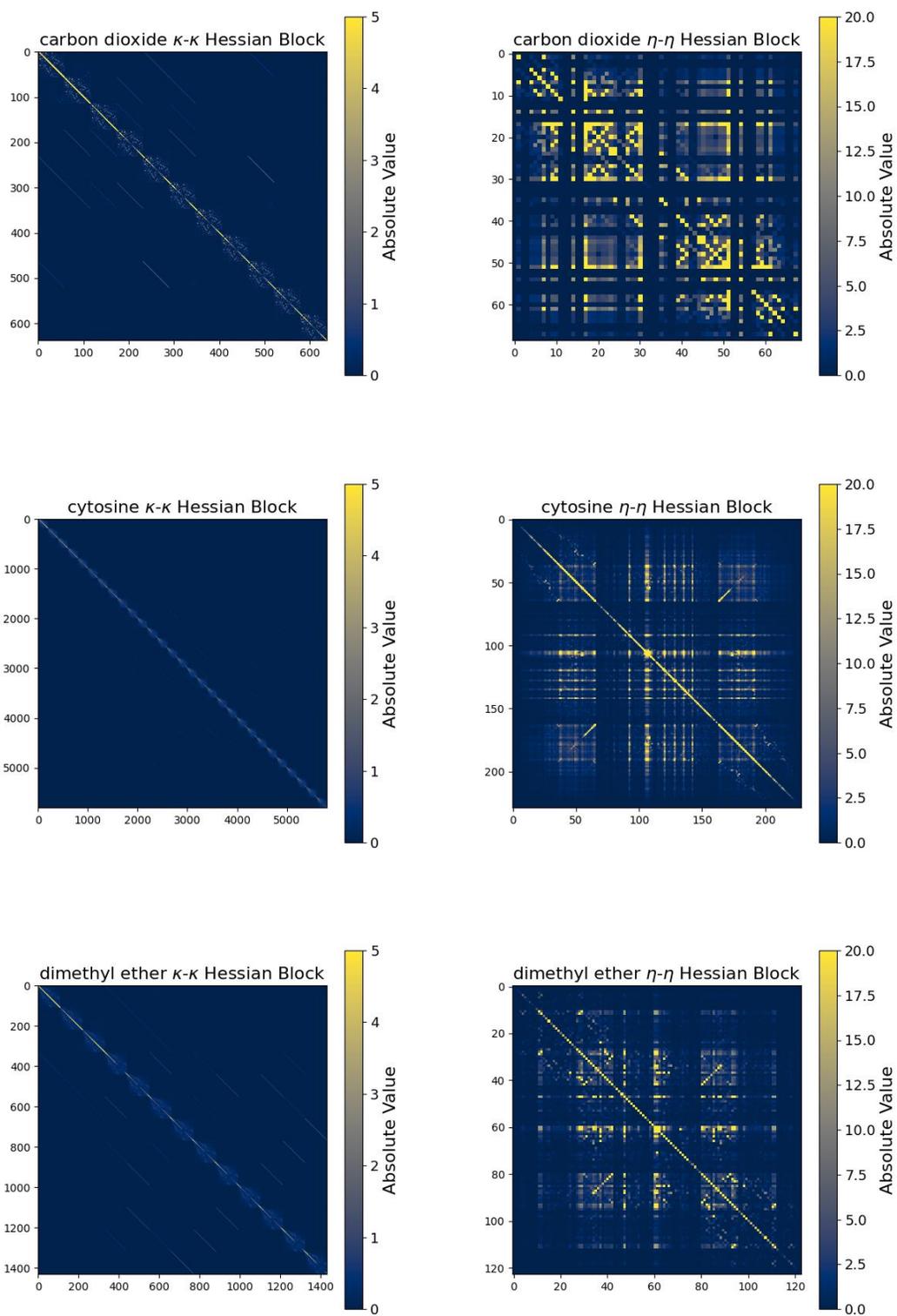

Figure S23: $\kappa$-$\kappa$ and $\eta$-$\eta$ Hessian heat map representations (part 3)

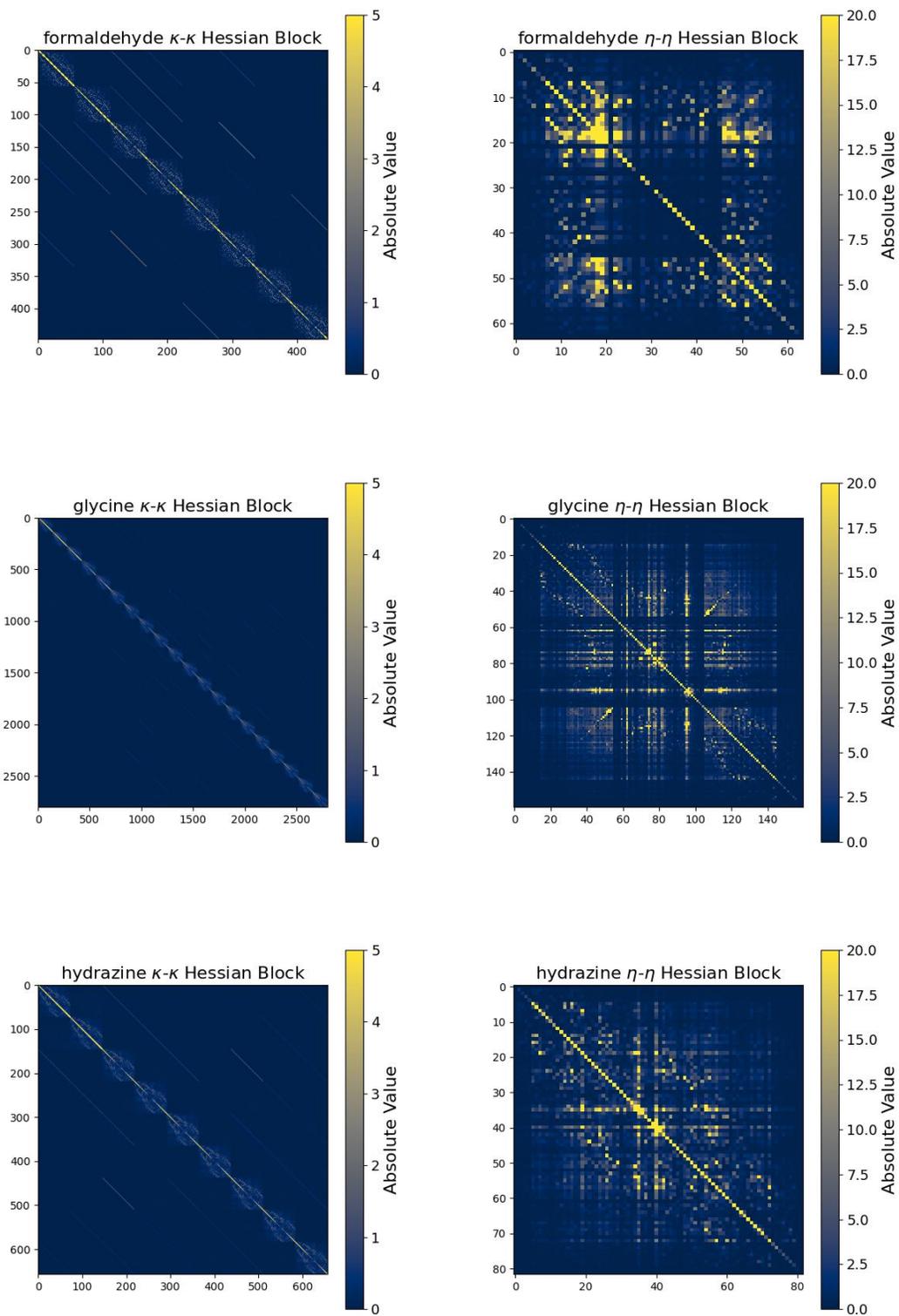

Figure S24: $\kappa$-$\kappa$ and $\eta$-$\eta$ Hessian heat map representations (part 4)

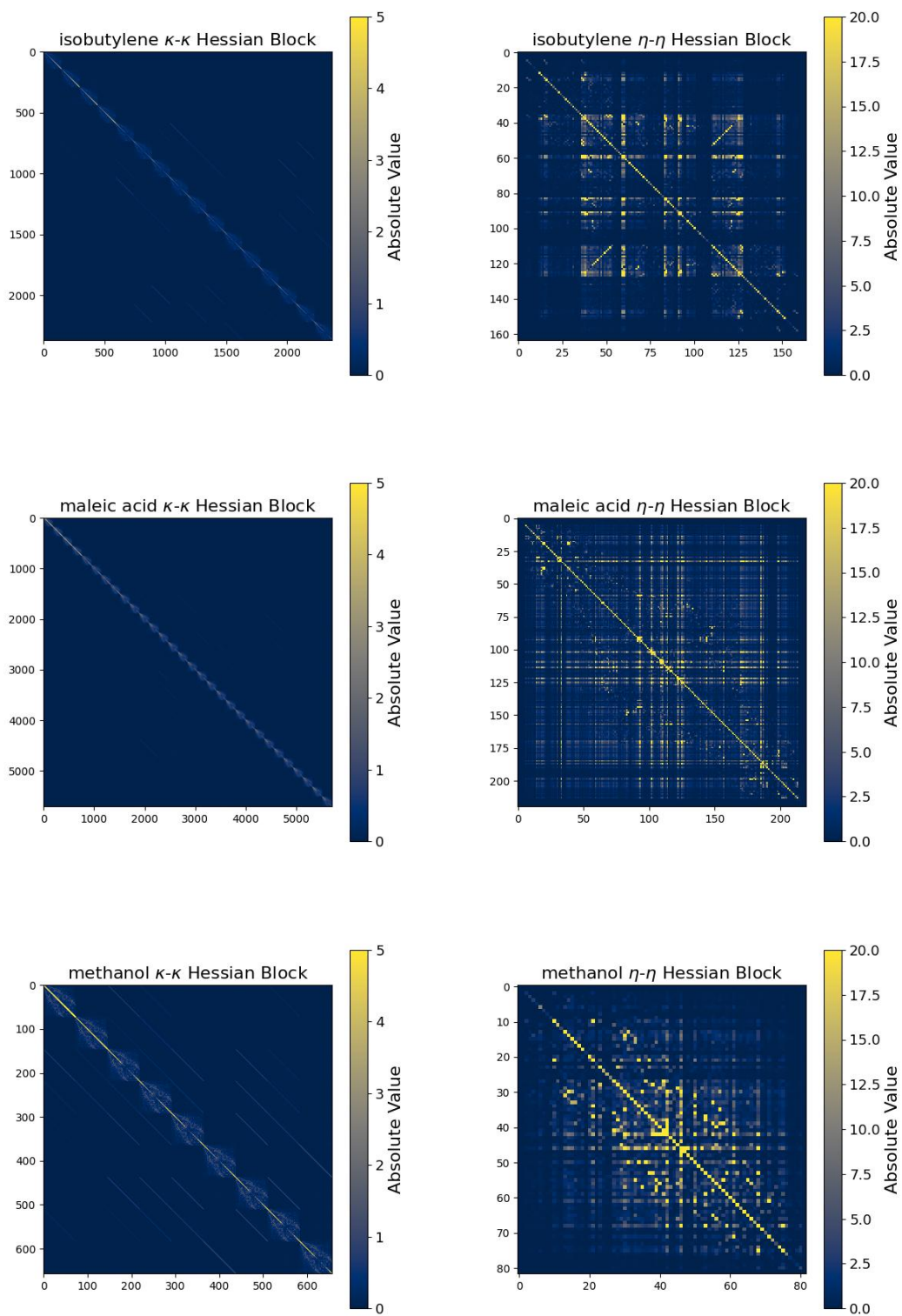

Figure S25: $\kappa$-$\kappa$ and $\eta$-$\eta$ Hessian heat map representations (part 5)

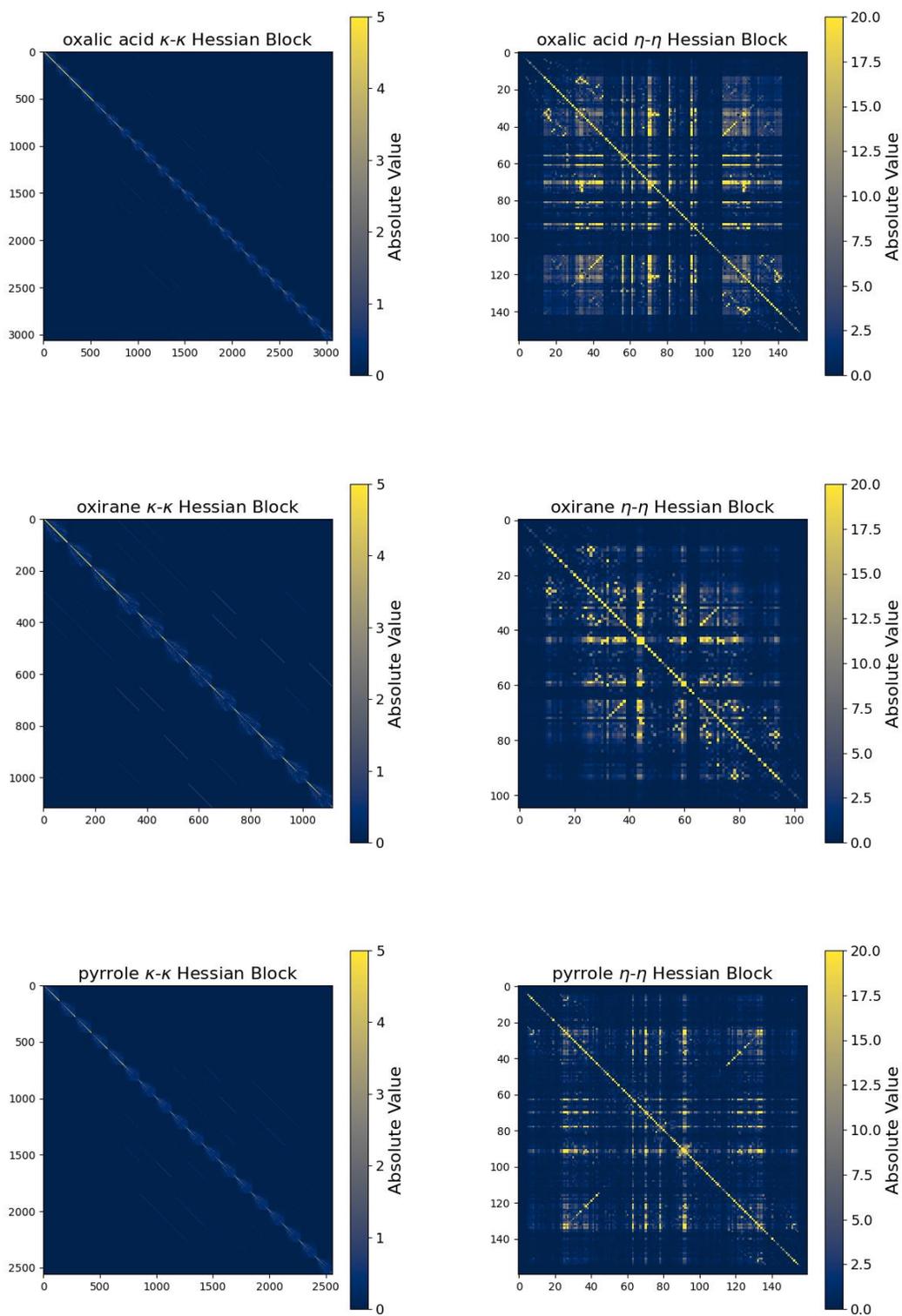

Figure S26: $\kappa$-$\kappa$ and $\eta$-$\eta$ Hessian heat map representations (part 6)

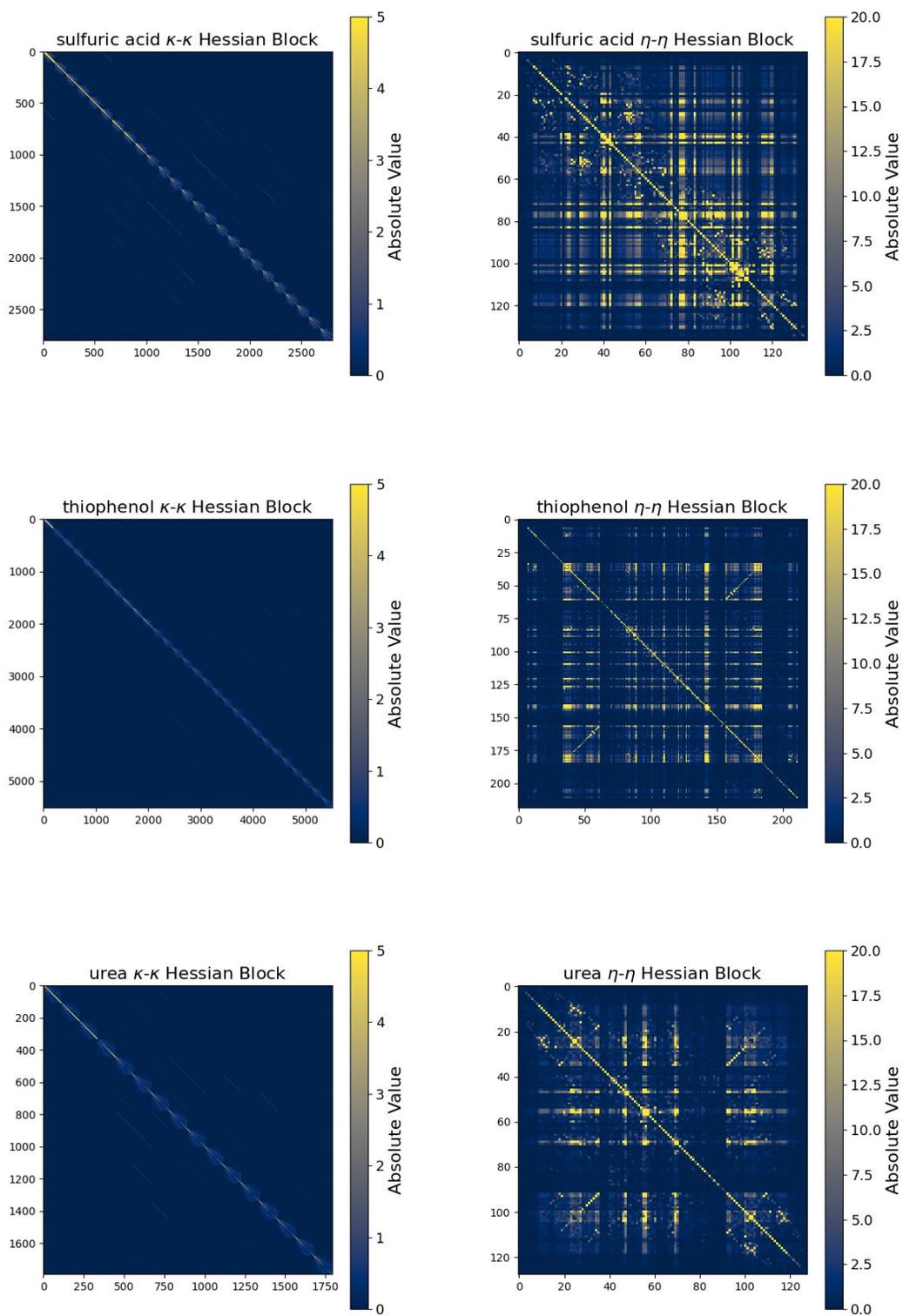

Figure S27: $\kappa$-$\kappa$ and $\eta$-$\eta$ Hessian heat map representations (part 7)

## S7 Non-necessety of the mixed bloks

Table S1: Wall time comparison between the tr-NR with and without considering the mixed blocks.

| Molecule | tr-NR$_{\kappa\kappa}^m$ | full tr-NR | Molecule | tr-NR$_{\kappa\kappa}^m$ | full tr-NR |
|---|---|---|---|---|---|
| alanine | 22.0 m | 178 m | hydrazine | 26.8 s | 2.07 m |
| ammonia | 5.91 s | 40.1 s | isobutylene | 6.37 s | 54.1 m |
| aniline | 23.7 m | 238 m | maleic acid | 36.2 m | 300 m |
| benzoquinone | 25.1 m | 203 m | methanol | 31.2 s | 2.53 m |
| boron trifluoride | 69.1 s | 6.38 m | oxalic acid | 5.37 m | 37.8 m |
| carbon dioxide | 20.2 s | 1.95 m | oxirane | 83.3 s | 7.65 m |
| cytosine | 47.8 m | 268 m | pyrrole | 5.23 m | 35.5 m |
| dimethyl ether | 2.09 m | 12.4 m | sulfuric acid | 8.76 m | 70.6 m |
| formaldehyde | 10.2 s | 51.6 s | thiophenol | 26.0 m | 193 m |
| glycine | 6.94 m | 46.5 m | urea | 2.68 m | 17.3 m |

Considering the mixed parameters blocks significantly increase the computational time.

## S8 Fullerene HOMO-1 and LUMO+1 reshapings

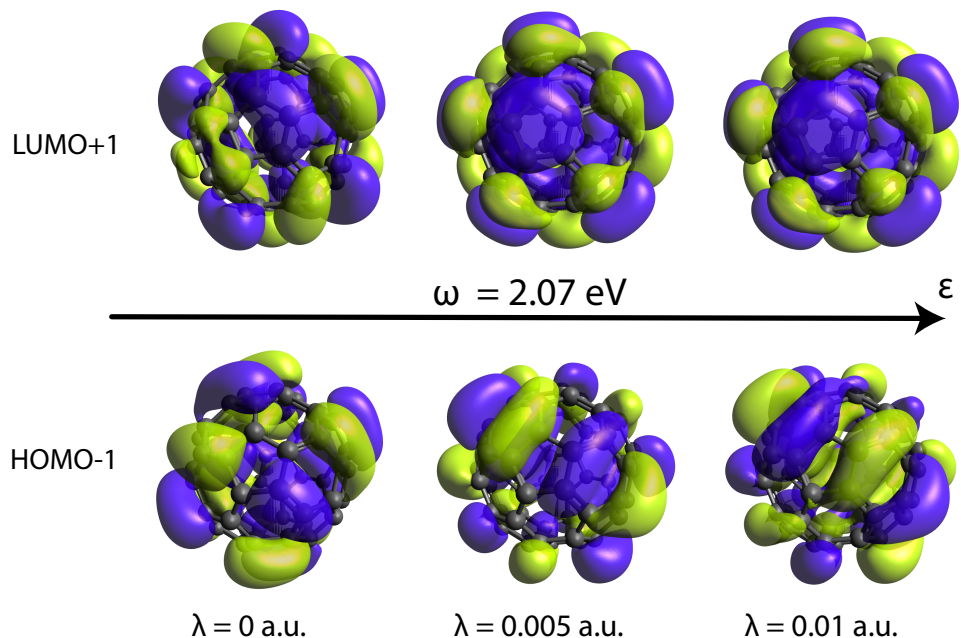

Figure S28: $C_{60}$ HOMO-1 and LUMO+1 at various couplings and frequency $\omega = 2.07$ eV.

# S9 Molecular geometries

## S9.1 Benchmark molecules

Table S2: alanine

| Atom | x / Å | y / Å | z / Å |
|---|---|---|---|
| O | 1.50753958114233 | -0.9029374701238 | 0.52054030256044 |
| O | 1.12221682236161 | 0.88332274641426 | -0.78112521053633 |
| N | -1.39540561214278 | 1.15983356228106 | 0.30075703084875 |
| C | -0.74512877528783 | -0.13676621947484 | 0.37697192544942 |
| C | -1.36340268958411 | -1.24559794663811 | -0.50418028806638 |
| C | 0.71361012877607 | 0.03946686293961 | -0.02223640526031 |
| H | -0.75284178901618 | -0.49045770582094 | 1.42313486090934 |
| H | -0.83177433524497 | -2.20336160316772 | -0.38594943757755 |
| H | -2.41677553294187 | -1.40412648176633 | -0.22104454535856 |
| H | -1.33106130560111 | -0.95414217030916 | -1.56732063426856 |
| H | -2.40884688413076 | 1.05498299638853 | 0.32109140324016 |
| H | -1.14505244036518 | 1.6143346655389 | -0.57924783924883 |
| H | 2.40272283203477 | -0.75465123626146 | 0.16840883730841 |

Table S3: ammonia

| Atom | x / Å | y / Å | z / Å |
|---|---|---|---|
| N | 3.786477015e-05 | 1.529907129e-05 | 0.31539639529544 |
| H | -0.46739972094214 | -0.80950919569646 | -0.10517015927472 |
| H | -0.46742951902069 | 0.80951619955967 | -0.10517027313719 |
| H | 0.93479137519269 | -2.23029345e-05 | -0.10515596288353 |

Table S4: aniline

| Atom | x / Å | y / Å | z / Å |
|---|---|---|---|
| N | -2.39697479260141 | -0.00014792296814 | 0.00275592032993 |
| C | -1.0214390544481 | -2.940659656e-05 | 0.00120960182057 |
| C | -0.29450905196247 | 1.21072502088442 | 0.00079053568694 |
| C | -0.29430055594105 | -1.21066037017928 | 0.00081803715932 |
| C | 1.10009392661827 | 1.20342497869635 | 7.445541682e-05 |
| C | 1.10030178218867 | -1.20312019059304 | 0.00010197259921 |
| C | 1.81432545418967 | 0.00021338511241 | -0.00028619441974 |
| H | -0.83629886730104 | 2.16180798159883 | 0.00110915183523 |
| H | -0.83592412918912 | -2.16183813269173 | 0.0011582159523 |
| H | 1.63659926728504 | 2.15687311088885 | -0.00025432723956 |
| H | 1.63696612932574 | -2.15647825037689 | -0.00020462567067 |
| H | 2.90697576154526 | 0.00031134050761 | -0.00090342650369 |
| H | -2.91493638919997 | -0.86575876011242 | -0.00342178356454 |
| H | -2.91507948050948 | 0.86537721582959 | -0.00344753340211 |

Table S5: benzoquinone

| Atom | x / Å | y / Å | z / Å |
|---|---|---|---|
| O | 2.67084480116988 | -5.221129452e-05 | -0.0001191050417 |
| O | -2.67089532135356 | 5.476825017e-05 | -0.00028364430711 |
| C | 1.45178071465234 | -2.999141965e-05 | -0.00023061377989 |
| C | -1.45183088256141 | 2.983846744e-05 | 0.00010175340409 |
| C | -0.67278346054122 | -1.26952063023965 | -0.00013010974846 |
| C | 0.67268214997036 | -1.26954701176556 | -0.00019078381802 |
| C | -0.67273172117467 | 1.26954679561817 | -0.00012174526235 |
| C | 0.67273269375586 | 1.26951918333657 | -0.00018139446467 |
| H | -1.26393891328834 | -2.18958008040806 | -0.00017620771278 |
| H | -1.2638486088068 | 2.18963104105478 | -0.00016070144168 |
| H | 1.263800367527 | -2.18963024259184 | -0.00026205174719 |
| H | 1.26388818065058 | 2.18957854099213 | -0.00024539608023 |

Table S6: boron trifluoride

| Atom | x / Å | y / Å | z / Å |
|---|---|---|---|
| F | -1.14641330751968 | -0.66138524600286 | -2.500794e-08 |
| F | 1.14641332413075 | -0.66138527062785 | -2.500794e-08 |
| F | -1.307361e-08 | 1.32290139350453 | -2.498917e-08 |
| B | -3.53746e-09 | -0.00013087687382 | 7.500505e-08 |

Table S7: carbon dioxide

| Atom | x / Å | y / Å | z / Å |
|---|---|---|---|
| C | 5.81258e-09 | -6.54397e-09 | -1.039e-11 |
| O | -2.90629e-09 | 3.27198e-09 | 1.16764877610542 |
| O | -2.90629e-09 | 3.27198e-09 | -1.16764877609503 |

Table S8: cytosine

| Atom | x / Å | y / Å | z / Å |
|---|---|---|---|
| O | 2.2134 | 1.1188 | 0.0001 |
| N | -0.0595 | 0.9816 | 0.0001 |
| N | -2.4422 | 0.9214 | 0.0001 |
| N | 1.2935 | -0.9744 | 0.0000 |
| C | -1.2406 | 0.2676 | -0.0001 |
| C | -1.1643 | -1.0726 | -0.0002 |
| C | 1.2088 | 0.4056 | -0.0003 |
| C | 0.1908 | -1.6478 | 0.0003 |
| H | -0.1331 | 1.9946 | 0.0003 |
| H | -2.0223 | -1.7317 | -0.0003 |
| H | 0.2351 | -2.7487 | 0.0005 |
| H | -3.3089 | 0.3984 | 0.0003 |
| H | -2.4533 | 1.9332 | 0.0006 |

Table S9: dimethyl ether

| Atom | x / Å | y / Å | z / Å |
|---|---|---|---|
| O | 3.66128307e-06 | 0.51112290643144 | -2.999968447e-05 |
| C | 1.17081361331246 | -0.25673130886892 | 4.14728915e-06 |
| C | -1.17079227694214 | -0.25675265797246 | -2.880408448e-05 |
| H | 1.24481186848303 | -0.90869784880785 | 0.89651386167065 |
| H | 2.02852882602686 | 0.43241084109005 | -1.03905564e-06 |
| H | 1.24483537822226 | -0.90874145526379 | -0.89647192938793 |
| H | -1.24480567073375 | -0.90871761620846 | 0.89648095516092 |
| H | -1.24477566004411 | -0.90876709802071 | -0.89650515291614 |
| H | -2.02851973960768 | 0.43237423762069 | -6.203899206e-05 |

Table S10: formaldehyde

| Atom | x / Å | y / Å | z / Å |
|---|---|---|---|
| O | 2.0 | -0.56 | 0.0 |
| C | 2.866 | -0.06 | 0.0 |
| H | 3.403 | -0.37 | 0.0 |
| H | 2.866 | 0.56 | 0.0 |

Table S11: glycine

| Atom | x / Å | y / Å | z / Å |
|---|---|---|---|
| O | -1.64635674767162 | 0.69666778713316 | -0.00334182128516 |
| O | -0.61190647304033 | -1.29751144451648 | -0.00897361260389 |
| N | 1.95564054620037 | -0.04998330502742 | -0.00367325674711 |
| C | 0.73525913452538 | 0.72177975311312 | 0.00189850557701 |
| C | -0.55335072638898 | -0.09291280869991 | -0.00419744441106 |
| H | 0.70062978735425 | 1.40127983531095 | -0.86870516782063 |
| H | 0.70016139540602 | 1.38698180960039 | 0.88351855270123 |
| H | 1.97062545610749 | -0.66416637873456 | -0.81911636975786 |
| H | 1.96877634739106 | -0.67848116135667 | 0.80085392619531 |
| H | -2.42217871988366 | 0.10934591317741 | -0.00676331184784 |

Table S12: hydrazine

| Atom | x / Å | y / Å | z / Å |
|---|---|---|---|
| N | 0.70860837953212 | -0.03218959337804 | -0.07356202980306 |
| N | -0.7084990570299 | -0.08023395228274 | 0.00283009152726 |
| H | 1.08800520911457 | -0.32912410869742 | 0.82528846910161 |
| H | 1.09050966707597 | -0.66641587032257 | -0.78161491031548 |
| H | -1.09054315549919 | -0.99313749504662 | -0.26207577244996 |
| H | -1.08808104319356 | 0.6014010197274 | -0.65396584806037 |

Table S13: isobutylene

| Atom | x / Å | y / Å | z / Å |
| --- | --- | --- | --- |
| C | 4.980472053e-05 | -0.07528002952911 | -7.654786715e-05 |
| C | -1.27649632620912 | 0.72800835536229 | -6.10075971e-05 |
| C | 1.27595031191592 | 0.72902672039391 | -6.100321455e-05 |
| C | 0.00057799931762 | -1.41577573781627 | -0.00010916141183 |
| H | -1.32792121764212 | 1.38901358275792 | 0.88429815741164 |
| H | -1.32793614774829 | 1.38901964206031 | -0.88441514151027 |
| H | -2.16939711761342 | 0.08521717189449 | -5.6654597e-05 |
| H | 1.32686005517176 | 1.39008316379982 | -0.88441271879614 |
| H | 1.32684804165912 | 1.39007187625966 | 0.88429949699484 |
| H | 2.16936630029022 | 0.08695536543351 | -5.972399418e-05 |
| H | 0.93396788960593 | -1.9870598645323 | -0.00012320228475 |
| H | -0.93236959346814 | -1.98778024608422 | -0.00012249313352 |

Table S14: maleic acid

| Atom | x / Å | y / Å | z / Å |
| --- | --- | --- | --- |
| O | 1.66960729814715 | -0.47043385601196 | 1.17854048038806 |
| O | -2.79148310658494 | -0.07256962639802 | 0.14873966666856 |
| O | 2.24098824521042 | -0.20776764721045 | -0.98150571517622 |
| O | -0.9206553384183 | -1.30340927363636 | -0.04484066412311 |
| C | 0.56470300783914 | 1.21380268483585 | -0.07098532165556 |
| C | -0.76973998734556 | 1.09199943133127 | -0.06507145354673 |
| C | 1.56078543804338 | 0.08980239955934 | -0.03325034521337 |
| C | -1.45876225140215 | -0.22258974015487 | 0.0138443063211 |
| H | 1.01195465085034 | 2.20777110624459 | -0.18202154486465 |
| H | -1.40673456465497 | 1.97647897218841 | -0.13906296708194 |
| H | 2.31646347928954 | -1.1943174239794 | 1.09822378593623 |
| H | -3.17212687097405 | -0.96716702676839 | 0.18558977234764 |

Table S15: methanol

| Atom | x / Å | y / Å | z / Å |
| --- | --- | --- | --- |
| H | 0.7399715323734 | -1.38235366551991 | -0.33023236437175 |
| O | 0.62611968268962 | -1.02887478951203 | -1.22132136661531 |
| C | 0.036585032519 | 0.24528547286065 | -1.11848590460926 |
| H | 0.67769368224915 | 0.9831542858221 | -0.59457107143463 |
| H | -0.9488144908505 | 0.23292209565651 | -0.60894466780252 |
| H | -0.12645543898066 | 0.61996660069268 | -2.14134462516651 |

Table S16: oxalic acid

| Atom | x / Å | y / Å | z / Å |
| --- | --- | --- | --- |
| O | 1.27882218548722 | -1.20774224846163 | 1.482722825e-05 |
| O | -1.27882232783038 | 1.20776691042176 | -2.13683262e-06 |
| O | 1.41293623399688 | 1.04139328928934 | -7.9240558e-06 |
| O | -1.41293612949503 | -1.04136743034375 | -0.00010073019257 |
| C | 0.77071305382101 | 0.02685660029597 | 8.002118598e-05 |
| C | -0.77071311085077 | -0.026831202192 | 0.00011364806898 |
| H | 2.24781230406452 | -1.11679771160483 | -6.04933413e-05 |
| H | -2.24781220919346 | 1.11682179259513 | -0.00013721206092 |

Table S17: oxirane

| Atom | x / Å | y / Å | z / Å |
| --- | --- | --- | --- |
| O | -0.00058471720591 | 0.78395285441668 | -4.322031081e-05 |
| C | 0.73516659848268 | -0.42950894787735 | 1.21501162e-06 |
| C | -0.73446927475546 | -0.43063271496722 | 1.23329293e-06 |
| H | 1.27950649278931 | -0.65953884020187 | 0.92676766588479 |
| H | 1.27950876432482 | -0.65960557437501 | -0.92674731378622 |
| H | -1.27846504260857 | -0.66156675999871 | -0.92674308556764 |
| H | -1.27846282102687 | -0.66150001699653 | 0.92676350547533 |

Table S18: pyrrole

| Atom | x / Å | y / Å | z / Å |
| --- | --- | --- | --- |
| N | 0.0032 | -1.158 | -0.0002 |
| C | -1.1187 | -0.3708 | 0.0002 |
| C | 1.1207 | -0.3648 | 0.0001 |
| C | -0.711 | 0.9448 | -0.0002 |
| C | 0.7058 | 0.9487 | 0.0 |
| H | 0.006 | -2.1687 | -0.0003 |
| H | -2.1038 | -0.8157 | 0.0003 |
| H | 2.1083 | -0.8041 | 0.0002 |
| H | -1.3638 | 1.8064 | -0.0003 |
| H | 1.3538 | 1.8139 | 0.0 |

Table S19: sulfuric acid

| Atom | x / Å | y / Å | z / Å |
| --- | --- | --- | --- |
| S | -5.625152911e-05 | 3.206172935e-05 | 0.00233411221782 |
| O | 1.26952591394871 | -0.01615530145921 | 1.04809977400958 |
| O | -1.27215920339628 | 0.01948359622497 | 1.0449629206123 |
| O | -0.02221283777656 | 1.29786308832917 | -0.67132745181754 |
| O | 0.02372057611632 | -1.29991717076664 | -0.66716375198493 |
| H | 1.46506005138462 | 0.91826666614824 | 1.25259148394876 |
| H | -1.4676782487477 | -0.91427294020588 | 1.25250291301401 |

Table S20: thiophenol

| Atom | x / Å | y / Å | z / Å |
| --- | --- | --- | --- |
| S | -2.72866564544314 | -0.04663050980193 | -0.0013273473747 |
| C | -0.94869923052328 | 0.02285092621988 | -0.00079969870173 |
| C | -0.22744247421651 | 1.22799267170112 | -0.00065877989291 |
| C | -0.24761404088989 | -1.19558962937262 | -0.00034597420843 |
| C | 1.16982352700033 | 1.20991880782021 | -0.00010353014917 |
| C | 1.14858243982194 | -1.20271352793089 | 0.00027091857853 |
| C | 1.86626857728308 | -0.00217394824555 | 0.00038737331671 |
| H | -0.75635975989472 | 2.18487618929624 | -0.00096157150485 |
| H | -0.79731648561778 | -2.14092480415403 | -0.00051166162546 |
| H | 1.71665630536956 | 2.15695347197674 | -2.666117455e-05 |
| H | 1.67911777011471 | -2.15894015335134 | 0.00063478709336 |
| H | 2.95923625025082 | -0.01148087864139 | 0.0008602545019 |
| H | -2.94218723325511 | 1.28806138448354 | -0.0028181088587 |

Table S21: urea

| Atom | x / Å | y / Å | z / Å |
| --- | --- | --- | --- |
| O | -1.25559968098163 | -0.00081369063439 | 1.583804951e-05 |
| N | 0.70496806919806 | -1.16029577309685 | 0.0004114573546 |
| N | 0.70344025576134 | 1.16124801359111 | 0.00016757054398 |
| C | -0.03727089884835 | -1.051746228e-05 | 0.00015841205984 |
| H | 1.71480091460371 | -1.18903947966413 | -0.00043723162763 |
| H | 0.19204822255911 | -2.03078766413551 | -0.00019274388568 |
| H | 0.18937808019146 | 2.03106585740785 | -0.00016546686953 |
| H | 1.7132350375163 | 1.1913332539942 | 0.00014216437491 |

## S9.2 Large molecular systems

Table S22: fullerene ($C_{60}$)

| Atom | x / Å | y / Å | z / Å | Atom | x / Å | y / Å | z / Å |
|---|---|---|---|---|---|---|---|
| C | 3.2742 | -1.4297 | -0.062 | C | 1.104 | 0.9078 | -3.275 |
| C | 3.5714 | -0.0192 | -0.085 | C | 0.6762 | 2.9475 | -1.9021 |
| C | 3.3119 | 0.7909 | 1.0789 | C | 0.266 | 3.5283 | 0.493 |
| C | 2.7182 | -2.0292 | 1.1233 | C | -0.6344 | 2.5573 | 2.4125 |
| C | 2.7702 | -1.799 | -1.3612 | C | -0.6207 | 1.3752 | 3.2375 |
| C | 3.2507 | 0.4831 | -1.3992 | C | -1.1802 | -3.3171 | 0.6015 |
| C | 2.7332 | 2.1042 | 0.93 | C | -1.5988 | -2.0068 | 2.487 |
| C | 2.7559 | 0.1913 | 2.2642 | C | -1.104 | -0.9078 | 3.275 |
| C | 1.6574 | -2.9989 | 1.0111 | C | -1.819 | -2.3154 | -2.0245 |
| C | 2.4586 | -1.2192 | 2.2872 | C | -0.612 | -0.6637 | -3.4559 |
| C | 1.7093 | -2.7687 | -1.4734 | C | -0.3141 | 0.748 | -3.4792 |
| C | 2.7559 | -0.6159 | -2.1872 | C | -0.742 | 2.7877 | -2.1063 |
| C | 2.6725 | 1.7952 | -1.5485 | C | -1.1533 | 3.3682 | 0.2881 |
| C | 1.819 | 2.3154 | 2.0245 | C | -1.7093 | 2.7687 | 1.4734 |
| C | 2.4129 | 2.6056 | -0.3832 | C | -1.6822 | 0.4043 | 3.1256 |
| C | 1.8327 | 1.1333 | 2.8495 | C | -2.4129 | -2.6056 | 0.3832 |
| C | 1.1533 | -3.3682 | -0.2881 | C | -2.6725 | -1.7952 | 1.5485 |
| C | 0.742 | -2.7877 | 2.1063 | C | -2.7332 | -2.1042 | -0.93 |
| C | 1.2367 | -1.6887 | 2.8943 | C | -1.8327 | -1.1333 | -2.8495 |
| C | 0.6344 | -2.5573 | -2.4125 | C | -1.2367 | 1.6887 | -2.8943 |
| C | 1.6822 | -0.4043 | -3.1256 | C | -1.6574 | 2.9989 | -1.0111 |
| C | 1.5988 | 2.0068 | -2.487 | C | -2.7702 | 1.799 | 1.3612 |
| C | 0.5863 | 3.027 | 1.8062 | C | -2.7559 | 0.6159 | 2.1872 |
| C | 1.1802 | 3.3171 | -0.6015 | C | -3.2507 | -0.4831 | 1.3992 |
| C | 0.612 | 0.6637 | 3.4559 | C | -3.3119 | -0.7909 | -1.0789 |
| C | -0.266 | -3.5283 | -0.493 | C | -2.7559 | -0.1913 | -2.2642 |
| C | -0.6762 | -2.9475 | 1.9021 | C | -2.4586 | 1.2192 | -2.2872 |
| C | 0.3141 | -0.748 | 3.4792 | C | -2.7182 | 2.0292 | -1.1233 |
| C | -0.5863 | -3.027 | -1.8062 | C | -3.2742 | 1.4297 | 0.06200 |
| C | 0.6207 | -1.3752 | -3.2375 | C | -3.5714 | 0.0192 | 0.08500 |

Table S23: heme group + proximal histidine + oxygen (part 1)

| Atom | x / Å | y / Å | z / Å | Atom | x / Å | y / Å | z / Å |
| --- | --- | --- | --- | --- | --- | --- | --- |
| C | 0.949 | -0.672 | 2.746 | C | -5.064 | 0.527 | 2.188 |
| C | 3.001 | -0.026 | -0.864 | Fe | 0.000 | 0.000 | 0.000 |
| C | -0.728 | 1.063 | -2.683 | N | 1.323 | -0.382 | 1.472 |
| C | -2.887 | 0.191 | 0.819 | N | 1.683 | 0.205 | -1.128 |
| C | 2.112 | -0.964 | 3.564 | N | -1.139 | 0.720 | -1.413 |
| C | 3.854 | 0.323 | -1.985 | N | -1.582 | -0.078 | 1.101 |
| C | -1.861 | 1.466 | -3.502 | O | 0.230 | 1.577 | 0.449 |
| C | -3.701 | -0.049 | 1.964 | O | 3.715 | -3.147 | 7.419 |
| C | 3.175 | -0.898 | 2.724 | O | -4.964 | -0.317 | 6.611 |
| C | 3.043 | 0.761 | -3.001 | O | -0.080 | 2.755 | 0.461 |
| C | -2.984 | 1.359 | -2.742 | O | 1.625 | -3.510 | 6.985 |
| C | -2.897 | -0.431 | 2.951 | O | -3.229 | 0.602 | 7.548 |
| C | 2.670 | -0.576 | 1.449 | C | -3.434 | -6.471 | -2.220 |
| C | 1.702 | 0.668 | -2.443 | C | -2.458 | -5.739 | -1.297 |
| C | -2.539 | 0.890 | -1.465 | C | -1.476 | -4.938 | -2.117 |
| C | -1.553 | -0.434 | 2.442 | C | -1.085 | -2.512 | -1.284 |
| C | 2.046 | -1.695 | 4.893 | C | 0.646 | -2.838 | 0.073 |
| C | 3.279 | 0.897 | -4.500 | C | -0.775 | -3.834 | -1.314 |
| C | -4.398 | 1.734 | -3.083 | N | -1.678 | -6.660 | -0.420 |
| C | -3.357 | -0.622 | 4.368 | N | 0.293 | -4.064 | -0.458 |
| C | 2.937 | -2.893 | 5.152 | N | -0.206 | -1.875 | -0.439 |
| C | 4.540 | 0.931 | -5.105 | O | -4.517 | -5.941 | -2.435 |
| C | -4.759 | 3.038 | -3.338 | H | 1.037 | -2.032 | 5.009 |
| C | -3.371 | 0.660 | 5.191 | H | 2.409 | -0.967 | 5.588 |
| C | 2.763 | -3.227 | 6.631 | H | 2.768 | 0.070 | -4.948 |
| C | -3.900 | 0.294 | 6.560 | H | 2.956 | 1.904 | -4.660 |
| C | -0.396 | -0.670 | 3.217 | H | -5.117 | 0.995 | -3.127 |
| C | 3.470 | -0.472 | 0.347 | H | -2.701 | -1.321 | 4.844 |
| C | 0.587 | 1.044 | -3.154 | H | -4.367 | -0.972 | 4.322 |
| C | -3.377 | 0.641 | -0.417 | H | 3.958 | -2.651 | 4.942 |
| C | 4.644 | -0.985 | 2.991 | H | 2.675 | -3.721 | 4.527 |
| C | 5.317 | -0.011 | -1.992 | H | 5.001 | -0.031 | -5.024 |
| C | -1.804 | 1.605 | -4.990 | H | 5.151 | 1.661 | -4.616 |

Table S24: heme group + proximal histidine + oxygen (part 2)

| Atom | x / Å | y / Å | z / Å |
|------|-------|-------|-------|
| H | 4.433 | 1.189 | -6.138 |
| H | -4.050 | 3.787 | -3.297 |
| H | -5.738 | 3.266 | -3.570 |
| H | -4.009 | 1.386 | 4.731 |
| H | -2.391 | 1.084 | 5.259 |
| H | -0.538 | -0.859 | 4.221 |
| H | 4.464 | -0.738 | 0.427 |
| H | 0.736 | 1.341 | -4.131 |
| H | -4.392 | 0.787 | -0.536 |
| H | 5.180 | -0.890 | 2.070 |
| H | 4.870 | -1.930 | 3.439 |
| H | 4.933 | -0.197 | 3.655 |
| H | 5.757 | 0.332 | -2.905 |
| H | 5.441 | -1.071 | -1.912 |
| H | 5.796 | 0.468 | -1.164 |
| H | -2.761 | 1.912 | -5.357 |
| H | -1.540 | 0.664 | -5.426 |
| H | -1.070 | 2.338 | -5.252 |
| H | -5.514 | 0.756 | 1.245 |
| H | -4.982 | 1.421 | 2.770 |
| H | -5.670 | -0.184 | 2.710 |
| H | 4.569 | -2.927 | 7.102 |
| H | -2.424 | 1.069 | 7.440 |
| H | -3.056 | -5.108 | -0.673 |
| H | -0.732 | -5.604 | -2.501 |
| H | -2.030 | -4.460 | -2.898 |
| H | -1.100 | -7.271 | -0.997 |
| H | -2.320 | -7.225 | 0.136 |
| H | 0.717 | -4.932 | -0.266 |
| H | 1.416 | -2.669 | 0.739 |
| H | -1.852 | -2.061 | -1.807 |
| O | -3.255 | -7.537 | -2.712 |
| H | -2.747 | -8.355 | -2.711 |



\end{document}